\documentclass[journal=jcisd8, manuscript=article]{achemso}
\usepackage{chemformula} % Formula subscripts using \ch{}
\usepackage[T1]{fontenc} % Use modern font encodings

\usepackage{float} % used for H placement
\usepackage{caption} 
\usepackage{geometry}
\usepackage{natbib}
\usepackage{setspace}
\usepackage{xkeyval}

\usepackage{array}
\usepackage{booktabs}
\usepackage{listings}
\usepackage{lmodern}
\usepackage{microtype}

\usepackage{graphicx,dblfloatfix} % Added to fit table in page
\usepackage{tabularx} % Add tables per column, also wrap text in tables
\newcolumntype{b}{X}
\newcolumntype{j}{>{\hsize=1.3\hsize}X}
\newcolumntype{s}{>{\hsize=.6\hsize}X}
\newcolumntype{m}{>{\hsize=.4\hsize}X}
\usepackage{multicol}
\usepackage{cleveref}
\setkeys{acs}{articletitle = true}

\usepackage{subcaption} % for subfigures
\usepackage{mwe} % for subfigures
\usepackage{multirow}
\usepackage{multicol}
\usepackage{booktabs}
\usepackage{xr}
\usepackage{color}
\usepackage[export]{adjustbox}

\newcommand*{\thead}[1]{%
\multicolumn{1}{l}{\begin{tabular}{@{}l@{}}#1\end{tabular}}}

\author{Maritza Hernandez}
\affiliation[1QBit]{\footnotesize{1QB Information Technologies (1QBit), 458-550 Burrard Street, Vancouver, BC, V6C 2B5, Canada}}
\email{maritza.hernandez@1qbit.com}
\author{Guo Liang Gan}
\affiliation[1QBit]{\footnotesize{1QB Information Technologies (1QBit), 458-550 Burrard Street, Vancouver, BC, V6C 2B5, Canada}}
\alsoaffiliation{\footnotesize{School of Computing Science, Simon Fraser University, 8888 University Drive, \\ Vancouver, BC, V5A 1S6, Canada}}
\author{Kirby Linvill}
\author{Carl Dukatz}
\affiliation[Accenture Labs]{\footnotesize{Accenture Labs, Accenture PLC, 415 Mission Street, Suite 3300, San Francisco, CA, 94105, USA}}
\author{\\Jun Feng}
\author{Govinda Bhisetti}
\affiliation[Biogen]{\footnotesize{Biogen, 225 Binney Street, Cambridge, MA, 02142, USA}}

\title{A Quantum-Inspired Method for Three-Dimensional Ligand-Based Virtual Screening}

\begin{document}

%%% ==========================================================================
%%%								ABSTRACT
%%% ==========================================================================

\begin{abstract}
Measuring similarity between molecules is an important part of virtual screening (VS) experiments deployed during the early stages of drug discovery. Most widely used methods for evaluating the similarity of molecules use molecular fingerprints to encode structural information. While similarity methods using fingerprint encodings are efficient, they do not consider all the relevant aspects of molecular structure. In this paper, we describe a quantum-inspired \emph{graph-based molecular similarity} (GMS) method for ligand-based VS. The GMS method is formulated as a quadratic unconstrained binary optimization problem that can be solved using a quantum annealer, providing the opportunity to take advantage of this nascent and potentially groundbreaking technology. 
In this study, we consider various features relevant to ligand-based VS, such as pharmacophore features and three-dimensional atomic coordinates, and include them in the GMS method. We evaluate this approach on various datasets from the \texttt{DUD\_LIB\_VS\_1.0} library. 
Our results show that using three-dimensional atomic coordinates as features for comparison yields higher early enrichment values. In addition, we evaluate the performance of the GMS method against conventional fingerprint approaches. The results demonstrate that the GMS method outperforms fingerprint methods for most of the datasets, presenting a new alternative in ligand-based VS with the potential for future enhancement. 
\end{abstract}

%%% ==========================================================================
%%%								INTRODUCTION
%%% ==========================================================================
\section{Introduction}
\label{sec:introduction}
The continued need for the development of innovative new medicines faces major challenges due to the increasing costs of drug development and the high failure rate of potential drug candidates. For example, only $19\%$ of new drugs that enter clinical trials eventually get FDA approval~\cite{FISHER2015322}. Thus, there remains high demand for innovation and opportunities to improve the drug discovery process. In recent years, the VS of small molecules has become a routine and integral part of the drug discovery process.

In silico VS is a critical, early phase drug discovery approach that enables the identification of potential candidate molecules from a large molecular database, in a high-throughput manner. A number of high-throughput screening (HTS) methods have been developed that offer alternative or complementary strategies for VS. They can be divided into two categories: structure based and ligand based. Structure-based approaches, such as docking algorithms, are based on 3D structural information of the protein target, and a scoring function is used to measure how well a ligand binds to the active site~\cite{Tuccinardi2009}. In situations where structural information of the binding site is unknown, ligand-based approaches such as similarity searching and pharmacophore mapping are utilized for VS~\cite{Willett2006}. The conventional molecular representation used for ligand-based VS methods is based on 2D fingerprint representations. A \textit{fingerprint} representation is a binary vector in which each entry indicates the presence or absence of a substructure in a given molecule. In general, they are computationally inexpensive and simple to use; however, they do not consider all of the relevant molecular features. One of the most widely used fingerprint representations is extended-connectivity fingerprints (ECFP)~\cite{Rogers2010}. ECFPs are circular fingerprints constructed by encoding an atom's neighbourhood while expanding iteratively into adjacent neighbourhoods until a given number of iterations is reached. 

Alternative molecular representations are intended to characterize the 3D nature of molecules~\cite{Rush2005}. These methods are based either on molecular shape or pharmacophore shape alignment. Molecular shape methods aim to measure similarity based on the maximum degree of overlap of molecular shapes (or volumes). Examples of algorithms that use a shape-based approach are OpenEye Scientific's rapid overlay of chemical structures (ROCS)~\cite{Hawkins2007, rocs} and Schr\"{o}dinger's Phase~\cite{Dixon2006, phase}. Graph matching algorithms have also been considered for the molecular similarity problem. In a molecular graph representation, atoms and bonds are represented by nodes and edges, respectively. In similarity searching applications, graph matching algorithms are commonly based on optimization problems, for instance, the maximum common subgraph (MCS) problem~\cite{Raymond2002-MCS} or the optimal assignment problem~\cite{Jahn2009}. Three-dimensional approaches take into account conformational properties; hence, they are more computationally expensive than 2D approaches. More recently, there has been a growing interest in applying machine learning methods to ligand-based VS. One example is the molecular graph convolutions~\cite{Kearnes2016} machine learning architecture. This method represents molecules as graphs in a deep learning system. Although it does not outperform all fingerprint-based methods, the authors introduce an alternative ligand-based VS method that is still being explored.

Hernandez et al.~\cite{Hernandez2016} proposed a graph-based molecular similarity approach that can be implemented with the use of a quantum annealer~\cite{Johnson2011}. The algorithm finds the maximum weighted common subgraph (MWCS) of two molecular graphs. The main difference between their algorithm and previous methods is that it finds the MWCS by solving the maximum weighted co-$k$-plex problem of an induced graph. This method has been reported to yield improvements in accuracy over conventional fingerprint methods when predicting mutagenicity in small molecules. The maximum co-$k$-plex problem is, in general, NP-hard, which means the time to solve the problem increases exponentially with the number of variables~\cite{Balasundaram2009}. Quantum annealing is a promising approach, with the potential to harness quantum mechanical effects, to solve hard optimization problems: it may be able to address the co-$k$-plex problem and, consequently, the molecular similarity problem more effectively than classical approaches. A study assessing the performance of a quantum annealer in solving the molecular similarity problem was performed by Hernandez and Aramon~\cite{Hernandez2017}, and provides useful insights into new techniques for using the quantum annealer and addressing some of its hardware limitations. 

The purpose of this study is to extend the applicability of the GMS approach introduced by Hernandez~et~al.~\cite{Hernandez2016} to ligand-based VS experiments. In this paper, we study how various molecular features, including 3D coordinates, affect similarity score and, therefore, the performance of the VS experiments. The GMS methodology described in this paper has been implemented in the 1QBit SDK~\cite{QDK}. The performance of VS, using the GMS method, is compared against conventional fingerprint methods. Previous studies comparing various methods for VS have, in general, concluded that ligand-based 2D methods tend to yield better performance than docking methods, and that 2D fingerprint approaches generally outperform 3D-shape-based methods~\cite{Scior2012,Guoping2012}. Remarkably, the 3D approach implemented in the GMS method generally outperforms its 2D counterpart for the 13 target proteins used in this study. 

%%% ==========================================================================
%%%								METHODS
%%% ==========================================================================

\section{Similarity Methods}
\label{sec:methods}
\begin{figure*}[ht]
\begin{center}
\includegraphics[width=6.5 in]{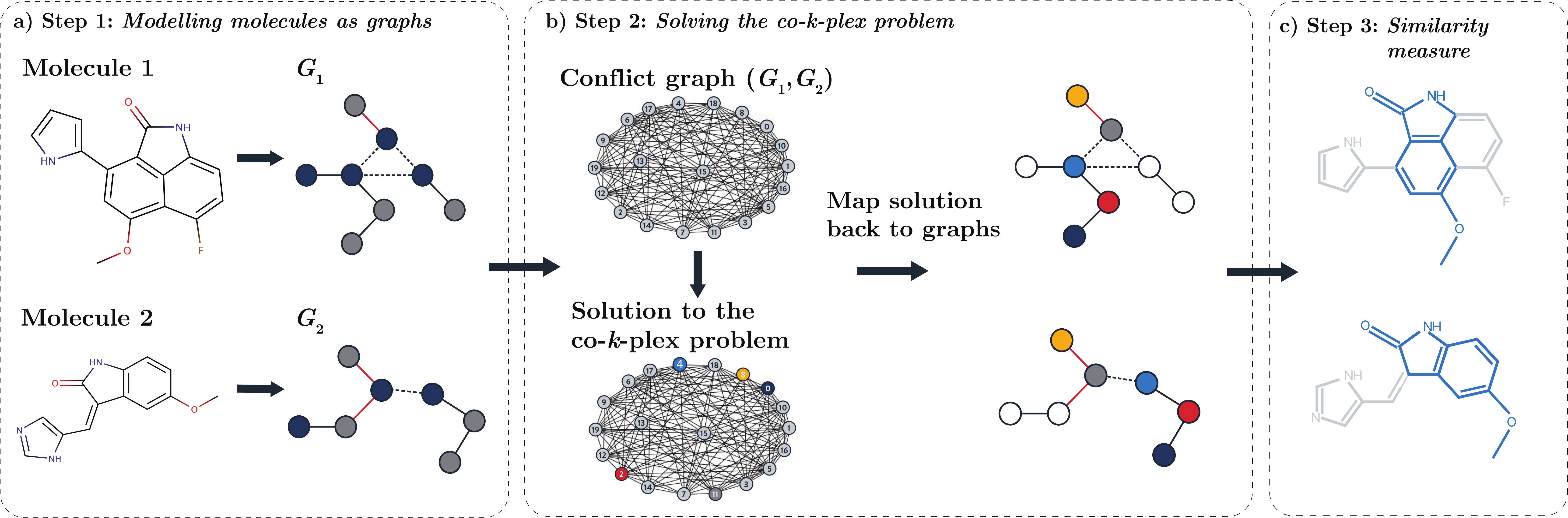}
\caption{Illustration of the GMS method: a)~Two molecules modelled as graphs. b)~A conflict graph is built, the co-$k$-plex problem of the conflict graph is solved, and the solution is mapped back to the molecules. c)~The similarity score is calculated.}
\label{fig:molsim}
\end{center}
\end{figure*}
Similarity is a fundamental concept in the field of chemoinformatics. Developing methods to evaluate a measure of similarity is, however, a difficult task due to the nature of the concept of similarity: ``like beauty, it is in the eye of the beholder''~\cite{Maggiora2004}. The criteria used to define similarity can vary accross applications. For instance, organic chemists may be interested in classifying molecules in terms of the core scaffold fragment and its substructures, whereas physical chemists may focus on physicochemical properties, such as excluded volume and electrostatic properties. The similarity method introduced by Hernandez~et~al.~\cite{Hernandez2016} compares molecules according to chemical descriptors. In this work, we have modified the criteria for considering two molecules as similar by including weighted pharmacophore features. Additionally, the new similarity criteria allow the inclusion of \emph{partial} matches. In this section, we describe how features and their relevance are incorporated in the GMS method.

%%=====================================================================
\subsection*{The Graph-Based Molecular Similarity Method} 
\label{subsec:gms_method}
The GMS method consists of three steps: 1)~modelling molecules as graphs; 2) solving the co-$k$-plex problem; and 3) calculating a coefficient that measures the similarity between the two input molecules. We describe these three steps in an earlier paper~\cite{Hernandez2016}. In the sections that follow, we present an overview of these processes with a special focus on the variations implemented in this study. A general scheme of this method is shown in Fig.~\ref{fig:molsim}. 
%%========================
\subsubsection*{Step 1: Modelling Molecules as Graphs}
\label{subsec:molsim}
\begin{figure}[H]
\begin{center}
\includegraphics[width=2.3 in]{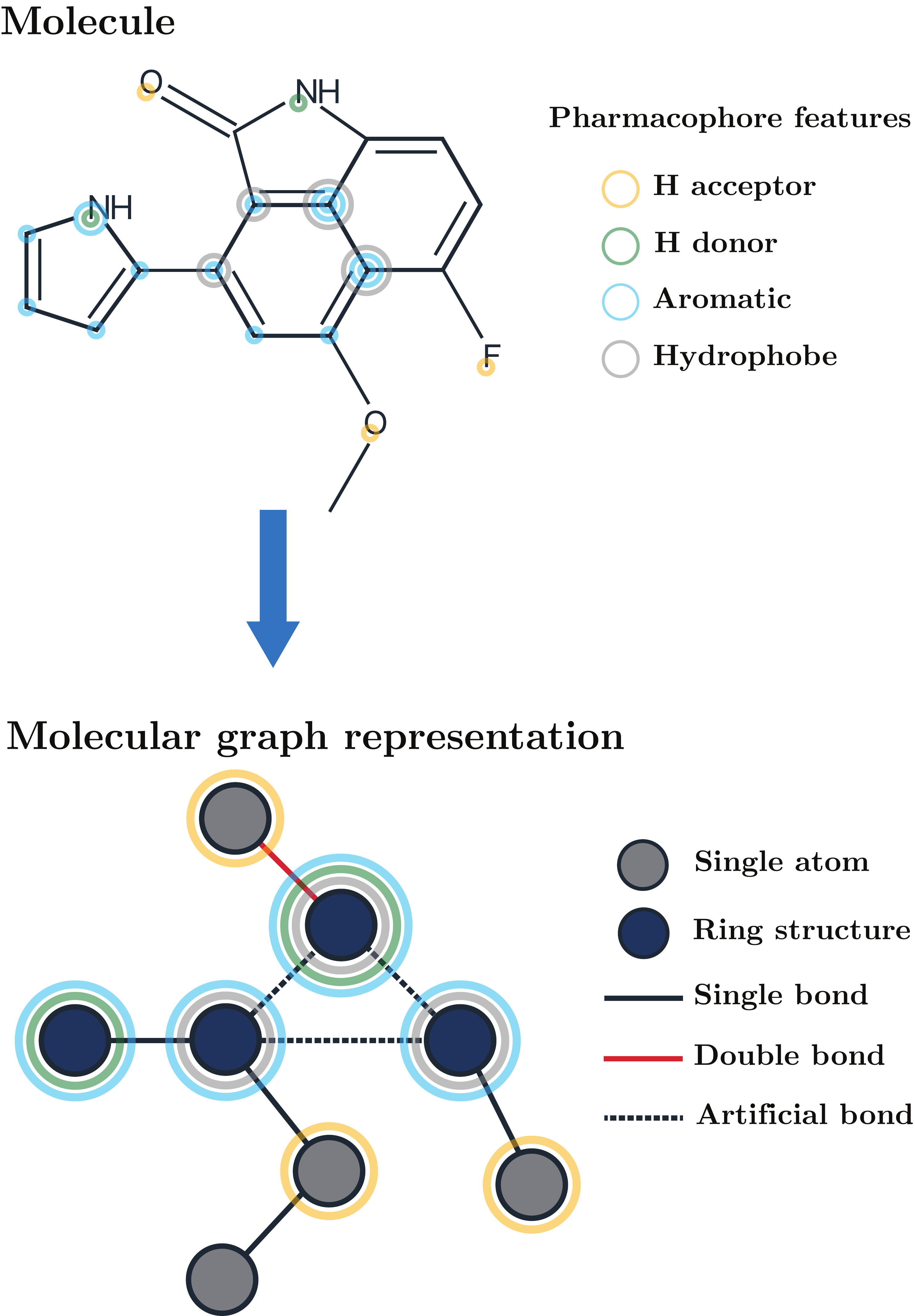}
\caption{ 
Modelling a molecule as a graph. Individual atoms and ring structures are mapped to individual vertices in the graph. Two ring structures that share atoms are represented by two vertices connected by an edge labelled an \textit{artificial bond}. Molecular features are generated using \mbox{RDKit} and stored in the respective label of each vertex. In this example, we consider pharmacophore features. A label is also generated for bonds, indicating whether the edge represents an artificial link, or a single, double, or triple covalent bond.}
\label{fig:mol2graph}
\end{center}
\end{figure}

The first step of the GMS method is to model molecules as graphs. Individual atoms and ring structures in the molecule are represented by individual vertices in the graph. Here, atoms connected cyclically are referred to as ring structures, without differentiating whether they are aromatic. Bonds connecting a pair of atoms or rings are represented by edges that connect the respective pairs of vertices. If two vertices representing rings share one or more atoms, an edge between these vertices is then added and labelled an \emph{artificial bond}, emphasizing that it is not intended to represent a natural chemical bond. An illustration of a molecular graph representation is shown in Fig.~\ref{fig:mol2graph}.

Formally, let $G=(V,E,\mathcal{L}_{V},\mathcal{L}_{E})$ be a labelled graph representing a molecule, where $V$ is the set of vertices, $E$ is the set of edges,  $\mathcal{L}_{V}$ is the set of labels assigned to each vertex, and $\mathcal{L}_{E}$ is the set of edge labels. Each label encodes a specific property or feature of an atom, ring, or bond. The set of features considered in this work is summarized in Table~\ref{table:mol_features}. It is evident that not every feature carries the same relevance; hence, we use weights to represent the relevance of each feature, which can be determined by an expert in the field. The weighting schemes used in this work are shown in Table~\ref{table:weights}. All features have been generated using RDKit~\cite{rdkit}. 

%%========================
\subsubsection{Step 2: Solving the Maximum Co-$k$-Plex Problem}
\begin{figure*}[ht]
\begin{center}
\includegraphics[width=5 in]{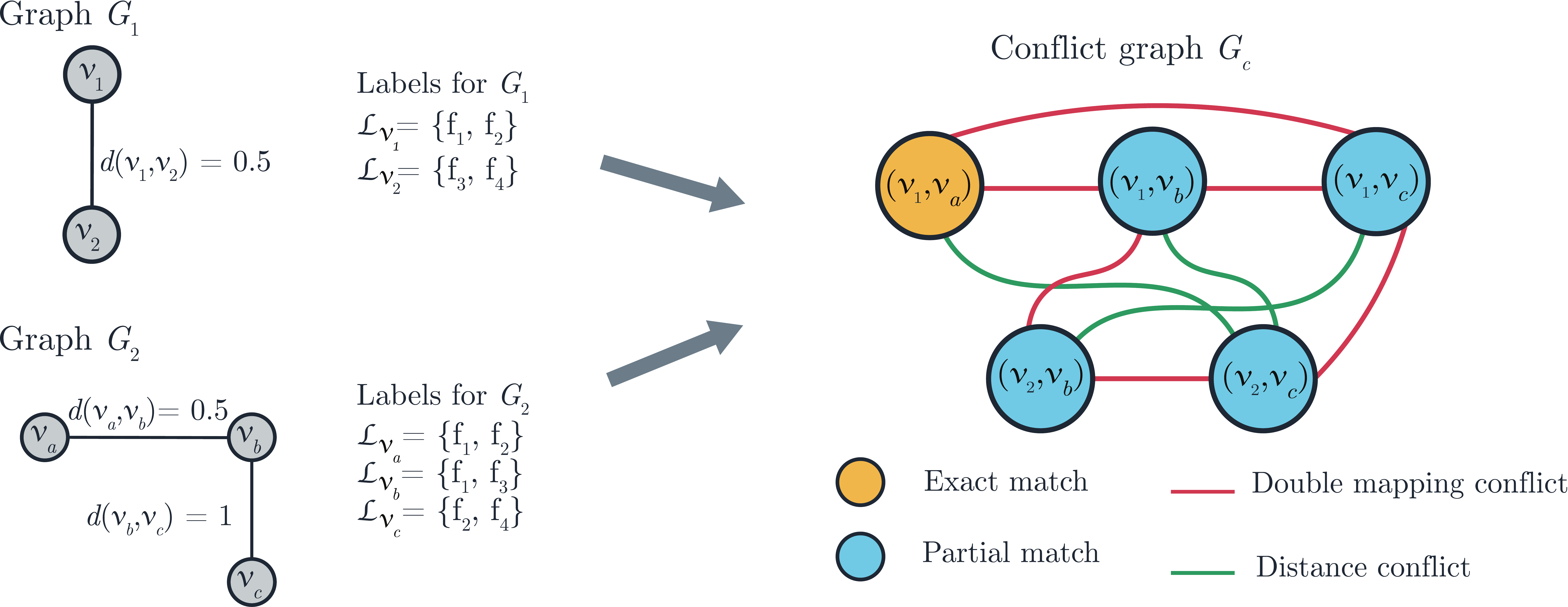}
\caption{Illustration of the conflict graph construction $G_{c}$ from two given graphs $G_1$ and $G_2$. The vertices $v_1$ from $G_1$ and $v_a$ from $G_2$ are added as a vertex $(v_1,v_a)$ in $G_{c}$ since their set of labels $L_{v_{1}}$ and $L_{v_{a}}$ match. The rest of the vertices in the conflict graph are added in the same way. Edges are added according to two conditions: bijective mapping and distance violations. Bijective mapping is violated if one of the nodes has been matched twice (represented by a red edge). Distance violation aims to incorporate 3D molecular information (represented by a blue edge). An edge between two vertices (e.g., between $(v_1,v_b)$ and $(v_2,v_c)$) is added if the distance between $v_1$ and $v_2$ is not comparable to the distance between $v_b$ and $v_c$. Formally, an edge is added if $|d(v_1,v_2) - d(v_b,v_c) | >  \epsilon $ ($\epsilon=0.4$ in this example). }
\label{fig:conflict_graph}
\end{center}
\end{figure*}

Given two molecular graphs, we are interested in finding the maximum co-$k$-plex of a third graph. This third graph is called a \textit{conflict graph} and can be induced from the graphs being compared. Its construction is illustrated in Fig.~\ref{fig:conflict_graph}. 

\begin{table}
\centering
\begin{tabularx}{\textwidth}{ s@{\hskip 0.2in} b} 
\toprule
Feature & Description \\
\addlinespace[0.1cm]
\midrule 
Atomic number & Atomic number for individual atoms or set of atomic numbers for atoms in ring structures  \\
\addlinespace[0.1cm]
Implicit hydrogen & Total number of implicit hydrogen bonds associated with the atom or atoms in the ring \\
\addlinespace[0.1cm]
Formal charge & Formal charge of individual atoms \\
\addlinespace[0.1cm]
Degree & Number of edges (bonds) incident to a vertex. The degree does not depend on the bond order, but on whether hydrogens atoms are explicit in the graph\\
\addlinespace[0.1cm]
Bond order & List of bonds incident to an atom or ring, distinguishing the covalent bond order (single, double, or triple) \\
\addlinespace[0.1cm]
3D coordinates & 3D vector indicating the position of an atom or the geometrical centre of a ring \\
\addlinespace[0.1cm] 
Pharmacophore features & Indicate the pharmacophore features present in an atom or ring. The features considered are: H acceptor, H donor, acidic, basic, aromatic, hydrophobic, and zinc binder \\
\addlinespace[0.1cm]
\bottomrule
\end{tabularx}
\caption{Set of features of each node representing atoms or rings}
\label{table:mol_features}
\end{table}

\textit{Conflict Graph.}
Formally, in a conflict graph, vertices represent possible mappings (or matchings), and edges represent conflicts between vertices. The notions of matches and conflicts are based on the criteria used to define similarity. Our algorithm allows the incorporation of different similarity criteria. For example, one criterion can be that two nodes match if they have the same atomic number; another criterion can be that two nodes match if they have the same pharmacophore feature, regardless of whether they are the same element. When comparing two molecular graphs to form a conflict graph, our algorithm compares only atoms to atoms and rings to rings. In addition, we allow both exact and partial matching. The term ``partial match'' refers to the matching of two nodes, where at least one of their labels matches exactly. An example of such matches is shown in Fig.~\ref{fig:conflict_graph}. 
In determining partial matches, we may wish to consider only a subset of matching features. To that end, we introduce the concept of criticality for each of the features described in Table~\ref{table:mol_features}. One is able to designate a feature as being ``critical'' or ``non-critical'' based on screening requirements. When comparing two nodes, if both nodes contain at least one matching feature marked as critical, they are included in the conflict graph. In the event that no feature is marked as critical, but both sets of features are an exact match, they will still be included in the conflict graph.

 \textit{Maximum co-$k$-plex solution.}
The objective of the maximum weighted co-$k$-plex problem is to identify the largest weighted set of vertices such that there are at most $k-1$ edges in the conflict graph. In this work, we set $k=1$, that is, the maximum weighted co-$1$-plex problem is equivalent to the maximum weighted independent set. The solution to the maximum weighted co-$1$-plex problem is a binary vector whose size is equivalent to the number of nodes in the conflict graph. The non-zeros in the solution vector identify a pair of vertices from the molecular graph. With this information, the solution can be mapped to the original molecular graphs to identify the common substructures. 

The maximum weighted co-$k$-plex problem is formulated as a quadratic unconstrained binary optimization (QUBO) problem~\cite{Hernandez2016}. Having the problem formulated as a QUBO problem  allows us to use a quantum annealer; however, it is not our aim in this paper to study the annealer's performance. Our focus is on investigating the performance of the GMS method in the context of ligand-based VS. In the ``Results and Discussion'' section we discuss in more detail the use of quantum annealers in ligand-based VS experiments. The solutions to the maximum weighted \mbox{co-$1$-plex} problems in this paper have been obtained using the classical heuristic parallel tempering Monte Carlo with isoenergetic cluster moves (PTICM) solver, also known as the ``borealis'' algorithm~\cite{Zhu2016}. The parameters used for the solver are given in Section S2 of the Supporting Information document.
%Section~\Cref{sec:solver} 
%%========================
\subsubsection{Step 3: Similarity Measure}
\label{similarity_measure}

There exists a large number of similarity measures in the literature for graph-based molecular similarity methods. These measures are formulated in terms of subset relations between  two graphs being compared and their common substructure. The similarity metric used in this work is based on the convex combination of two existing similarity measures---Bunke and Shearer, and asymmetric~\cite{Raymond2002}. In addition, we incorporate the information regarding the weights of each molecular feature. Given two molecules $A$ and $B$, we denote the set of features of molecule $A$ by $F_A$ and the set of features of molecule $B$ by $F_B$, and associate each feature $f$ with a weight $w_f$. After solving the maximum weighted co-$1$-plex problem, we recover the common features between molecules $A$ and $B$ and denote the set of common features by $F_{AB}$. Hence, the similarity score is defined by
\begin{equation}
\text{Sim}(A,B) = \delta \min(\alpha, \beta) + (1-\delta) \max(\alpha, \beta), \\
\end{equation}
with $ \delta \in [0,1]$,  
\begin{equation}
\alpha = \frac{\displaystyle \sum_{f \in F_{AB}} w_f}{\displaystyle \sum_{f \in F_A} w_f}, \hspace{1cm} \text{and} \hspace{1cm} \beta = \frac{\displaystyle \sum_{f \in F_{AB}}w_f}{\displaystyle \sum_{f \in F_B} w_f}.
\end{equation}
%%========================

%%% ==========================================================================
%%%								EXPERIMENTAL PROCEDURES
%%% ==========================================================================

\section{Experimental Procedures}

In order to perform an experimental validation of the GMS method for ligand-based VS, we have adopted standardized experimental procedures. This involves selecting an appropriate dataset and performance metric for a ligand-based VS approach. The experiments reported in this paper and the dataset used are based on the work of Jahn~et~al.~\cite{Jahn2009}, in which they tested the optimal assignment approach on molecular graphs as a ligand-based VS method. To evaluate the performance of VS using the GMS application, we report four metrics: the receiver operating characteristics (ROC) enrichment (ROCE), the area under the curve (AUC) of ROC, and their arithmetic weighted versions: awROCE and awAUC. In the sections that follow, we present further details regarding the dataset and performance evaluation used in our work. 

%%%========================
\subsection{Dataset}
%\label{}

The chemical library used for screening is a modified version of the Directory of Useful Decoys (DUD): Release~2~\cite{Mysinger2012, Huang2006}, a standard dataset for benchmarking virtual screening, which contains a set of active structures for 40 target proteins. For each active compound, there are 36 inactive structures referred to as ``decoys''. Decoys have similar physical properties but a dissimilar topology. We conduct experiments on the 13 target classes reported by Jahn~et~al.~\cite{Jahn2009}. The original DUD dataset is not suitable for ligand-based VS experiments, as it was designed for the evaluation of docking techniques. Good and Oprea~\cite{Good2008} suggested and modified the active structures of the dataset using a lead-like filter and a clustering algorithm, aiming to reduce retrieval bias due to the presence of analogous structures. 

The set of decoys was modified in a similar way by Jahn et al.~\cite{Jahn2009} to reduce the introduction of artificial enrichment due to their having similar physical property values. As these actives and decoys contain only topological information, the researchers generated 3D atomic coordinates for each structure initially using CORINA~\cite{Gasteiger1990}, then optimized using MacroModel 9.6~\cite{MacroModel}. We use the same modified set of actives and decoys in our experiments. 

\begin{table*}[ht]
\centering
%\begin{tabular}{|| c | c | c | c | c | c ||}
\begin{tabular}{ l  l  l  l  l  l }
\toprule
\thead{Target \\class} & \thead{Number of \\ actives }& \thead{Number of \\ decoys} & \thead{Number of \\ clusters}  & \thead{Target \\ PDB code} & \thead{Query \\ PubChem CID}\\
\addlinespace[0.1cm]
\midrule
ACE         & 46         & 1796     & 18 & 1o86 & 5362119  \\
AChE     & 100     & 3859     & 18 & 1eve & 3152 \\
CDK2     & 47         & 2070     & 32 & 1ckp & 448991  \\
COX-2     & 212     & 12606     & 44 & 1cx2 & 1396  \\
EGFr     & 365     & 15560     & 40 & 1m17 & 176870 \\
FXa         & 63         & 2092     & 19 & 1f0r & 445480 \\
HIVRT     & 34         & 1494     & 17 & 1rt1 & 65013  \\
InhA         & 57         & 2707     & 23 & 1p44 & 447767 \\
P38 MAP     & 137     & 6779     & 20 & 1kv2 & 156422  \\
PDE5     & 26         & 1698     & 22 & 1xp0 & 110634 \\
PDGFrb     & 124     & 5603     & 22 & 1t46 & 5291 \\
Src     & 98         & 5679     & 21 & 2src & 44462678 \\
VEGFr-2     & 48         & 2712     & 31 & 1fgi & 5289418 \\
\addlinespace[0.1cm]
\bottomrule
\end{tabular}
\caption{The number of actives, decoys, and clusters for each target class, target RCSB Protein Data Bank (PDB) code, and PubChem Compound ID number (CID) for each search query of 13 target classes. ACE: angiotensin-converting enzyme; AChE: acetylcholinesterase; CDK2: cyclin-dependent kinase 2; COX-2: cyclooxygenase-2; EGFr: epidermal growth factor receptor; FXa: factor Xa; HIVRT: HIV reverse transcriptase; InhA: Enoyl-acyl carrier protein reductase; P38 MAP: P38 mitogen-activated protein; PDE5: phosphodiesterase type 5; PDGFrb: platelet-derived growth factor receptor beta; Src: protein-tyrosine kinase; VEGFr-2: vascular endothelial growth factor receptor 2.}

\label{table:data}
\end{table*}
In order to screen for actives, ligand-based VS experiments require a query---a known biologically active structure that reacts with a target protein. Previous works utilized bound ligands of complexed crystal structure, extracted directly from the RCSB Protein Data Bank (PDB)~\cite{Jahn2009,Cheeseright2008,Berman2000}. In situations where ligand-based VS is performed, the conformation adopted by a ligand upon binding to a receptor is often unknown; therefore, it seems advisable to test our ligand-based VS method using ligands with conformations generated by standard conformer models. In this work, we retrieve the 3D conformer of the query ligand from the \mbox{PubChem} website~\cite{pubchem}. The PubChem repository~\cite{Bolton2011} generates a 3D conformer model that represents all possible biologically relevant conformations for a given molecule. Table~\ref{table:data} presents the 13 target classes; the number of actives, decoys, and clusters for each target class; the PDB code of the complexed crystal structure which contains the ligand; and the PubChem Compound ID number (CID) for the query. For all molecules in our dataset, we use RDKit to generate the molecular information. Note that for the target class \textit{FXa}, there is a total of 64 actives, but RDKit is unable to generate one of the molecules.

%%========================
\subsection{Fingerprint Methods}
%\label{}
To assess whether the GMS method has an advantage over traditional similarity methods in ligand-based VS experiments, we evaluate the performance of molecular fingerprint methods. Generally, there are two types of 2D fingerprints: dictionary-based and hash-based fingerprint methods~\cite{Khanna2010}. Dictionary-based methods have a fixed number of bits and each bit represents a certain type of feature of a substructure. Unlike dictionary-based fingerprints, hash-based fingerprints can be used to encode any new types of substructure features. 

In this work, we select one of the most widely used fingerprints from each category. The dictionary-based fingerprint selected is the Molecular ACCess System (MACCS), which has 166 bits. The MACCS fingerprint is calculated using RDKit. The hash-based fingerprint is called a circular fingerprint, also known as an ECFP (extended-connectivity fingerprint). It uses a fixed number of iterations to generate identifiers for each atom based on its neighbours. At each iteration, only neighbours connected through a certain number of bonds are considered. After there are no further iterations, all atomic identifiers are collected, and all duplicated identifiers are removed. The resulting set of identifiers is then converted to a bit string with a hash function. In order to consider ECFP fingerprints, we use Morgan fingerprints generated by RDKit. Morgan fingerprints are built by applying the Morgan algorithm, whose default atom invariants use similar connectivity information as the one used for ECFP fingerprints. We also consider feature-based invariants information, similar to that used for functional-class fingerprints (FCFP). Landrum~\cite{FP_RDKit} presents details regarding the use of Morgan fingerprints as being equivalent to ECFPs/FCFPs.

%%========================
\subsection{Experimental Setup}
\label{subsubsec:exp_setup}
As discussed in previous sections, our algorithm operates on a set of molecular features as defined in Table~\ref{table:mol_features}, where each feature is assigned a criticality and a weighting value. Details on the use of criticality and weighting values are given in the section ``Similarity Methods''.

In order to gain a comprehensive understanding of the effect of the features considered in this work, we generate 12 criticality schemes (CS) and 10 weighting schemes (WS). A particular configuration of CS and WS is referred to as a \textit{similarity criteria setting}. The detailed similarity criteria settings used in this work are shown in Section S1 of the Supporting Information document. 
In both~\Cref{table:crit,table:weights}, we present three selected CSs and WSs, respectively. 
% Section~\Cref{sec:similarity_criteria} 

The CSs presented in Table~\ref{table:crit} consider the features ``Atomic number'' and ``Pharmacophore features'' critical, that is, two nodes are considered a match if at least one of these features match. The rest of the features are considered non-critical. Three-dimensional coordinates are considered in CS$_7$ and CS$_9$, whereas CS$_3$ does not consider this feature. The only difference between CS$_7$ and CS$_9$ is in how the feature that strictly compares atoms in the rings is set. CS$_7$ imposes a critical constraint on ring comparison, whereas this condition is not imposed in CS$_9$.

\begin{table}
\centering
\begin{tabularx}{0.7\textwidth}{j@{\hskip 0.2in} m m m }
%\begin{tabularx}{\columnwidth}{j@{\hskip 0.2in} m m m }
\toprule 
Feature & CS$_{3}$ & CS$_{7}$ & CS$_{9}$  \\
\addlinespace[0.1cm]
\midrule
Atomic number (single atom) & C & C & C \\
\addlinespace[0.1cm]
Atomic number (ring)  & OFF & C & OFF \\
\addlinespace[0.1cm]
Implicit hydrogens & NC & NC & NC \\
\addlinespace[0.1cm]
3D coordinates & OFF & ON & ON \\
\addlinespace[0.1cm]
\bottomrule
\end{tabularx}
\caption{For each criticality scheme, we determine whether a feature is considered critical (C), non-critical (NC), or is ignored (OFF). In this table we show three selected criticality schemes as used in the VS experiment. ``Atomic number'' and ``Pharmacophore features''  are considered C (i.e., two nodes are considered a match if at least one of these features match), and the rest of the features are considered NC. The  additional criticality schemes used in this work are detailed in Section S1 of the Supporting Information document.} %~\Cref{sec:similarity_criteria}
\label{table:crit}
\end{table}

We consider a baseline weighting scheme WS (WS$_{B}$) where each feature is weighted equally; hence, WS$_{B}$ acts a control. Also interested in studying the relevance of assigning a higher weight to rings than to individual atoms, we introduce WS$_{B-5}$, where rings are given a weight of 5, which is proportional to the average number of atoms in a typical ring structure in our dataset. The rest of the WSs used in this work are selected based on previous knowledge; they have not been optimized. 
\begin{table}
\centering
\begin{tabularx}{0.7\textwidth}{j@{\hskip 0.3in} m  m  m}
%\begin{tabularx}{\columnwidth}{j@{\hskip 0.3in} m  m  m}
\toprule
Features & WS$_{B}$ & WS$_{B-5}$ & WS$_{4}$ \\
\addlinespace[0.1cm]
\midrule
Atom & 1 & 1 & 0.1 \\
Ring & 1 & 5 & 0.1 \\
Implicit hydrogens & 1 & 1 & 0.1 \\
Formal charge & 1 & 1 & 0.1 \\
Bond order & 1 & 1 & 0.1 \\
Degree & 1 & 1 & 0.1 \\
\addlinespace[0.1cm]
Basic & 1 & 1 & 3 \\
Acidic & 1 & 1 & 3 \\
H donor & 1 & 1 & 2 \\
H acceptor & 1 & 1 & 2 \\
Aromatic & 1 & 1 & 2 \\
Hydrophobic & 1 & 1 & 1 \\
Zinc binder & 1 & 1 & 3 \\
\addlinespace[0.1cm]
\bottomrule
\end{tabularx}
\caption{For each weighting scheme, we assign a weight to each feature. Three selected weighting schemes as used in the VS experiment are shown. Additional weighting schemes considered in this work are presented in Section S1 of the Supporting Information document.}%~\Cref{sec:similarity_criteria}
\label{table:weights}
\end{table}
For example, to set weighting scheme WS$_{4}$, we consider that there are various types of non-covalent interactions between a ligand and receptor, like a salt bridge interaction, a hydrogen bond, an aromatic interaction, and a hydrophobic interaction. Generally, a salt bridge interaction has the highest interaction energy, followed by a hydrogen bond, an aromatic interaction, and a hydrophobic interaction. Therefore, we assign the highest weight to basic and acidic pharmacophore features; less weight to the hydrogen bond donors, acceptors, and aromatic centres; and the lowest weight to hydrophobic groups. In Table~\ref{table:weights}, we present these three selected WSs. 

The following VS experimental procedure is performed for all similarity criteria settings: for each target class, a query is obtained from the \mbox{PubChem} database and compared against the respective set of active and decoy structures using the GMS method. For each comparison, we obtain the similarity score. The similarity scores between the query and the active and decoy molecules are then sorted to produce a ranking of molecules. Based on the ranking of molecules, we compute the VS performance measures awROCE, ROCE, awAUC, and AUC.

%%========================
\subsection{Performance Evaluation}
%\label{}

Enrichment has traditionally been the standard measure used to characterize the performance of VS methods. It can be defined as ``the ratio of the observed fraction of active compounds in the top few percent of a virtual screen to that expected by random selection''~\cite{Jain2008}. The enrichment factor is, however, considered a poor performance measure because it depends on an extrinsic variable, that is, the ratio of active to decoy molecules. To address this issue, Jain and Nicholls~\cite{Jain2008} suggested two alternative measures. One is AUC, which is commonly used in other fields such as machine learning. Formally, AUC is defined as 
\begin{equation}
\text{AUC} = 1 - \frac{1}{N_{\text{actives}}} \sum_{i=1}^{N_{\text{ actives}}} \frac{N_{\text{decoys seen}}^{i}}{N_{\text{ decoys}}}\,,
\end{equation}
where $N_{\text{ actives}}$ and $N_{\text{ decoys}}$ is the number of actives and decoys in the dataset, respectively, and $N_\text{decoys seen}^{i}$ is the number of decoy molecules that are ranked higher than the $i$-th active structure in the ranking list. 

AUC values are a global measure that considers the whole dataset and, therefore, does not represent the concept of ``early enrichment''. Hence, the other measure (ROCE) incorporates the concept. Specifically, it reports the ratio of true positive rates to false positive rates at different enrichment percentages. For example, ``enrichment at $1\%$'' is the ratio of actives observed with the highest $1\%$ of known decoys (multiplied by 100). The ROC enrichment for a false positive rate (FPR) of  $x\%$ is given by the expression
\begin{equation}
\text{ROCE @ x\% } =  \frac{  \frac{N_\text{actives selected}^\text{x\%} }{ N_\text{ actives}} }{ \frac{N_\text{decoys selected}^\text{x\%} }{N_\text{ decoys}} }\,,
\end{equation}
where $ N_\text{actives selected}^\text{x\%}$ and $N_\text{decoys selected}^\text{x\%}$ is the number of actives (true positives) and decoys (false positives) retrieved in the range containing a false positive rate of $x\%$, respectively. The enrichment percentages used in this work are 0.5\%, 1.0\%, 2.0\%, and 5.0\%. 

Another important aspect to consider when evaluating the performance of a VS method is the retrieval of new scaffolds~\cite{Jahn2009}. Ligand-based methods could generate artificially higher enrichment results when the dataset contains analogous structures. To reduce this bias induced by structurally similar structures, Clark and Webster-Clark~\cite{Clark2008} proposed two weighted schemes for the standard ROC and AUC calculations: the \textit{harmonic weighted scheme} and the \textit{arithmetic weighted scheme}. In our work, we implement the arithmetic weighting scheme as done by Jahn~et~al.~\cite{Jahn2009}, denoted by awROCE. In order to use this metric, the actives in the dataset used for the VS experiment need to be clustered according to analogous structures. awRoce is given by the expression 
\begin{equation}
\text{awROCE @ x\%} = \frac{ \frac{ \sum\limits_{i}^{N_\text{c}} w_{i} A_{i} ^{x\%} }{N_\text{c}} }{ \frac{N_\text{decoys selected} ^{x\%}}{N_\text{ decoys}} }\,,
\label{awroce}
\end{equation}
where $N_{\text{c}}$ is the number of clusters, $w_{i} = \frac{1}{N_i}$ is the weight of each structure in the $i$-th cluster with $N_i$ molecules, and $A_i^{x\%}$ is the total number of active structures retrieved from the $i$-th cluster at $x\%$ of the FPR.

Likewise, the arithmetic weighting scheme for AUC is denoted by awAUC and given by the equation
\begin{equation}
\text{awAUC} = 1 - \frac{1}{N_\text{c}} \sum_{j=1}^{N_\text{c}} \sum_{i}^{N_{j}} \frac{N_\text{decoys seen}^{ij}}{N_\text{ decoys}}\,,
\end{equation}
where $N_\text{decoys seen}^{ij}$ is the number of decoys retrieved earlier than the $i$-th active molecule in cluster $j$, and $N_j$ is the number of molecules in cluster $j$.

%%% ==========================================================================
%%%								RESULTS AND DISCUSSION
%%% ==========================================================================
\section{Results and Discussion}
\label{sec:results}

The performance of the GMS method was evaluated for each combination of $12$ CSs and $10$ WSs. The extensive set of results for the GMS method is presented in Sections S3 and S4 of the Supporting Information document. 

In this section, we present the results for a selected set of similarity criteria settings of the GMS method that are representative of our analysis. Specifically, we include nine similarity criteria settings which consist of CS$_{3}$, CS$_{7}$, and CS$_{9}$, each of them with the WSs WS$_{B}$, WS$_{B-5}$, and WS$_{4}$. %Sections~\Cref{sec:overall} and~\Cref{sec:results_classes}

\begin{table}[ht]
\centering
\begin{tabular}{l l l l l l l l l l l}
\toprule
\multirow{2}{*}{\raisebox{-\heavyrulewidth}{VS Performance }}
	& \multicolumn{3}{c}{CS$_3$} & \multicolumn{3}{c}{ CS$_7$}  & \multicolumn{3}{c}{ CS$_9$}  \\   \cmidrule{2-10} 
	&  WS$_4$ & WS$_B$ & WS$_{B-5}$	 &      WS$_4$ & WS$_B$ & WS$_{B-5}$	  &      WS$_4$ & WS$_B$ & WS$_{B-5}$\\ 
	\midrule
\addlinespace[0.1cm]
%auc
AUC 				& 0.52 	& 0.50	& 0.52	& 0.62	& 0.60	& 0.65	& 0.62	& 0.62	&\textbf{0.67}	 \\
\addlinespace[0.05cm]
awAUC  				& 0.48  	& 0.48	& 0.50	& 0.59	& 0.58	&0.61	& 0.59	&0.59	&\textbf{0.65}	 \\
\addlinespace[0.15cm]
%roce
ROCE $0.5 \%$ 		& 32.11 	& 32.18	&34.60	&50.99	&43.46	&48.83	&49.10	&42.93	&\textbf{50.93}	 \\
\addlinespace[0.05cm]
ROCE $1 \%$ 			& 18.97 	& 18.64	&20.28	&27.84	&24.95	&27.52	&27.45	&25.04	&\textbf{29.03}	 \\
\addlinespace[0.05cm]
ROCE $2 \%$ 			& 10.64 	& 10.58	& 11.63	& 15.25	&13.97	&15.20	&14.95	&14.11	&\textbf{15.57}	 \\
\addlinespace[0.05cm]
ROCE $5 \%$ 			& 5.74	&  4.83	& 5.35	&6.86	&6.63	&\textbf{7.39}	&6.74	&6.61	&7.02	\\
\addlinespace[0.15cm]
% awroce
awROCE 	$0.5 \%$ 		& 23.78 	& 28.32	& 27.71	&39.96	&36.94	&38.68	&38.67	&33.88	&\textbf{43.39}	 \\
\addlinespace[0.05cm]
awROCE $1 \%$ 		& 14.57 	& 16.51	&17.19	&22.01	&21.32	&22.79	&21.74	&22.80	&\textbf{26.10}	 \\
\addlinespace[0.05cm]
awROCE 	$2 \%$ 		& 8.33 	& 9.34	&9.46	&12.68	&12.38	&13.42	&12.24	&13.08	&\textbf{14.06}	 \\
\addlinespace[0.05cm]
awROCE $5 \%$ 		& 4.67 	& 4.36	&4.64	&6.06	&6.22	&\textbf{6.60}	&5.94	&6.15	& 6.36	 \\
\addlinespace[0.1cm]
\bottomrule
\end{tabular}
\caption{Ligand-based VS performance of GMS methods. We show the mean awAUC, mean AUC, mean awROCE, and mean ROCE at 0.5$\%$, 1.0$\%$, 2.0$\%$, and 5.0$\%$ over the 13 targets.}
\label{table:overal}
\end{table}
In Table~\ref{table:overal}, we present the mean values for awAUC, awROCE, AUC, and ROCE over all 13 target classes shown in Table~\ref{table:data}. The mean awROCE and mean ROCE are reported with decoy rates of 0.5\%, 1.0\%, 2.0\%, and 5.0\%. As  mentioned earlier, awAUC and awROCE were introduced in order to reduce the inflated enrichments by considering structurally analogous molecules in the dataset. The similarity criterion CS$_{9}$WS$_{B-5}$ yielded the highest value for each of the metrics considered in this study. In particular, we observe that similarity criteria with WS$_{B-5}$ generally result in higher scores than similarity criteria with the baseline WS$_B$, suggesting it would be advisable to assign higher weights to rings. 

\begin{table*}[ht]
\centering
\begin{tabular}{ l  l  l  l  l  l }
	\toprule
	\multirow{2}{*}{  \raisebox{-\heavyrulewidth}{Target Class }  } & \multicolumn{2}{c}{GMS} & \multicolumn{2}{c}{Morgan Fingerprint}  \\  \cmidrule{2-5}
	& ROCE~$0.5 \%$ & awROCE~$0.5 \%$ & ROCE~$0.5 \%$ & awROCE~$0.5 \%$ \\ 
	\midrule
\addlinespace[0.1cm]
ACE 		& \textbf{52.06} 		& 58.53 	& 47.72	& \textbf{63.02}		\\
AChE 	&\textbf{46.78}	&\textbf{24.06}	&42.88	&21.16	\\
CDK2 	&\textbf{36.03}	&\textbf{29.99}	&20.02	&9.41		\\
COX-2 	&\textbf{154.2}	&\textbf{103}	&103.13	&38.55		\\
EGFr 		&\textbf{135}	&\textbf{126.3}	&133.36	&123.88		\\
FXa 		&6.038	&20.02	&\textbf{18.11}	&\textbf{30.7}		 \\
HIVRT 		&\textbf{43.94}	&\textbf{32.96}	&38.45	&31.12		\\
InhA 		&74.63	&55.11	&\textbf{88.2}	&\textbf{66.32}	\\
P38 MAP 		&\textbf{27.65}	&\textbf{9.969}	&13.1	&4.72		 \\
PDE5 	&\textbf{29.03}	&\textbf{10.72}	&21.77	&8.55		\\
PDGFrb 	&4.67	&3.764	&\textbf{17.14}	&\textbf{43.91}		\\
Src 		&\textbf{39.96}	&\textbf{80.22}	&25.98	&9.33		\\
VEGFr-2 	&\textbf{12.11}	&\textbf{9.37}	&8.07	&6.25		 \\
\addlinespace[0.1cm]
\bottomrule
\end{tabular}
\caption{Ligand-based VS performance comparison of the GMS and Morgan fingerprint methods. We show ROCE and awROCE at $0.5\%$ enrichment for each target. The similarity criterion selected for the GMS method is CS$_9$WS$_{B-5}$, and the settings for the Morgan fingerprint method are having 2048 bits, a radius of 4, and feature-based invariants set to ``False''.}
\label{table:target_results}
\end{table*}

%%% ================================================================================
\subsection{Comparison against Other Methods}

We also evaluated the performance of various fingerprint methods with the objective of comparing GMS against the most common method used for VS experiments. Specifically, we evaluated MACCS and Morgan fingerprints. In the case of Morgan fingerprints, we considered two numbers of bits, 1024 and 2048;  four different radii, 1, 2, 3, and 4; and we kept the default option for the atom-based invariant feature.  In summary, we evaluated 16 variations of Morgan fingerprints. The overall performance for fingerprint methods is presented in~\Cref{tab:overall_fingerprint} in the Supporting Information document. Additional sets of results for each target class are detailed in Section S4. %~\Cref{sec:results_classes}

\begin{figure*}[h!]
%\centering
	% figure A
	\begin{subfigure}[b]{0.46\textwidth}
	%\begin{subfigure}[b]{\columnwidth}
	\centering
	\includegraphics[width=0.85\columnwidth]{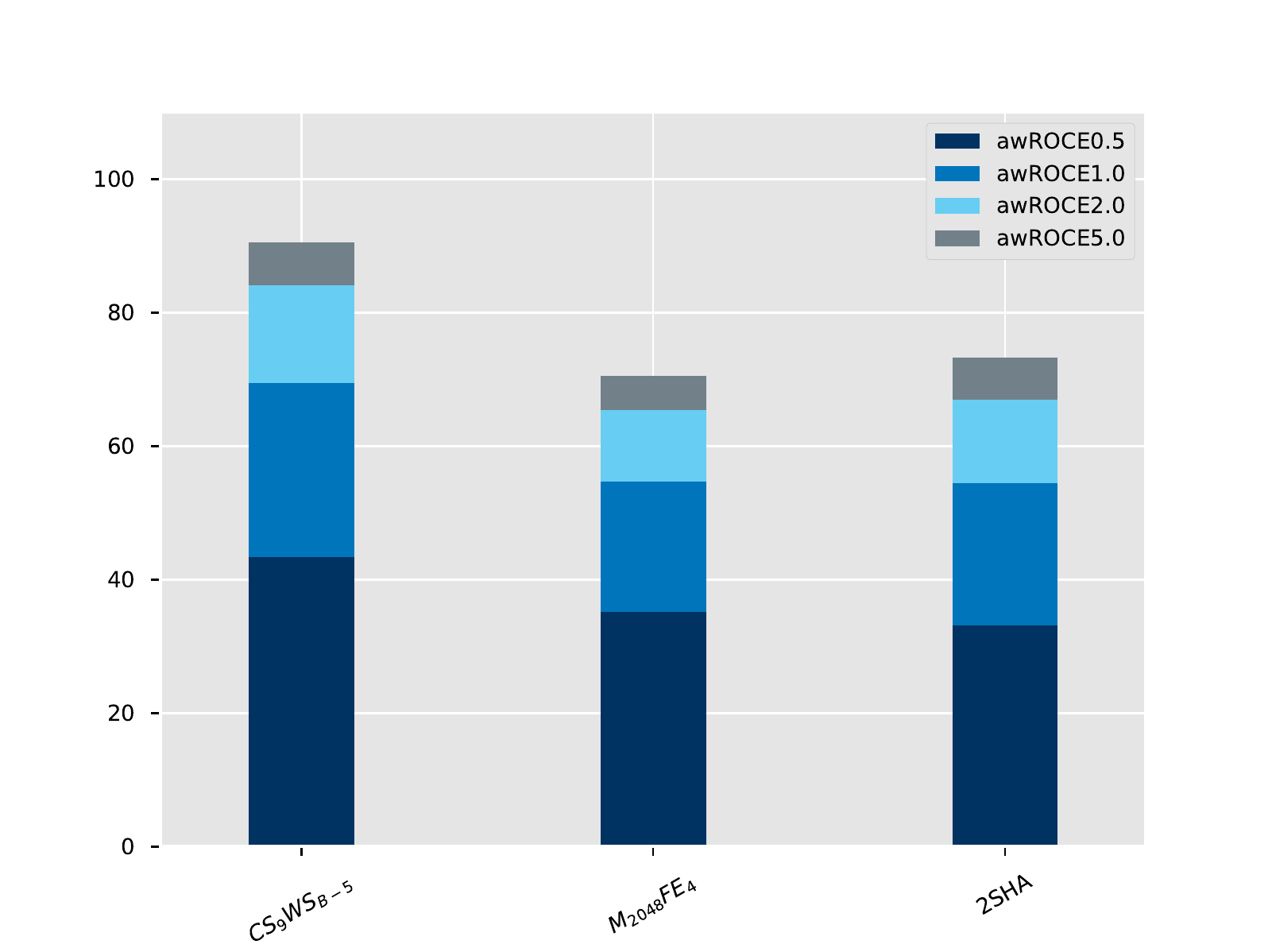}
	\caption{\small Mean awROCE} 
	\label{fig:mean_awroce}
        \end{subfigure}
        \quad
        	% figure B
	\begin{subfigure}[b]{0.46\textwidth}
	%\begin{subfigure}[b]{\columnwidth}
	\centering
	\includegraphics[width=0.85\columnwidth]{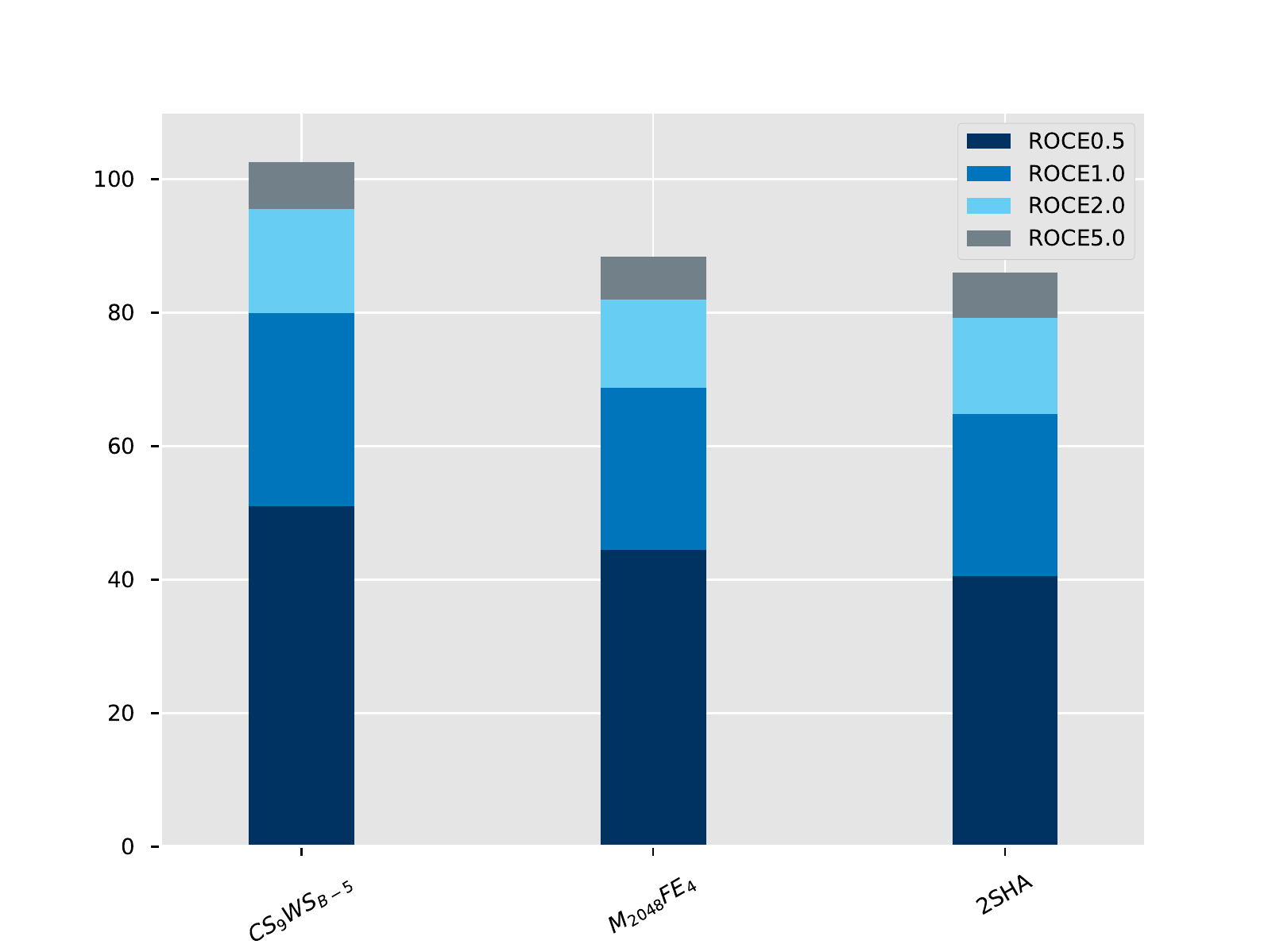}
	\caption{\small Mean ROCE} 
	\label{fig:mean_roce}
        \end{subfigure}
        \caption[]
        {\small Ligand-based VS performance comparison of the GMS, Morgan fingerprint, and optimal assignment methods.
        (a) Mean awROCE over the 13 targets. (b) Mean ROCE over the 13 targets. Each was measured at $0.5\%, 1\%, 2\%$, and $5\%$ enrichment.} 
        \label{fig:overal_comparison}
    \end{figure*}

Additionally, we considered the results of the optimal assignment methods presented by Jahn~et~al.~\cite{Jahn2009}. There are two main reasons for selecting these methods. First, optimal assignment methods act on molecular graphs; and second, the ligand-based VS results presented in this paper report awROCE as well as ROCE. The results for the optimal assignment methods were retrieved from their supplementary material, which is publicly available. We should note that their results were calculated with query molecules retrieved from the protein data bank and the authors corrected the bond lengths. The results using the optimal method could vary if the query were to have a different conformation. In Fig.~\ref{fig:overal_comparison}, we show the mean awROCE and the mean ROCE at four percentage values for one variation for each of the GMS, Morgan fingerprint, and optimal assignment methods. The selected variation we report for each method is the one with highest mean value of awROCE at $0.5\%$. The Morgan fingerprint method, with 2048 bits, a radius of 4, and atom-based invariants, was selected from among the various fingerprint methods. Among the optimal assignment methods reported by Jahn~et~al.~\cite{Jahn2009}, a two-step hierarchical assignment approach (2SHA) was selected. From among the different criticality and weighting schemes of the GMS method, CS$_9$WS$_{B-5}$ was selected. Overall, the GMS method yielded higher enrichment values than the fingerprint and optimal assignment methods. 
\Cref{table:target_results} shows the ROCE and awROCE at $0.5\%$ enrichment for each of the 13 targets for the GMS and fingerprint methods.

\textit{Src---Tyrosine kinase.}
In addition to the overall results already presented, we would like to highlight the results for the target kinase \mbox{c-Src} with the GMS method against fingerprint methods. In Table~\ref{table:src}, we present the results for a GMS method and a Morgan fingerprint method. The GMS method yielded  higher scores than the Morgan fingerprint method, in particular, awROCE at early enrichment at $0.5\%$. The awROCE metric was introduced to reduce the bias produced by comparing analogous structures. This high score of $80.22$ indicates that the GMS method is able to distinguish the subtle conformational difference of the inhibitor for the \mbox{c-Src} kinase. Although a number of kinase inhibitors have been developed, designing a highly selective kinase inhibitor remains a challenge. We anticipate that our approach can help in the design of such selective inhibitors. 

\begin{table}[H]
\centering
\begin{tabularx}{250pt}{b@{\hskip 0.5in} m m }
\toprule 
\thead{Performance \\ Metric} & \thead{GMS } & \thead{Morgan \\ Fingerprint} \\
\addlinespace[0.05cm]
\midrule
%roce
ROCE $0.5 \%$ 		& 39.96 & 29.97 \\
\addlinespace[0.05cm]
ROCE $1 \%$ 			& 23.38 & 15.24 \\
\addlinespace[0.05cm]
ROCE $2 \%$ 			& 13.21 & 9.14 \\
\addlinespace[0.05cm]
ROCE $5 \%$ 			& 6.32 &  4.69\\
\addlinespace[0.3cm]
% awroce
awROCE $0.5 \%$ 		& 80.22 & 9.99 \\
\addlinespace[0.05cm]
awROCE $1 \%$ 		& 44.93 & 5.08 \\
\addlinespace[0.05cm]
awROCE $2 \%$ 		& 23.42 & 2.79 \\
\addlinespace[0.05cm]
awROCE $5 \%$ 		& 9.57 & 2.21 \\
\addlinespace[0.05cm]
\bottomrule
\end{tabularx}
\caption{VS performance comparison of the GMS method with the similarity criterion CS$_9$WS$_{B-5}$, and the Morgan fingerprint method with 2048 bits, a radius of 3, and feature-based invariants set to ``False'', on the Src dataset.}
\label{table:src}
\end{table}

%%% ================================================================================
\subsection{Impact of Three-Dimensional Coordinates as a Matching Feature}

Previous studies comparing the performance of 2D versus 3D molecular similarity methods~\cite{Guoping2012} have shown that 2D methods outperform 3D methods both in terms of achieving early enrichment and in computational time. In the GMS application, we have introduced a simple approach to using the 3D coordinates of atoms. From our VS experimentation, we observed that CSs with the 3D feature set to ``ON'' yield better results than CSs with the 2D feature set to ``ON''. More specifically, the mean awROCE values at 0.5\% ranged from 21.9 to 30.02 for CS$_{1}$ to CS$_{6}$, and from 31.11 to 43.39 for CS$_{7}$ to CS$_{12}$. The same was true for the mean ROCE values at 0.5\%, whose values ranged from 30.37 to 37.66 for CS$_{1}$ to CS$_{6}$, and from 42.18 to 50.99 for CS$_{7}$ to CS$_{12}$. 
\begin{figure}[h]
%\centering
	% figure A
	% trim={<left> <lower> <right> <upper>}
	\begin{subfigure}[b]{0.375\textwidth}
	\centering
	\adjincludegraphics[width=\textwidth,]{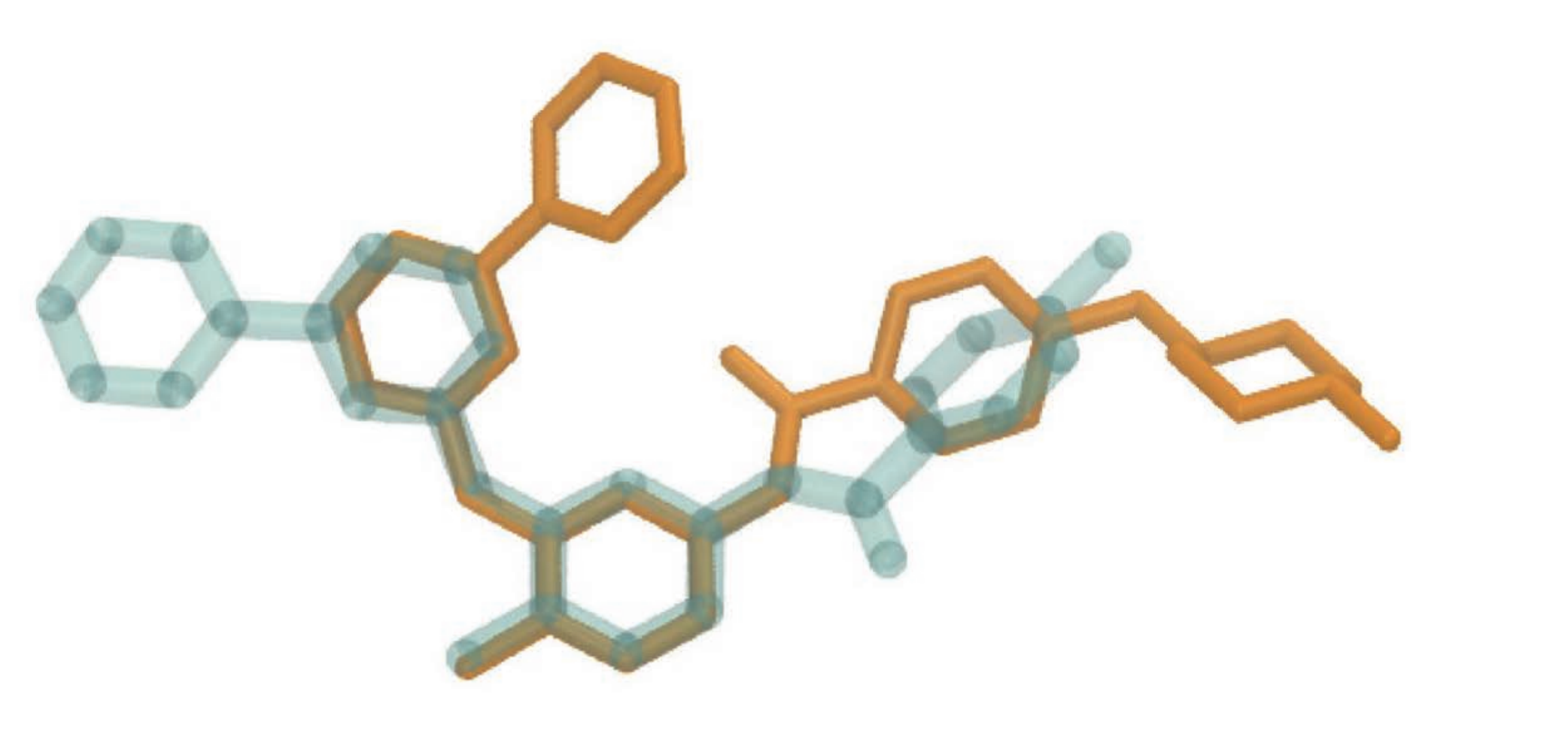}
	\caption{\small Overlap between query PDGFrb and ZINC03832219 molecules, with original conformers. Similarity score = 0.615.} 
	\label{overlap_original}
        \end{subfigure}
        \qquad
        %\vskip\baselineskip
%        \hfill
        % figure B
        \begin{subfigure}[b]{0.375\textwidth}  
	\centering 
	\includegraphics[width=\textwidth]{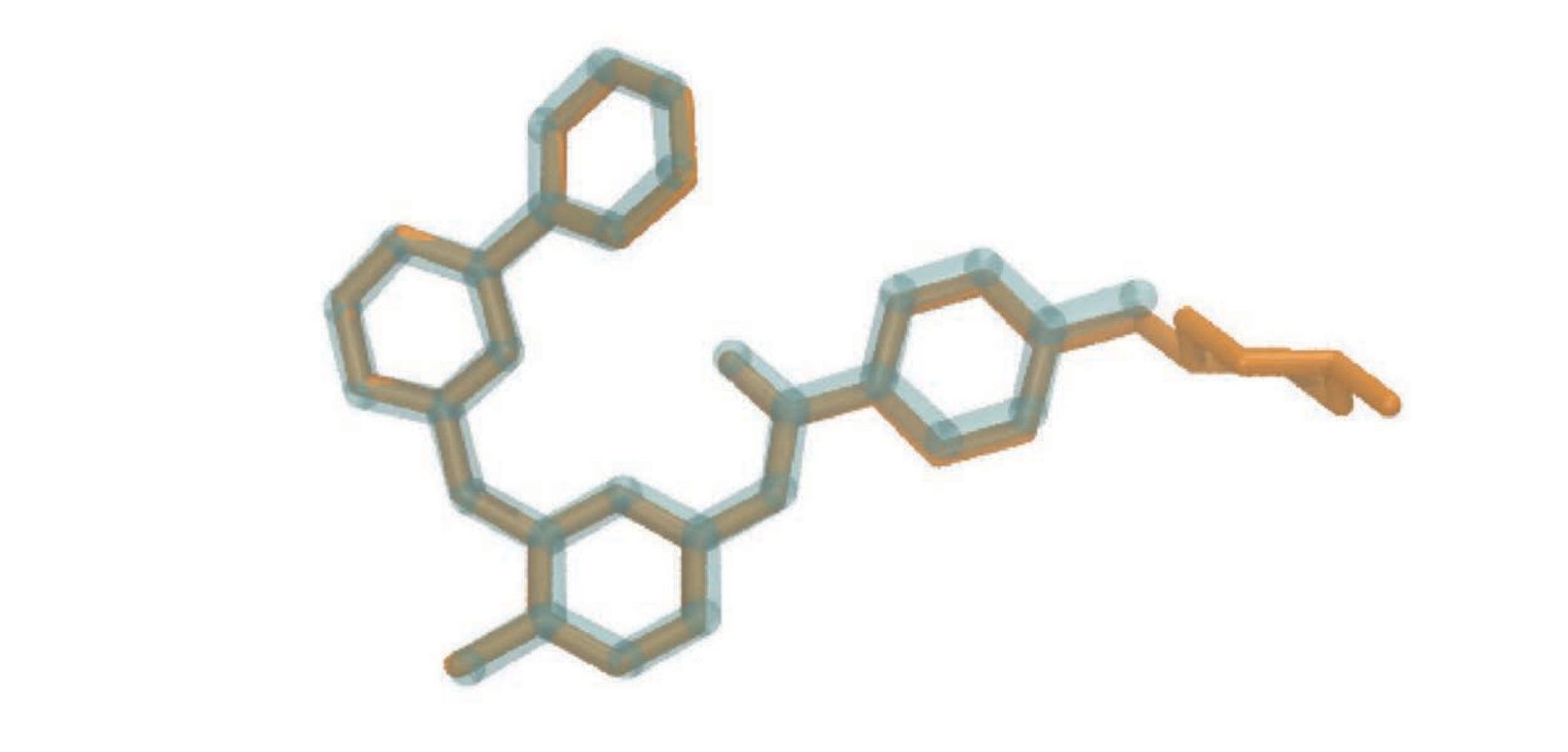}
	\caption{\small Overlap between query PDGFrb and ZINC03832219 molecules, with optimized conformers. Similarity score = 0.887.}
	\label{overlap_opt}
        \end{subfigure}
        \caption[]
        {\small Overlap between query PDGFrb and ZINC03832219 molecules. The query is represented by the yellow structure and the ZINC03832219 molecule is represented by the blueish-green structure.} 
        \label{overlap}
    \end{figure}
    
In particular, let us consider the target class PDGFrb. One possible reason that could explain the poor performance of 3D methods on the PDGFrb dataset is the conformer optimization performed on the molecules in this dataset. In order to understand how different molecular conformations affect the performance of VS experiments, we generated various low-energy conformations of 10 active molecules out of 124, and selected the conformer that had the best overlay score. No modification was performed on the query molecule.
In Fig.~\ref{overlap_original}, we show the overlap of the query PDGFrb and ZINC03832219 molecules with the conformation as retrieved from the  \texttt{DUD\_LIB\_VS\_1.0} library. The similarity score obtained using the GMS method is 0.615, whereas for the Morgan fingerprint it is 0.729. Fig.~\ref{overlap_opt} shows a conformation for the ZINC03832219 molecule with a higher degree of overlap: in this case, the similarity score using the GMS method increased to 0.887, whereas the Morgan fingerprint remained at 0.729.

\begin{table}
\centering
\begin{tabularx}{250pt}{j@{\hskip 0.5in} m m }
\toprule 
\thead{Performance \\ Metric} & \thead{Original \\ Dataset} & \thead{Optimized \\ Dataset} \\
\addlinespace[0.05cm]
\midrule
%auc
AUC 	& 0.42 	& 0.42 \\
\addlinespace[0.05cm]
awAUC  	& 0.48  	& 0.48 \\
\addlinespace[0.15cm]
%roce
ROCE $0.5 \%$ & 4.67 & 15.58 \\
\addlinespace[0.05cm]
ROCE $1 \%$ & 4.76 & 7.92 \\
\addlinespace[0.05cm]
ROCE $2 \%$ & 3.2 & 3.99 \\
\addlinespace[0.05cm]
ROCE $5 \%$ & 1.45 &  1.61\\
\addlinespace[0.15cm]
% awroce
awROCE $0.5 \%$ & 3.76 & 35.12 \\
\addlinespace[0.05cm]
awROCE $1 \%$ & 11.49 & 17.87 \\
\addlinespace[0.05cm]
awROCE $2 \%$ & 6.44 & 9.02 \\
\addlinespace[0.05cm]
awROCE $5 \%$ & 2.72 & 3.63 \\
\addlinespace[0.05cm]
\bottomrule
\end{tabularx}
\caption{VS performance of CS$_9$WS$_{B-5}$ on two PDGFrb datasets}
\label{table:pdgfrb}
\end{table}
Table~\ref{table:pdgfrb} shows the VS performance of CS$_9$WS$_{B-5}$ for two PDGFrb datasets. One dataset is the complete original set with the molecules curated by Jahn~et~al.~\cite{Jahn2009}. The second set replaced 10 active molecules and the query molecule with modified conformations; we refer to this set as an ``optimized dataset''. The optimized dataset significantly improved results for early enrichment; however, the overall scores for AUC and awAUC remained the same. In the future, it would be advisable to generate a certain number of conformations for all molecules, choosing the conformer that has the best similarity score.

The GMS method outperformed conventional techniques for most of the targets tested in this work by achieving higher early enrichment values, but a question remains regarding its computational cost. Finding the optimum solution for a GMS method comparison quickly grows impractical; however, heuristic methods can be used to find an optimal or near-optimal solution. A challenge that remains is to improve and scale these methods to more accurately and efficiently tackle molecular comparisons using large commercial databases. Quantum annealers have been argued to have the potential to take advantage of quantum phenomena such as quantum tunnelling~\cite{Boixo2016} and entanglement~\cite{Lanting2014} to solve optimization problems. The GMS algorithm has been formulated as a QUBO problem. Although this formulation is hardware agnostic, that is, classical optimization techniques can solve this problem, it also corresponds to the class of objective functions native to the quantum annealing hardware. This approach allows for implementation using the most effective solver (classical or quantum) available in a rapidly changing field.

Quantum annealers are a nascent, growing technology: every new generation shows a reduction in the amount of noise and an increase in the number of qubits (i.e., the basic unit of quantum information). The number of qubits is related to the number of variables an optimization problem can have. The most recent quantum annealer consists of 2048 qubits, and the number of variables (fully connected) that can be mapped to it is 66. The GMS problems generated using the DUD targets varied in size and density (a relation between the number of variables and their connections). 
\begin{figure}[H]
%\centering
	% figure A
	\begin{subfigure}[b]{0.37\textwidth}
	\centering
	\includegraphics[width=1.0\columnwidth]{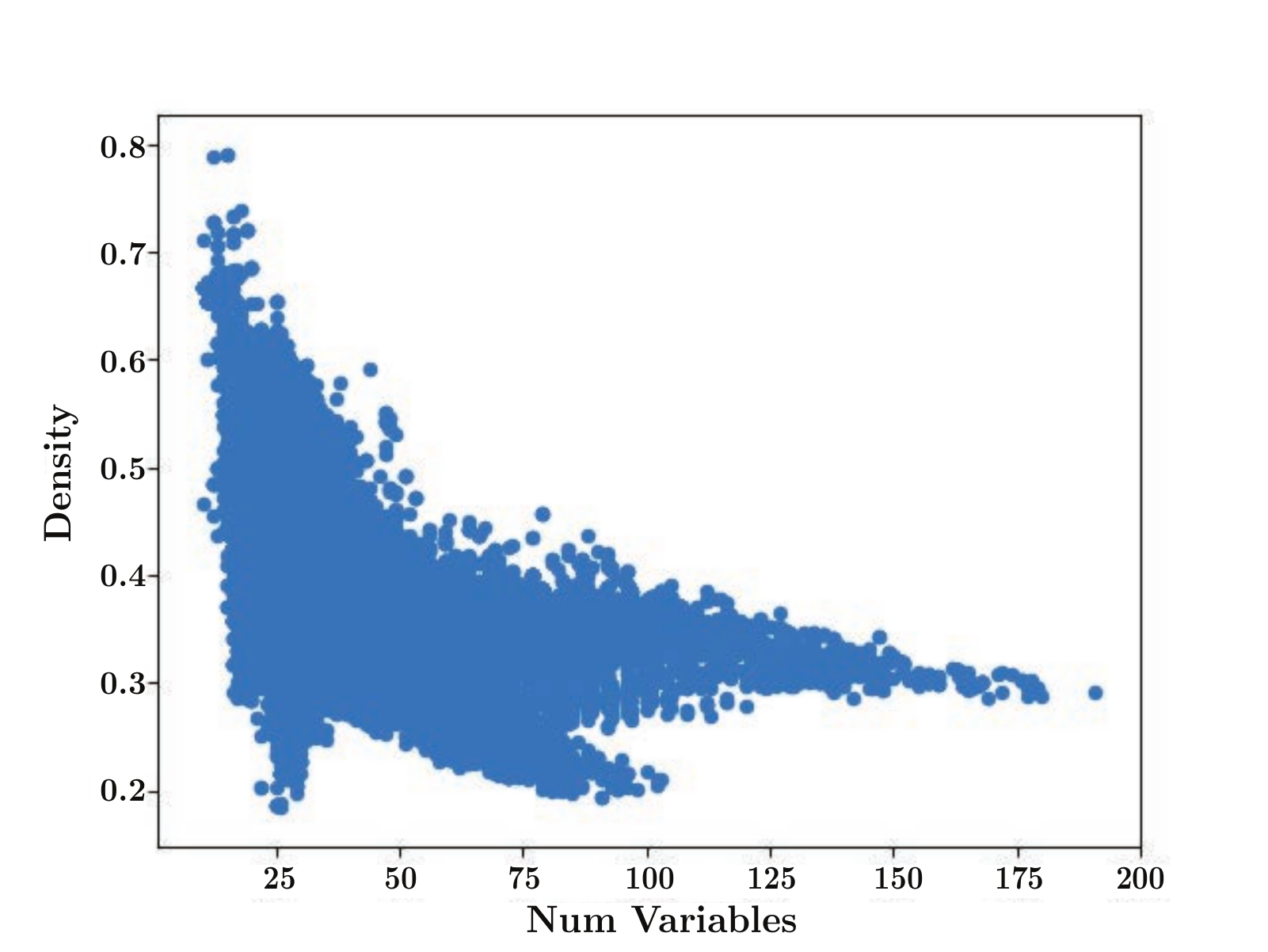}
	\caption{Density vs. number of variables for problems generated with criticality scheme \small CS$_3$} 
	\label{ds_exp3}
        \end{subfigure}
        \qquad
        % figure B
        \begin{subfigure}[b]{0.37\textwidth}  
	\centering 
	\includegraphics[width=1.0\columnwidth]{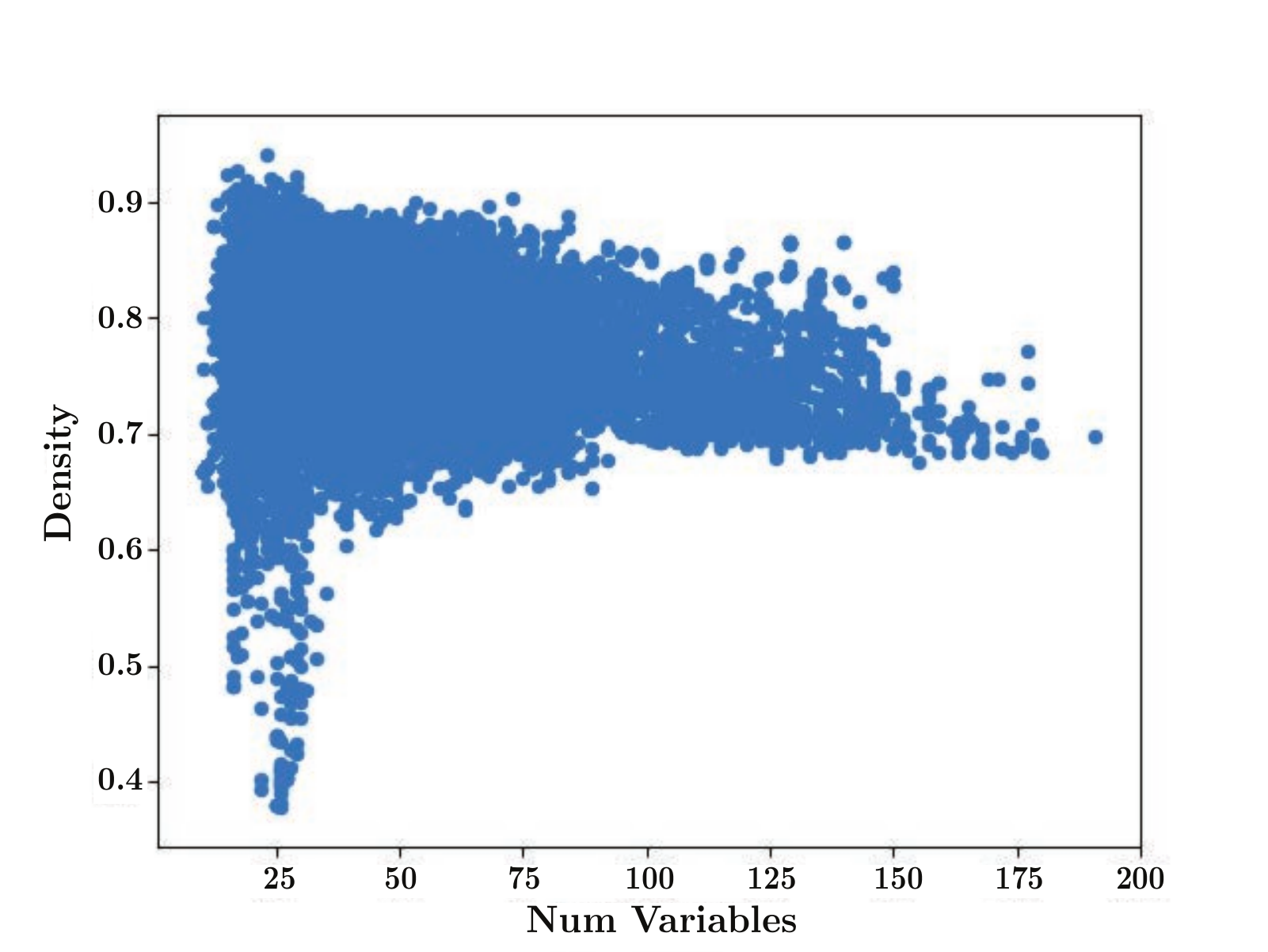}
	\caption{Density vs. number of variables for problems generated with criticality scheme \small CS$_9$ }
	\label{ds_exp9}
        \end{subfigure}
        \caption[]
        {\small Density vs. number of variables for problems generated with particular criticality schemes} 
        \label{ds}
    \end{figure}
In Fig.~\ref{ds}, we illustrate the relation between the density and the number of variables for each problem in two cases. Fig.~\ref{ds_exp3} and Fig.~\ref{ds_exp9} correspond to the criticality schemes CS$_3$ and CS$_9$, respectively. Both schemes yielded problem sizes ranging up to 200 variables. The difference lies in their densities. Criticality scheme CS$_9$ will generate denser problems, as it includes the feature of having 3D coordinates, whereas CS$_3$ does not. From these figures, we can observe that it is not yet possible to solve all the generated problems using a quantum annealer. Therefore, a rigorous, exhaustive study of the quantum annealer's performance is not yet feasible. The objective of this paper has been to evaluate the GMS formulation in terms of success metrics such as early enrichment. The study of the performance of this emergent and potentially groundbreaking technology is a subject for future work.

%%% ==========================================================================
%%%								CONCLUSION
%%% ==========================================================================

\section{Conclusion}

In this paper, we have presented a new methodology for determining molecular similarity, which enables the implementation of ligand-based virtual screening using a quantum computer. Our new \emph{graph-based molecular similarity} (GMS) method solves the maximum weighted co-$k$-plex problem of an induced graph to find the maximum weighted common subgraph (MWCS). Solving the MWCS problem tends to be time consuming on classical computers, as it is, in general, NP-hard. Recent advances in quantum computing, and the availability of quantum annealing devices, offer alternatives for solving these classically hard problems~\cite{lidar}. The GMS method has been formulated such that it can be implemented on quantum annealers. This formulation has enabled the future use of these devices when they have sufficiently improved (e.g., when their number of qubits and connectivity increase).

The advantage of using graphs is that they are able to encode any molecular information perceived as relevant. In our implementation of the method, we have incorporated both 2D and 3D descriptors into the molecular graphs. The highlight of the method is its flexibility, which allows the user to either include or exclude different molecular features according to their relevance to the problem. We tested this flexibility by implementing several combinations of descriptors. Various similarity criteria and combinations of molecular features used in the GMS method were evaluated on 13 datasets from the \texttt{DUD\_LIB\_VS\_1.0} library. We identified a particularly successful configuration of features that uses an equal weighting of 1 for all features except the rings; the rings were assigned a higher weight of 5. Our results demonstrate that the GMS method, where similarity criteria include three-dimensional coordinates, has, in general, a higher early enrichment value than the similarity criteria that do not include this feature. We also found the results to be sensitive to the 3D conformation of the molecules used. Further research is needed to find the optimal conformation of query molecules that do not have crystal structures in complex with target proteins in an unbiased manner. Overall, we conclude that the GMS method outperforms conventional fingerprint and optimal assignment methods for most of the 13 DUD targets.

%%% ==========================================================================
%%%								ACKNOWLEDGEMENT
%%% ==========================================================================

\begin{acknowledgement}
The authors thank Arman Zaribafiyan and Dominic Marchand for their early contributions to the project; Takeshi Yamazaki for valuable discussions, collaboration on molecular figures, and suggestions for revisions to the manuscript; Rudi Plesch, Nick Cond\'{e}, and SungYeon Kim for their contribution to the development of the GMS application; Jayne Drew for digitizing and editing the GMS method figures; Shawn Wowk for his support; Kausar N. Samli for his comments on an early version of the paper; and Marko Bucyk for his helpful comments and for editing the manuscript. Additionally, the authors thank Giuliano Berellini for insightful discussions at the beginning of the project; Jeff Elton for his expertise and guidance; Simon Sekhon, Brandon Hedrick, and Spencer Herath for their work on the proof-of-concept molecular comparison application; and Charles Rozea, Teresa Tung, and Daniel Garrison for their support.
\end{acknowledgement}

%%% ==========================================================================
%%%	BIBLIOGRAPHY
%%% ==========================================================================

\bibliography{Manuscript_ACS}

%%%%%%%%%%%%%%%%%%%%%%%%%%%%%%%%%%%%%%%%%%%%%%%%%%%%%%%%%%%%%%%%%%%%%
%% Supplementary Information
%%%%%%%%%%%%%%%%%%%%%%%%%%%%%%%%%%%%%%%%%%%%%%%%%%%%%%%%%%%%%%%%%%%%%
\newpage
\section{Supporting Information}

\setcounter{equation}{0}
\setcounter{figure}{0}
\renewcommand{\figurename}{Supplementary Figure}

\subsection{S1 Criticality and Weighting Schemes}
\label{sec:similarity_criteria}

For each feature presented in~\Cref{table:mol_features}, we have set its value to critical (C), non-critical (NC), or off (OFF). Bond order, formal charge, and degree have been set to NC and pharmacophore features have been set to C. For the remaining set of features, we have set various combinations of values as presented in~\Cref{tab:criticality_schemes}. Each combination is called a \textit{criticality scheme} (CS). In total, we have generated 12 CSs.

\begin{table}[H]
\centering
\resizebox{\textwidth}{!}{
\begin{tabular}{ l c c c c c c c c c c c c}
\hline
\thead{Features} & \thead{$CS_{1}$} & \thead{$CS_{2}$} & \thead{$CS_{3}$} & \thead{$CS_{4}$} & \thead{$CS_{5}$} & \thead{$CS_{6}$} & \thead{$CS_{7}$} & \thead{$CS_{8}$} & \thead{$CS_{9}$} & \thead{$CS_{10}$} & \thead{$CS_{11}$} & \thead{$CS_{12}$}\\
\addlinespace[0.1cm]
\hline
Atomic number	(single atom)	& C 		& C 		& C 		& NC 	& NC 	& NC 		& C 		& C 		& C 		& NC 	& NC 	& NC \\
Atomic number	(ring)			& C 		& C 		& OFF 	& NC 	& NC 	& OFF 		& C 		& C 		& OFF 	& NC 	& NC 	& OFF \\
Implicit hydrogen 			& NC 	& OFF 	& OFF 	& NC 	& OFF 	& OFF 		& NC 	& OFF 	& OFF 	& NC 	& OFF 	& OFF \\
3D						& OFF	& OFF 	& OFF 	& OFF	& OFF 	& OFF 		& ON 	& ON 	& ON 	& ON 	& ON 	& ON \\
\hline
\end{tabular}
}
\caption{Criticality Schemes}
\label{tab:criticality_schemes}
\end{table}

In addition to identifying the features as C or NC, we have assigned them a weighting value to reflect its relevance in the virtual screening (VS) experiments. Each combination of weighting values is called a \textit{weighting scheme} (WS). We set one WS as baseline $WS_{B}$, with every feature equally weighted to act as a control. In total we have 10 WSs, shown in~\Cref{tab:weighting_schemes}.

\begin{table}[H]
\centering
\resizebox{\textwidth}{!}{
\begin{tabular}{ l c c c c c c c c c c}
\hline
\thead{Features} & \thead{$WS_{B}$} & \thead{$WS_{1}$} & \thead{$WS_{2}$} & \thead{$WS_{3}$} & \thead{$WS_{4}$} & \thead{$WS_{B-5}$} & \thead{$WS_{1-5}$} & \thead{$WS_{2-5}$} & \thead{$WS_{3-5}$} & \thead{$WS_{4-5}$}\\
\addlinespace[0.1cm]
\hline
Atom	 & 1.0 & 0.1 & 0.1 & 0.1 & 0.1 & 1.0 & 1.0 & 1.0 & 1.0 & 1.0 \\
Ring		 & 1.0 & 0.1 & 0.1 & 0.1 & 0.1 & 5.0 & 5.0 & 5.0 & 5.0 & 5.0 \\
Degree & 1.0 & 0.1 & 0.1 & 0.1 & 0.1 & 1.0 & 0.1 & 0.1 & 0.1 & 0.1 \\
Implicit hydrogen & 1.0 & 0.1 & 0.1 & 0.1 & 0.1 & 1.0 & 0.1 & 0.1 & 0.1 & 0.1 \\
Bond orders & 1.0 & 0.1 & 0.1 & 0.1 & 0.1 & 1.0 & 0.1 & 0.1 & 0.1 & 0.1 \\
Formal charge & 1.0 & 0.1 & 0.1 & 0.1 & 0.1 & 1.0 & 0.1 & 0.1 & 0.1 & 0.1 \\
\addlinespace[0.1cm]
Basic & 1.0 & 1.0 & 3.0 & 1.0 & 2.0 & 1.0 & 1.0 & 3.0 & 1.0 & 2.0 \\
Acidic & 1.0 & 1.0 & 3.0 & 1.0 & 2.0 & 1.0 & 1.0 & 3.0 & 1.0 & 2.0 \\
H donor & 1.0 & 1.0 & 2.0 & 2.0 & 3.0 & 1.0 & 1.0 & 2.0 & 2.0 & 3.0 \\
H acceptor & 1.0 & 1.0 & 2.0 & 2.0 & 3.0 & 1.0 & 1.0 & 2.0 & 2.0 & 3.0 \\
Aromatic & 1.0 & 1.0 & 2.0 & 1.0 & 1.0 & 1.0 & 1.0 & 2.0 & 1.0 & 1.0 \\
Hydrophobic & 1.0 & 1.0 & 1.0 & 1.0 & 1.0 & 1.0 & 1.0 & 1.0 & 1.0 & 1.0 \\
Zinc binder & 1.0 & 1.0 & 3.0 & 1.0 & 2.0 & 1.0 & 1.0 & 3.0 & 1.0 & 2.0 \\
\addlinespace[0.1cm]
\hline
\end{tabular}
}
\caption{Weighting Schemes}
\label{tab:weighting_schemes}
\end{table}

%%%%%%%%%%%%%%%%%%%%%%%%%%%%%%%%%%%%
\subsection{S2 Solver Parameters}
\label{sec:solver}

The algorithm used to solve the maximum weighted co-$1$-plex problem is the parallel tempering Monte Carlo with isoenergetic cluster moves (PTICM) heuristic solver. The performance of replica-exchange algorithms, such as the PTICM algorithm, depends on the parameters used, especially the temperature schedule selected. In the main paper, the temperature schedule has been based on the geometric schedule for each replica. The low and high temperature values are determined based on the values of the coefficients of each quadratic unconstrained binary optimization problem instance, and the number of replicas has been chosen to be two. 

%%%%%%%%%%%%%%%%%%%%%%%%%%%%%%%%%%%%
\newpage
\subsection{S3 Overall VS Performance for the GMS and Fingerprint Methods}
\label{sec:overall}
\begin{figure}[H]
%\centering
	% figure A
	\begin{subfigure}[b]{0.48\textwidth}
	\centering
	\includegraphics[width=\textwidth]{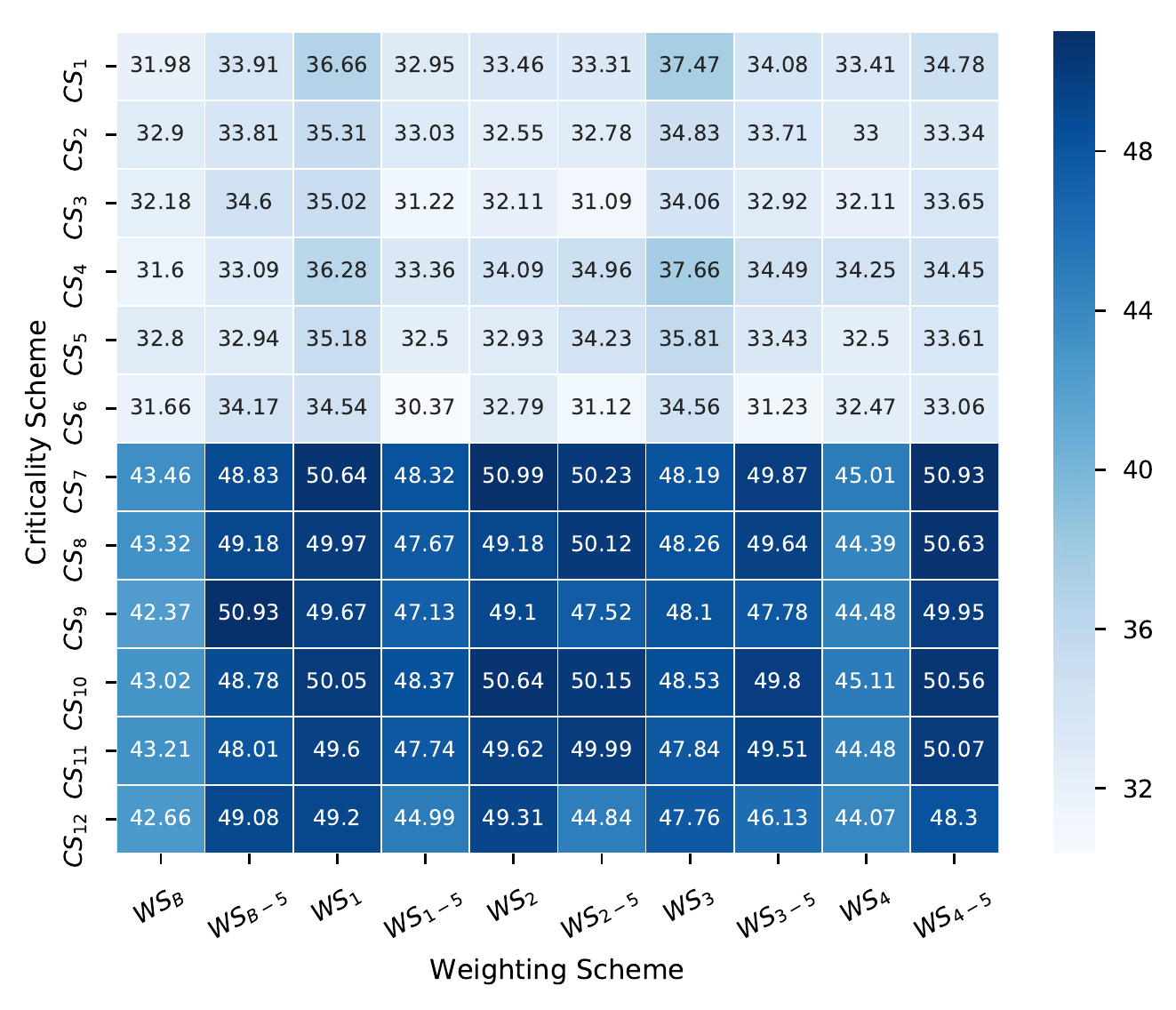}
	\caption{Mean ROCE values at 0.5\%}
	\label{fig:mean_ROCE_0.5}
	\end{subfigure}
        %\quad
        	% figure B
	\begin{subfigure}[b]{0.48\textwidth}
	\centering
	\includegraphics[width=\textwidth]{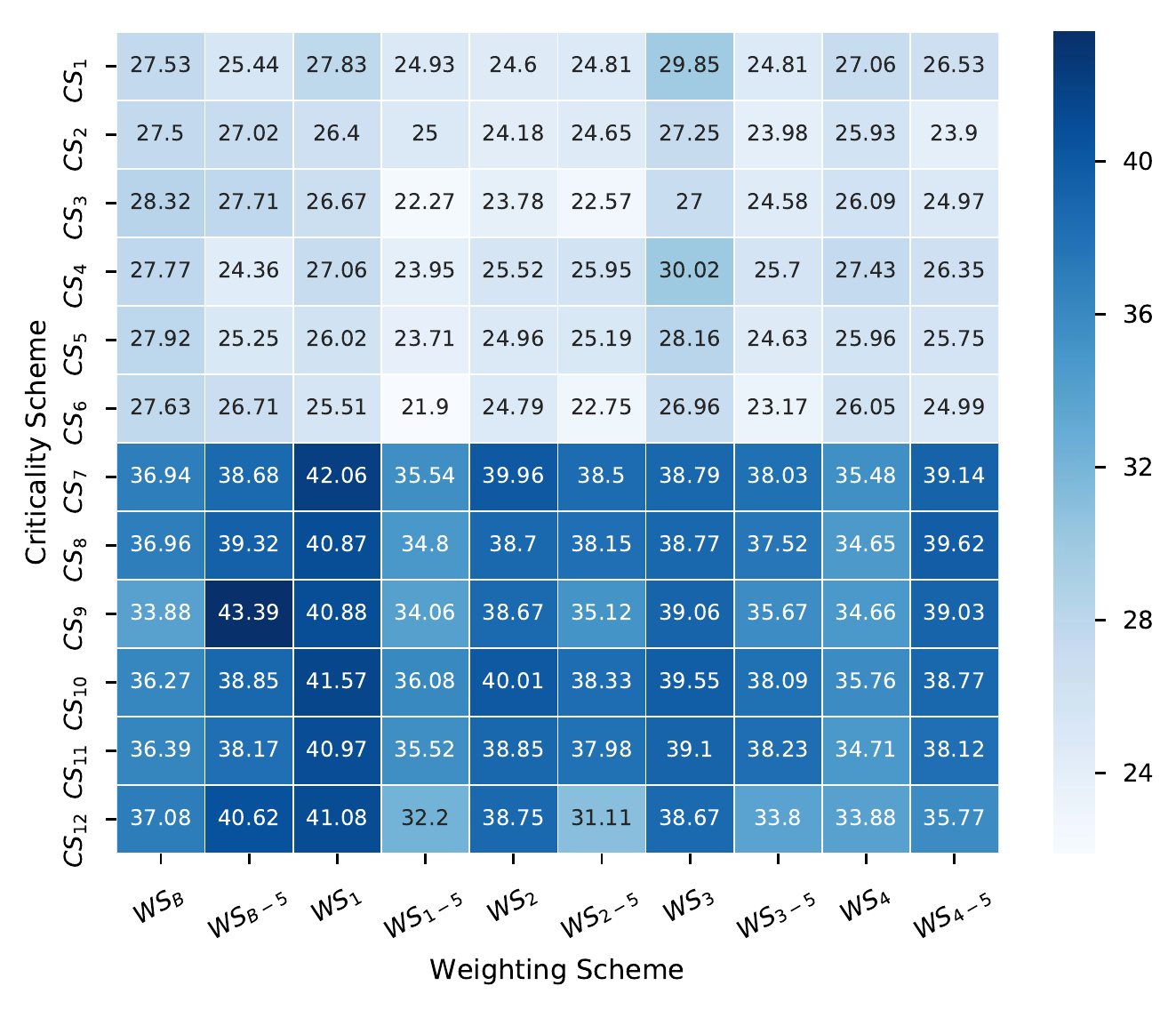}
	\caption{Mean awROCE values at 0.5\%}
	\label{fig:mean_awROCE_0.5}
	\end{subfigure}
\caption{Overall VS performance for each GMS method: a) mean ROCE and b) mean awROCE values at 0.5\% over 13 targets in the \texttt{DUD\_LIB\_VS\_1.0} library}
\end{figure}

\begin{table}[H]
\centering
\resizebox{\textwidth}{!}{
\begin{tabular}{ l l l l c c}
\hline
\thead{Fingerprint} & \thead{Bits} & \thead{Radius} & \thead{Use Features} & \thead{Mean ROCE 0.5\%} & \thead{Mean awROCE 0.5\%}\\
\addlinespace[0.1cm]
\hline
MACCS & N/A & N/A & N/A & 23.79 & 16.86 \\
Morgan & 1024 & 1 & False	&	41.095	&	29.160 \\
Morgan & 1024 & 1 & True	&	29.386	&	25.178 \\
Morgan & 1024 & 2 & False	&	43.981	&	33.899 \\
Morgan & 1024 & 2 & True	&	40.116	&	32.147 \\
Morgan & 1024 & 3 & False	&	43.463	&	32.502 \\
Morgan & 1024 & 3 & True	&	41.205	&	33.851 \\
Morgan & 1024 & 4 & False	&	41.786	&	30.422 \\
Morgan & 1024 & 4 & True	&	40.216	&	29.738 \\
Morgan & 2048 & 1 & False	&	41.910	&	29.445 \\ 
Morgan & 2048 & 1 & True	&	29.673	&	25.426 \\
Morgan & 2048 & 2 & False	&	\textbf{44.692}	&	32.659 \\
Morgan & 2048 & 2 & True	&	41.291	&	34.377 \\
Morgan & 2048 & 3 & False	&	44.530	&	34.868 \\
Morgan & 2048 & 3 & True	&	43.292	&	34.754 \\
Morgan & 2048 & 4 & False	&	44.455	&	\textbf{35.148} \\
Morgan & 2048 & 4 & True	&	42.008	&	33.794 \\
\hline
\end{tabular}
}
\caption{Overall VS performance for MACCS and Morgan fingerprints. For Morgan fingerprints, we consider four radii and two bit-vector sizes. We also compare the use of feature-based invariants. We report the mean ROCE and mean awROCE at values at 0.5\% over 13 targets in the \texttt{DUD\_LIB\_VS\_1.0} library}
\label{tab:overall_fingerprint}
\end{table}

\newpage
%%%%%%%%%%%%%%%%%%%%%%%%%%%%%%%%%%%%
\subsection{S4 VS Performance Across 13 targets in the \texttt{DUD\_LIB\_VS\_1.0} Library}
\label{sec:results_classes}

%==================================
%	ACE
%==================================

\subsubsection{ACE}
\begin{figure}[H]
%\centering
	% figure A
	\begin{subfigure}[b]{0.44\textwidth}
	\centering
	\includegraphics[width=\textwidth]{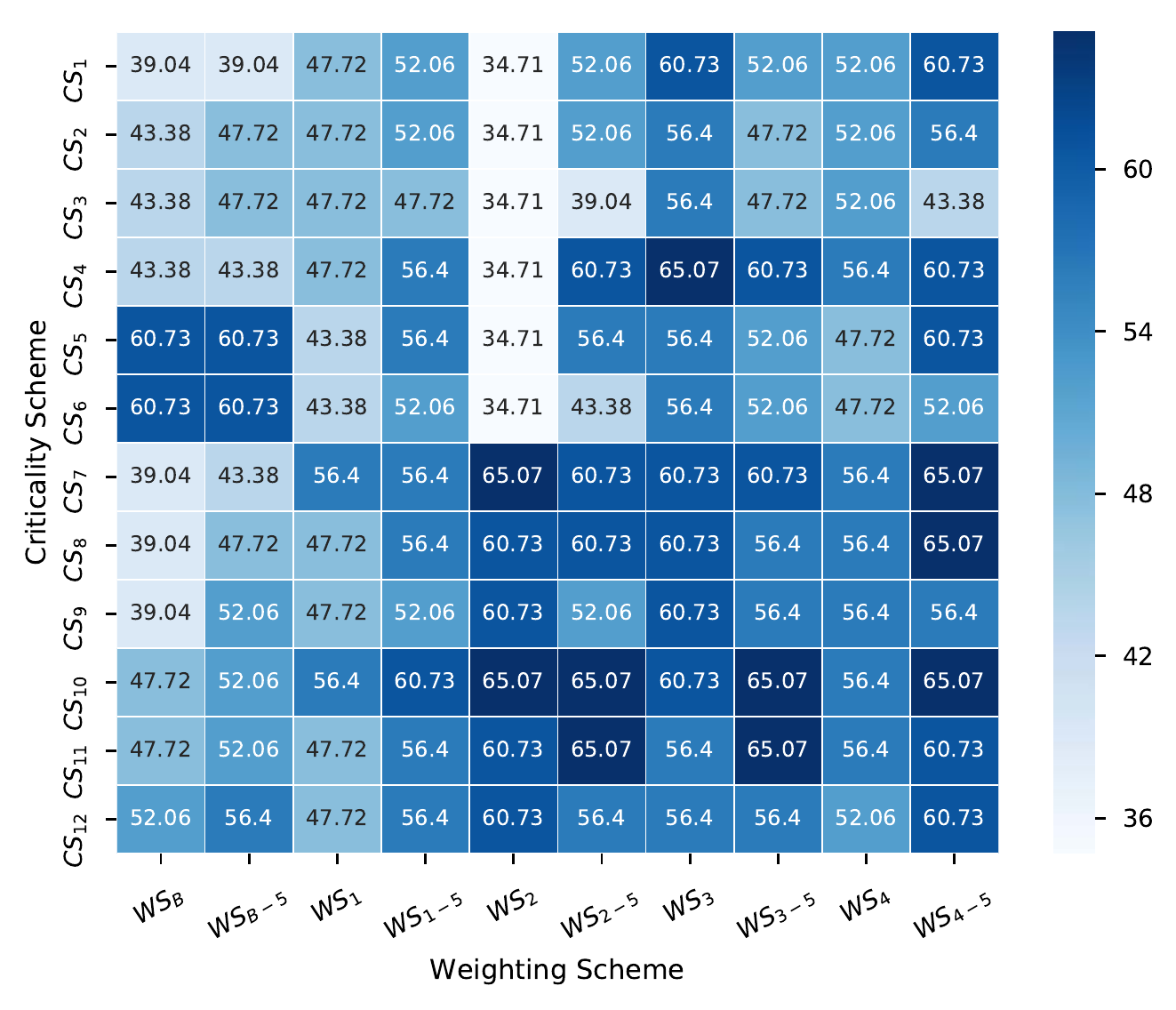}
	\caption{ROCE 0.5\% values}
	\label{fig:ace_ROCE_0.5}
	\end{subfigure}
        %\quad
        	% figure B
	\begin{subfigure}[b]{0.44\textwidth}
	\centering
	\includegraphics[width=\textwidth]{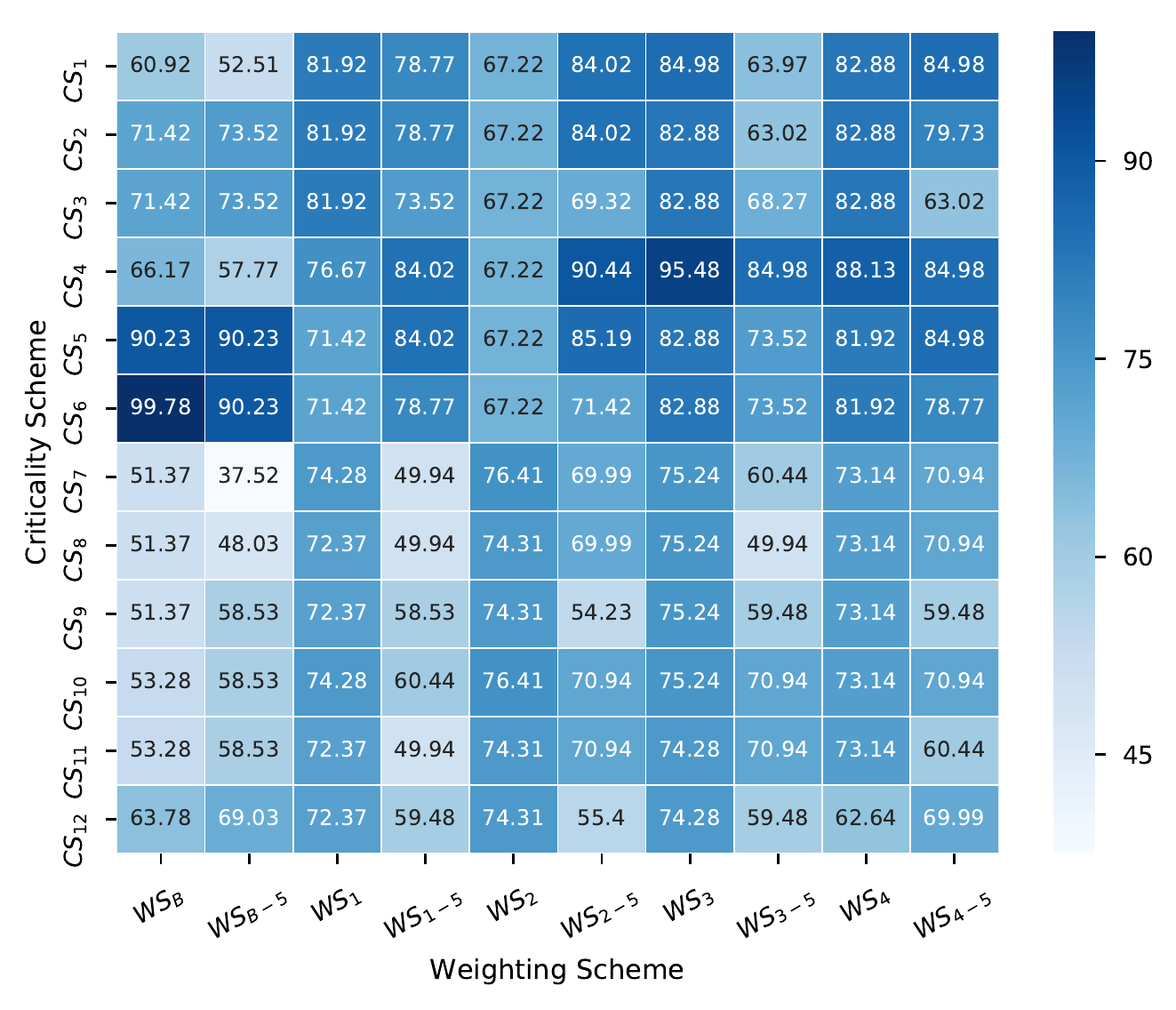}
	\caption{awROCE 0.5\% values}
	\label{fig:ace_awROCE_0.5}
	\end{subfigure}
\caption{ROCE and awROCE 0.5\% values for the ACE class}
\end{figure}

\begin{table}[H]
\centering
\resizebox{\textwidth}{!}{
\begin{tabular}{ l l l l c c}
\hline
\thead{Fingerprint} & \thead{Bits} & \thead{Radius} & \thead{Use Features} & \thead{ROCE 0.5\%} & \thead{awROCE 0.5\%}\\
\addlinespace[0.1cm]
\hline
MACCS & N/A & N/A & N/A & 17.35 & 8.40 \\
Morgan & 1024.0 & 1.0 & False		& 34.71 & 31.51 \\
Morgan & 1024.0 & 1.0 & True 		& 21.69 & 33.61 \\
Morgan & 1024.0 & 2.0 & False 	& 43.38 & 52.51 \\
Morgan & 1024.0 & 2.0 & True 		& 39.04 & 42.01 \\
Morgan & 1024.0 & 3.0 & False 	& \textbf{52.06} & 54.42 \\
Morgan & 1024.0 & 3.0 & True 		& 47.72 & 63.02 \\
Morgan & 1024.0 & 4.0 & False 	& 34.71 & 36.76 \\
Morgan & 1024.0 & 4.0 & True 		& 30.37 & 21.01 \\
Morgan & 2048.0 & 1.0 & False 	& 34.71 & 31.51 \\
Morgan & 2048.0 & 1.0 & True 		& 21.69 & 33.61 \\
Morgan & 2048.0 & 2.0 & False 	& 34.71 & 31.51 \\
Morgan & 2048.0 & 2.0 & True 		& 47.72 & 63.02 \\
Morgan & 2048.0 & 3.0 & False 	& 47.72 & 63.02 \\
Morgan & 2048.0 & 3.0 & True 		& \textbf{52.06} & \textbf{73.52} \\
Morgan & 2048.0 & 4.0 & False 	& 47.72 & 63.02 \\
Morgan & 2048.0 & 4.0 & True 		& 39.04 & 50.41 \\
\hline
\end{tabular}
}
\caption{ACE class -- Fingerprint results}
\label{tab:ace_fingerprint}
\end{table}

%==================================
%	AChE
%==================================
\subsubsection{AChE}
 \begin{figure}[H]
%\centering
	% figure A
	\begin{subfigure}[b]{0.48\textwidth}
	\centering
	\includegraphics[width=\textwidth]{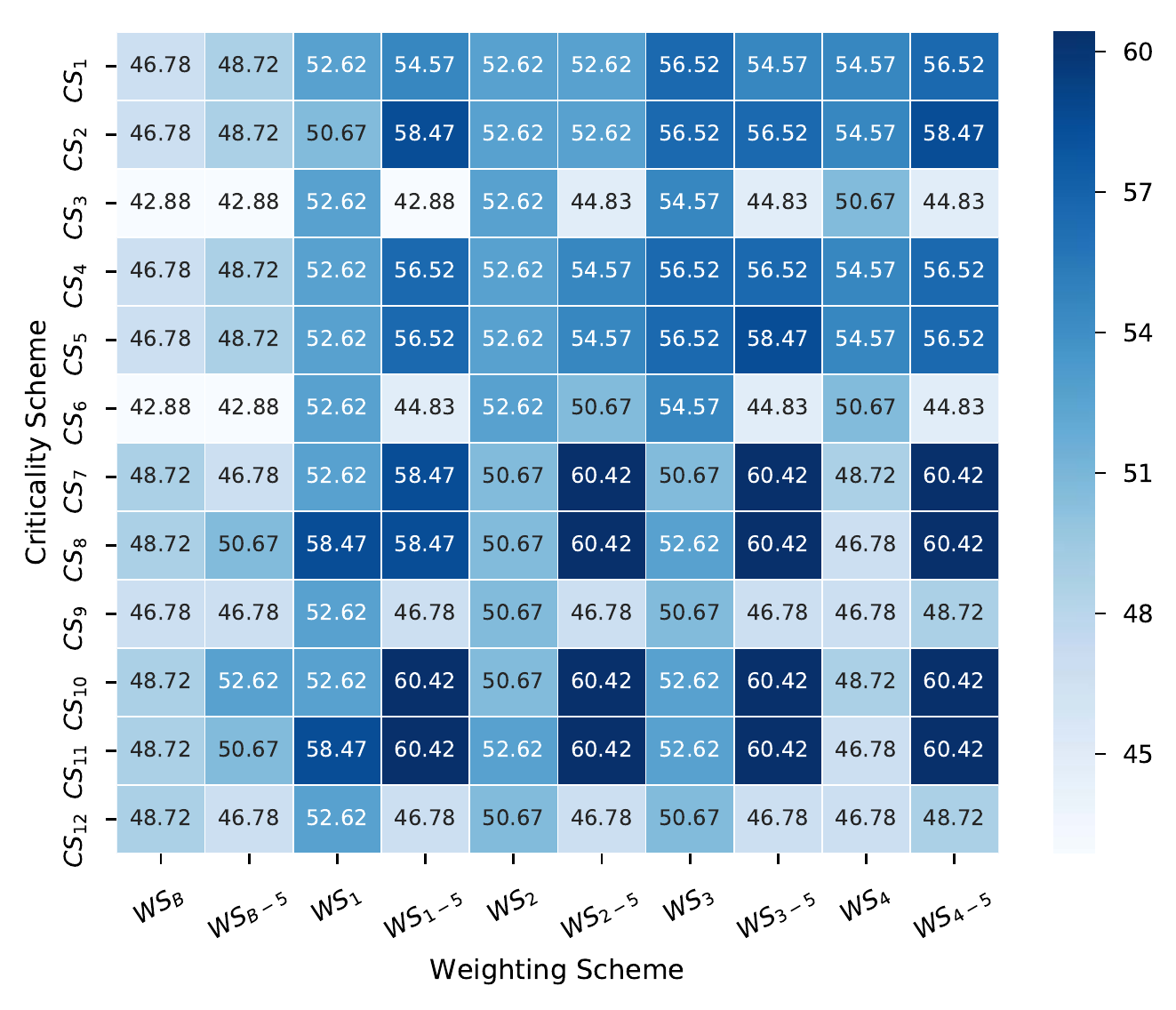}
	\caption{ROCE 0.5\% values}
	\label{fig:ache_ROCE_0.5}
	\end{subfigure}
        %\quad
        	% figure B
	\begin{subfigure}[b]{0.48\textwidth}
	\centering
	\includegraphics[width=\textwidth]{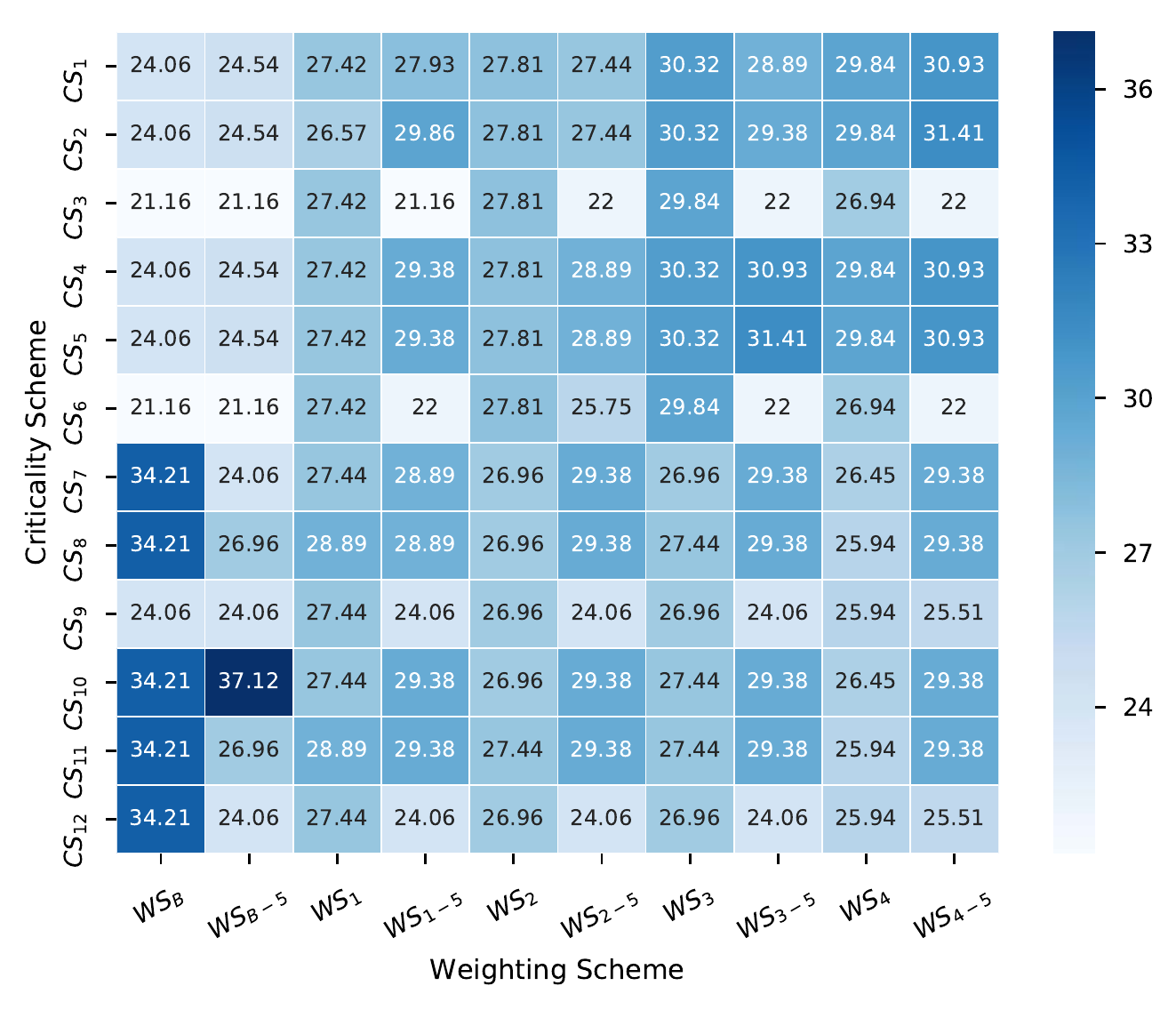}
	\caption{awROCE 0.5\% values}
	\label{fig:ache_awROCE_0.5}
	\end{subfigure}
\caption{ROCE and awROCE 0.5\% values for the AChE class}
\end{figure}

\begin{table}[H]
\centering
\resizebox{\textwidth}{!}{
\begin{tabular}{ l l l l c c}
\hline
\thead{Fingerprint} & \thead{Bits} & \thead{Radius} & \thead{Use Features} & \thead{ROCE 0.5\%} & \thead{awROCE 0.5\%}\\
\addlinespace[0.1cm]
\hline
MACCS & N/A & N/A & N/A & 35.08 & 19.13 \\
Morgan & 1024.0 & 1.0 & False & 44.83 & 22.00 \\
Morgan & 1024.0 & 1.0 & True & 37.03 & 21.50 \\
Morgan & 1024.0 & 2.0 & False & 44.83 & 22.00 \\
Morgan & 1024.0 & 2.0 & True & \textbf{48.72} & \textbf{33.00} \\
Morgan & 1024.0 & 3.0 & False & 42.88 & 21.16 \\
Morgan & 1024.0 & 3.0 & True & 46.78 & 22.85 \\
Morgan & 1024.0 & 4.0 & False & 42.88 & 21.16 \\
Morgan & 1024.0 & 4.0 & True & 46.78 & 22.85 \\
Morgan & 2048.0 & 1.0 & False & 44.83 & 22.00 \\
Morgan & 2048.0 & 1.0 & True & 37.03 & 21.50 \\
Morgan & 2048.0 & 2.0 & False & 44.83 & 22.00 \\
Morgan & 2048.0 & 2.0 & True & \textbf{48.72} & \textbf{33.00} \\
Morgan & 2048.0 & 3.0 & False & 44.83 & 22.00 \\
Morgan & 2048.0 & 3.0 & True & 46.78 & 22.85 \\
Morgan & 2048.0 & 4.0 & False & 42.88 & 21.16 \\
Morgan & 2048.0 & 4.0 & True & 46.78 & 22.85 \\
\hline
\end{tabular}
}
\caption{AChE class -- Fingerprint results}
\label{tab:ache_fingerprint}
\end{table}

%==================================
%	CDK2
%==================================
\subsubsection{CDK2}
 \begin{figure}[H]
%\centering
	% figure A
	\begin{subfigure}[b]{0.48\textwidth}
	\centering
	\includegraphics[width=\textwidth]{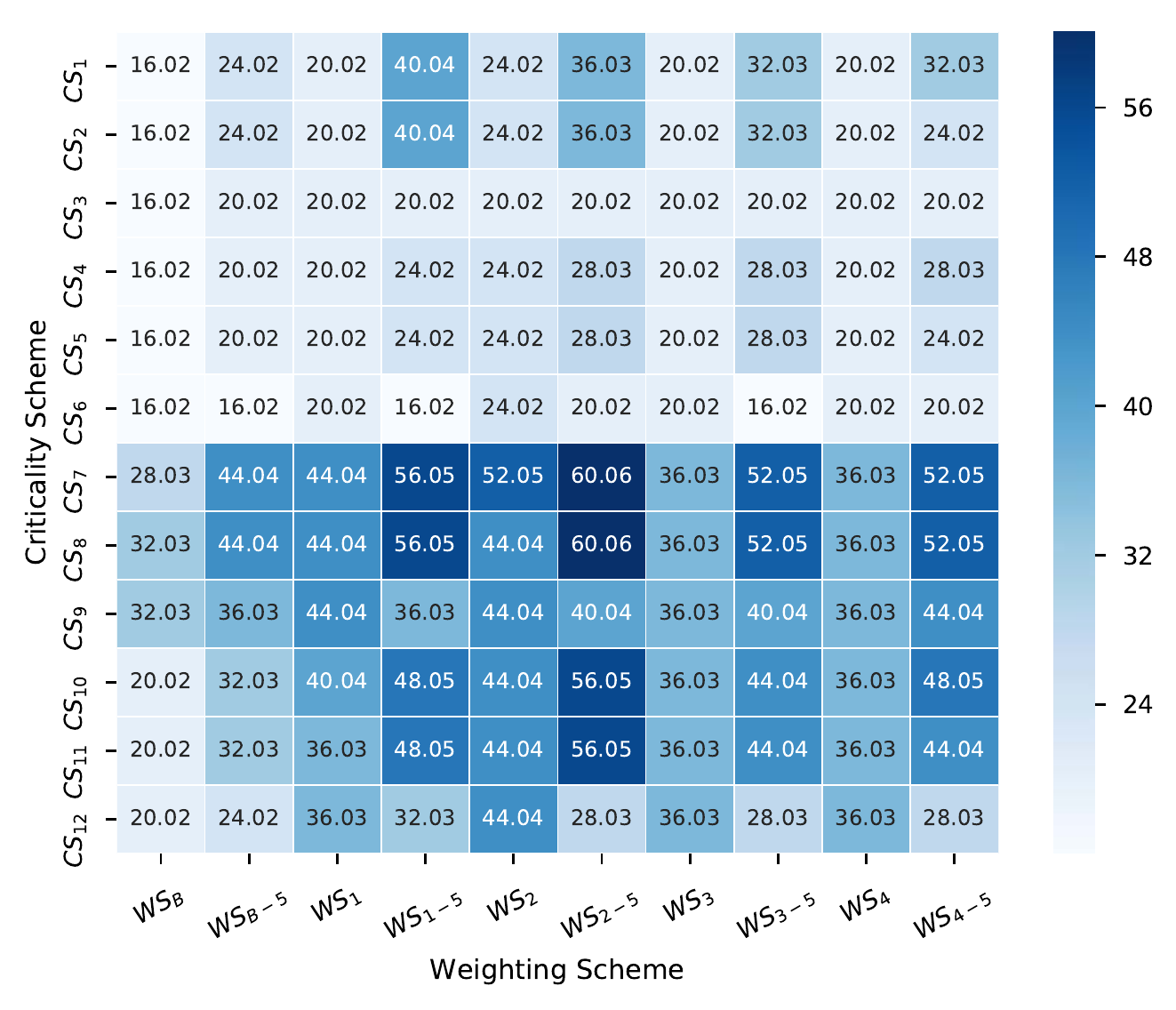}
	\caption{ROCE 0.5\% values}
	\label{fig:cdk2_ROCE_0.5}
	\end{subfigure}
        %\quad
        	% figure B
	\begin{subfigure}[b]{0.48\textwidth}
	\centering
	\includegraphics[width=\textwidth]{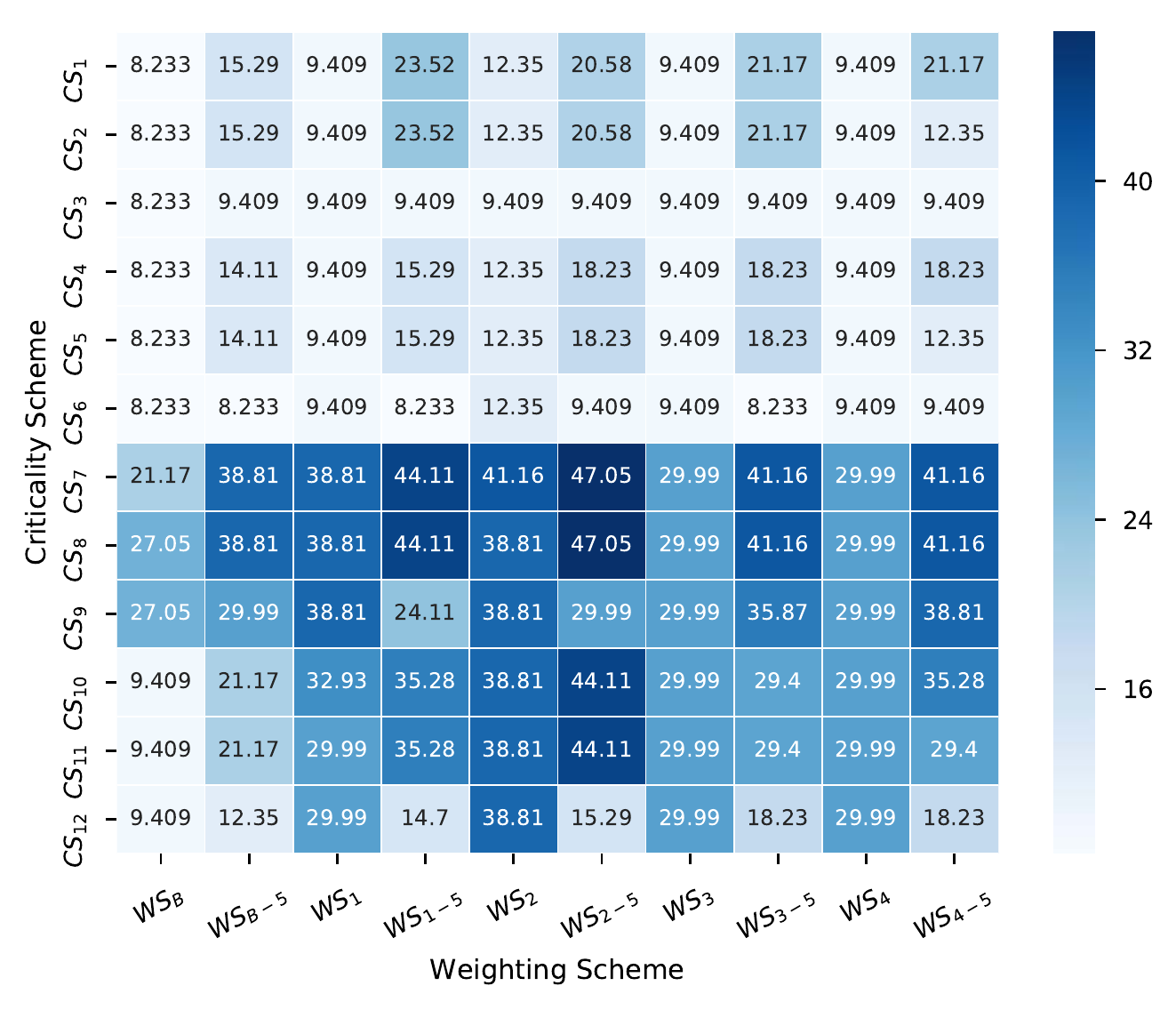}
	\caption{awROCE 0.5\% values}
	\label{fig:cdk2_awROCE_0.5}
	\end{subfigure}
\caption{ROCE and awROCE 0.5\% values for the CDK2 class}
\end{figure}

\begin{table}[H]
\centering
\resizebox{\textwidth}{!}{
\begin{tabular}{ l l l l c c}
\hline
\thead{Fingerprint} & \thead{Bits} & \thead{Radius} & \thead{Use Features} & \thead{ROCE 0.5\%} & \thead{awROCE 0.5\%}\\
\addlinespace[0.1cm]
\hline
MACCS & N/A & N/A & N/A & 20.02 & 9.41 \\
Morgan & 1024.0 & 1.0 & False & 20.02 & 9.41 \\
Morgan & 1024.0 & 1.0 & True & 32.03 & 27.05 \\
Morgan & 1024.0 & 2.0 & False & 20.02 & 9.41 \\
Morgan & 1024.0 & 2.0 & True & 32.03 & 27.05 \\
Morgan & 1024.0 & 3.0 & False & 20.02 & 9.41 \\
Morgan & 1024.0 & 3.0 & True & \textbf{36.03} & \textbf{32.93} \\
Morgan & 1024.0 & 4.0 & False & 20.02 & 9.41 \\
Morgan & 1024.0 & 4.0 & True & \textbf{36.03} & \textbf{32.93} \\
Morgan & 2048.0 & 1.0 & False & 20.02 & 9.41 \\
Morgan & 2048.0 & 1.0 & True & 32.03 & 27.05 \\
Morgan & 2048.0 & 2.0 & False & 20.02 & 9.41 \\
Morgan & 2048.0 & 2.0 & True & 32.03 & 27.05 \\
Morgan & 2048.0 & 3.0 & False & 24.02 & 15.29 \\
Morgan & 2048.0 & 3.0 & True & 32.03 & 27.05 \\
Morgan & 2048.0 & 4.0 & False & 20.02 & 9.41 \\
Morgan & 2048.0 & 4.0 & True & 32.03 & 27.05 \\
\hline
\end{tabular}
}
\caption{CDK2 class -- Fingerprint results}
\label{tab:cdk2_fingerprint}
\end{table}

%==================================
%	COX-2
%==================================
\subsubsection{COX-2}
 \begin{figure}[H]
%\centering
	% figure A
	\begin{subfigure}[b]{0.48\textwidth}
	\centering
	\includegraphics[width=\textwidth]{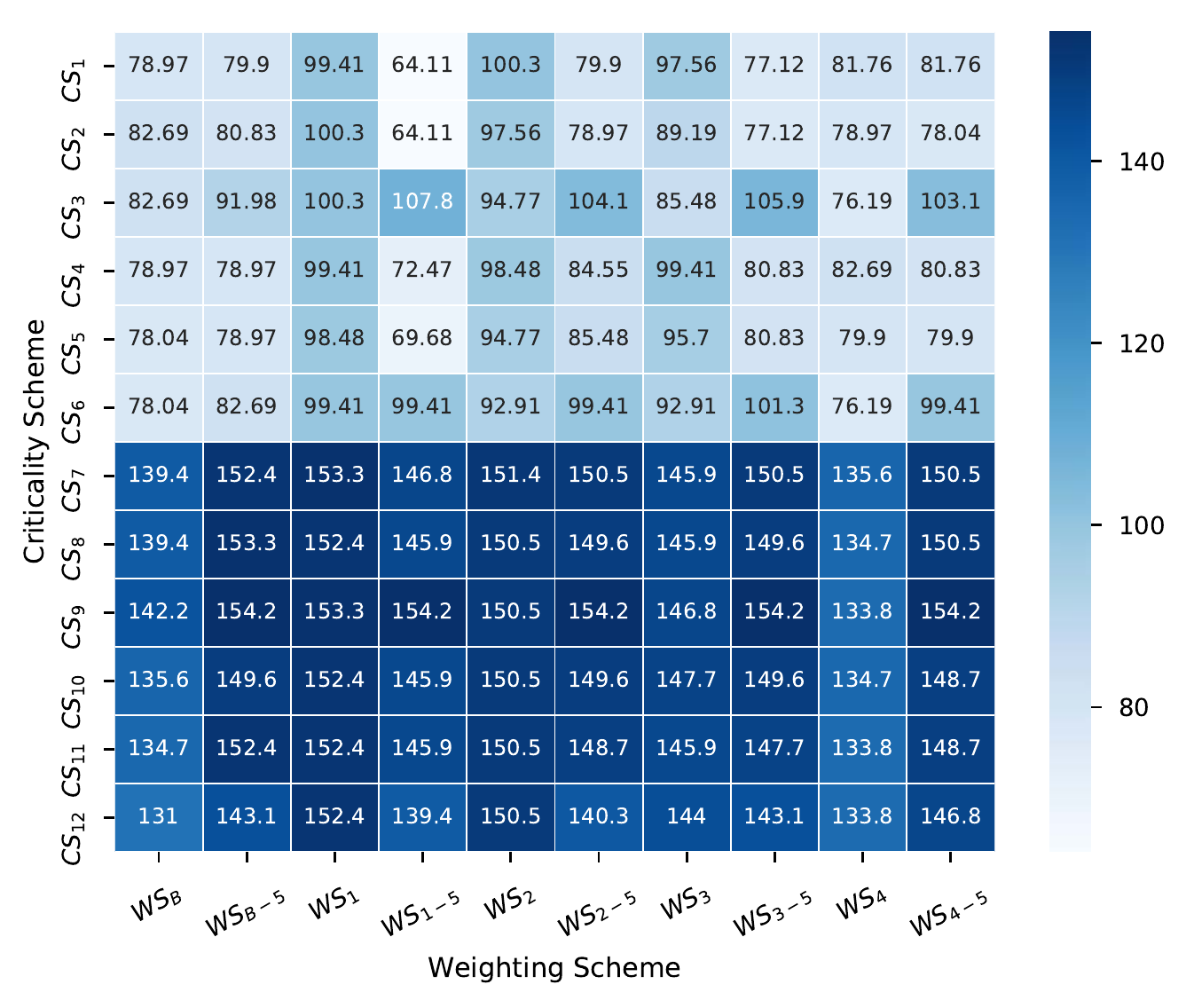}
	\caption{ROCE 0.5\% values}
	\label{fig:cox2_ROCE_0.5}
	\end{subfigure}
        %\quad
        	% figure B
	\begin{subfigure}[b]{0.48\textwidth}
	\centering
	\includegraphics[width=\textwidth]{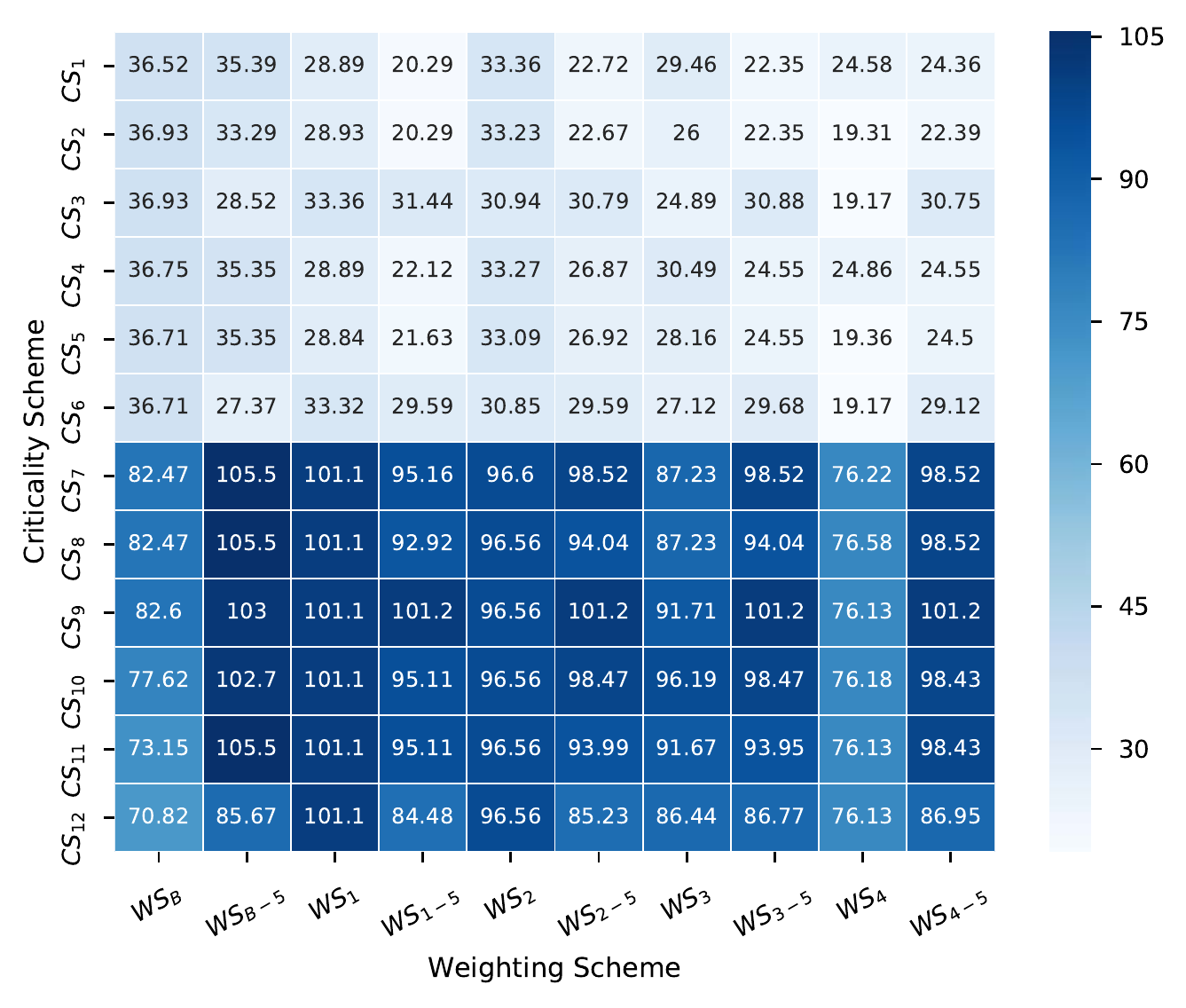}
	\caption{awROCE 0.5\% values}
	\label{fig:cox2_awROCE_0.5}
	\end{subfigure}
\caption{ROCE and awROCE 0.5\% values for the COX-2 class}
\end{figure}

\begin{table}[H]
\centering
\resizebox{\textwidth}{!}{
\begin{tabular}{ l l l l c c}
\hline
\thead{Fingerprint} & \thead{Bits} & \thead{Radius} & \thead{Use Features} & \thead{ROCE 0.5\%} & \thead{awROCE 0.5\%}\\
\addlinespace[0.1cm]
\hline
MACCS & N/A & N/A & N/A & 63.18 & 17.72 \\
Morgan & 1024.0 & 1.0 & False & 102.20 & 41.53 \\
Morgan & 1024.0 & 1.0 & True & 68.75 & 19.53 \\
Morgan & 1024.0 & 2.0 & False & 105.92 & 43.11 \\
Morgan & 1024.0 & 2.0 & True & 77.12 & 22.41 \\
Morgan & 1024.0 & 3.0 & False & 99.41 & 36.00 \\
Morgan & 1024.0 & 3.0 & True & 75.26 & 23.21 \\
Morgan & 1024.0 & 4.0 & False & 93.84 & 31.29 \\
Morgan & 1024.0 & 4.0 & True & 72.47 & 22.84 \\
Morgan & 2048.0 & 1.0 & False & 100.34 & 41.44 \\
Morgan & 2048.0 & 1.0 & True & 67.82 & 19.48 \\
Morgan & 2048.0 & 2.0 & False & \textbf{108.70} & \textbf{43.25} \\
Morgan & 2048.0 & 2.0 & True & 77.12 & 21.34 \\
Morgan & 2048.0 & 3.0 & False & 102.20 & 38.50 \\
Morgan & 2048.0 & 3.0 & True & 79.90 & 23.90 \\
Morgan & 2048.0 & 4.0 & False & 103.13 & 38.55 \\
Morgan & 2048.0 & 4.0 & True & 73.40 & 23.12 \\
\hline
\end{tabular}
}
\caption{COX-2 class -- Fingerprint results}
\label{tab:cox2_fingerprint}
\end{table}

%==================================
%	EGFr
%==================================
\subsubsection{EGFr}

 \begin{figure}[H]
%\centering
	% figure A
	\begin{subfigure}[b]{0.48\textwidth}
	\centering
	\includegraphics[width=\textwidth]{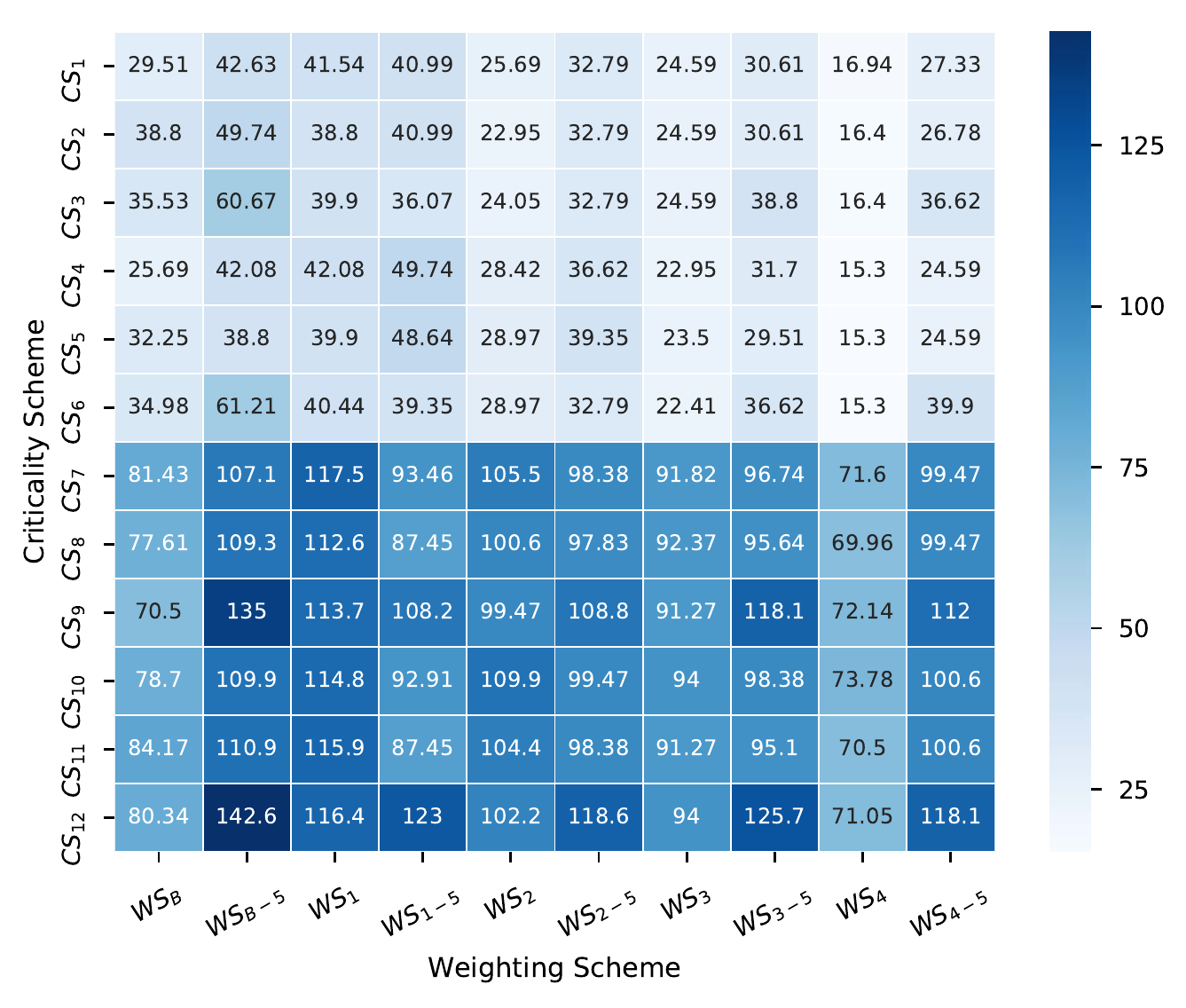}
	\caption{ROCE 0.5\% values}
	\label{fig:egfr_ROCE_0.5}
	\end{subfigure}
        %\quad
        	% figure B
	\begin{subfigure}[b]{0.48\textwidth}
	\centering
	\includegraphics[width=\textwidth]{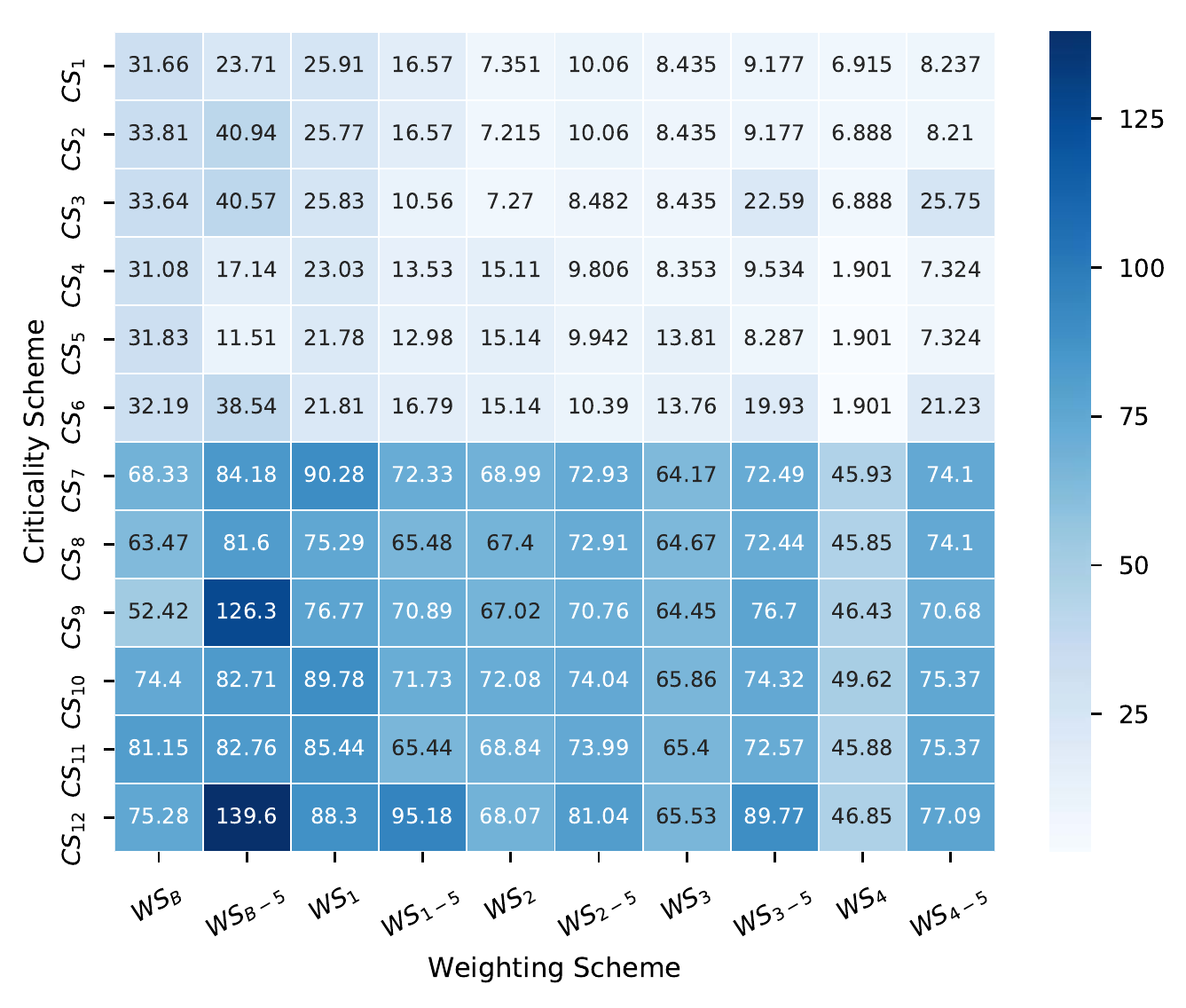}
	\caption{awROCE 0.5\% values}
	\label{fig:egfr_awROCE_0.5}
	\end{subfigure}
\caption{ROCE and awROCE 0.5\% values for the EGFr class}
\end{figure}

\begin{table}[H]
\centering
\resizebox{\textwidth}{!}{
\begin{tabular}{ l l l l c c}
\hline
\thead{Fingerprint} & \thead{Bits} & \thead{Radius} & \thead{Use Features} & \thead{ROCE 0.5\%} & \thead{awROCE 0.5\%}\\
\addlinespace[0.1cm]
\hline
MACCS & N/A & N/A & N/A & 40.44 & 39.81 \\
Morgan & 1024.0 & 1.0 & False & 120.79 & 97.61 \\
Morgan & 1024.0 & 1.0 & True & 59.03 & 52.44 \\
Morgan & 1024.0 & 2.0 & False & 131.72 & 120.80 \\
Morgan & 1024.0 & 2.0 & True & 121.33 & 110.11 \\
Morgan & 1024.0 & 3.0 & False & 128.44 & 112.21 \\
Morgan & 1024.0 & 3.0 & True & 125.16 & 114.32 \\
Morgan & 1024.0 & 4.0 & False & 124.61 & 100.24 \\
Morgan & 1024.0 & 4.0 & True & 120.24 & 104.77 \\
Morgan & 2048.0 & 1.0 & False & 120.24 & 96.78 \\
Morgan & 2048.0 & 1.0 & True & 57.39 & 51.87 \\
Morgan & 2048.0 & 2.0 & False & \textbf{134.45} & 123.57 \\
Morgan & 2048.0 & 2.0 & True & 119.15 & 101.63 \\
Morgan & 2048.0 & 3.0 & False & 132.81 & 123.05 \\
Morgan & 2048.0 & 3.0 & True & 127.34 & 114.38 \\
Morgan & 2048.0 & 4.0 & False & 133.36 & \textbf{123.88} \\
Morgan & 2048.0 & 4.0 & True & 127.89 & 116.29 \\
\hline
\end{tabular}
}
\caption{EGFr class -- Fingerprint results}
\label{tab:egfr_fingerprint}
\end{table}

%==================================
%	FXa
%==================================
\subsubsection{FXa}

\begin{figure}[H]
%\centering
	% figure A
	\begin{subfigure}[b]{0.48\textwidth}
	\centering
	\includegraphics[width=\textwidth]{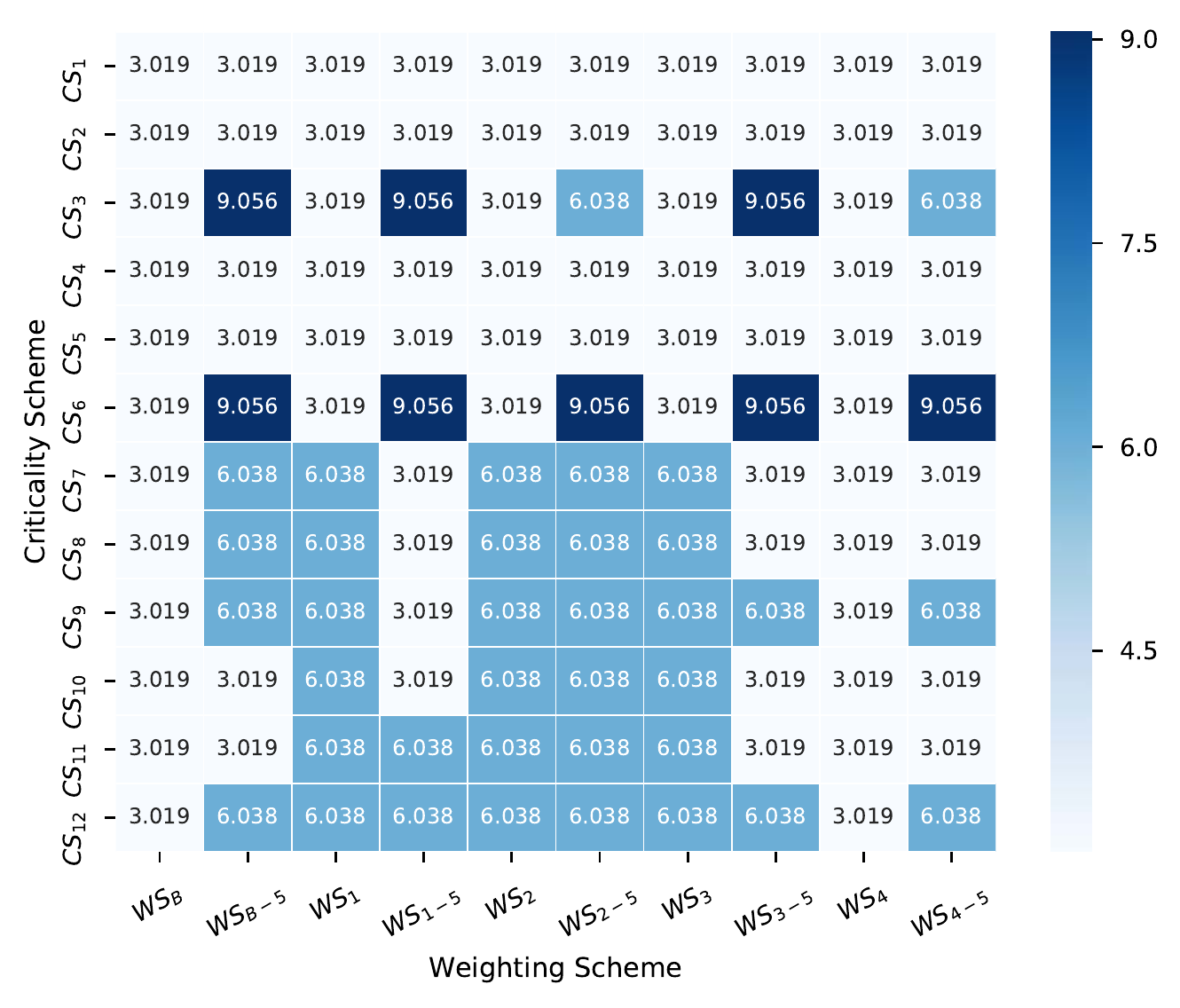}
	\caption{ROCE 0.5\% values}
	\label{fig:fxa_ROCE_0.5}
	\end{subfigure}
        %\quad
        	% figure B
	\begin{subfigure}[b]{0.48\textwidth}
	\centering
	\includegraphics[width=\textwidth]{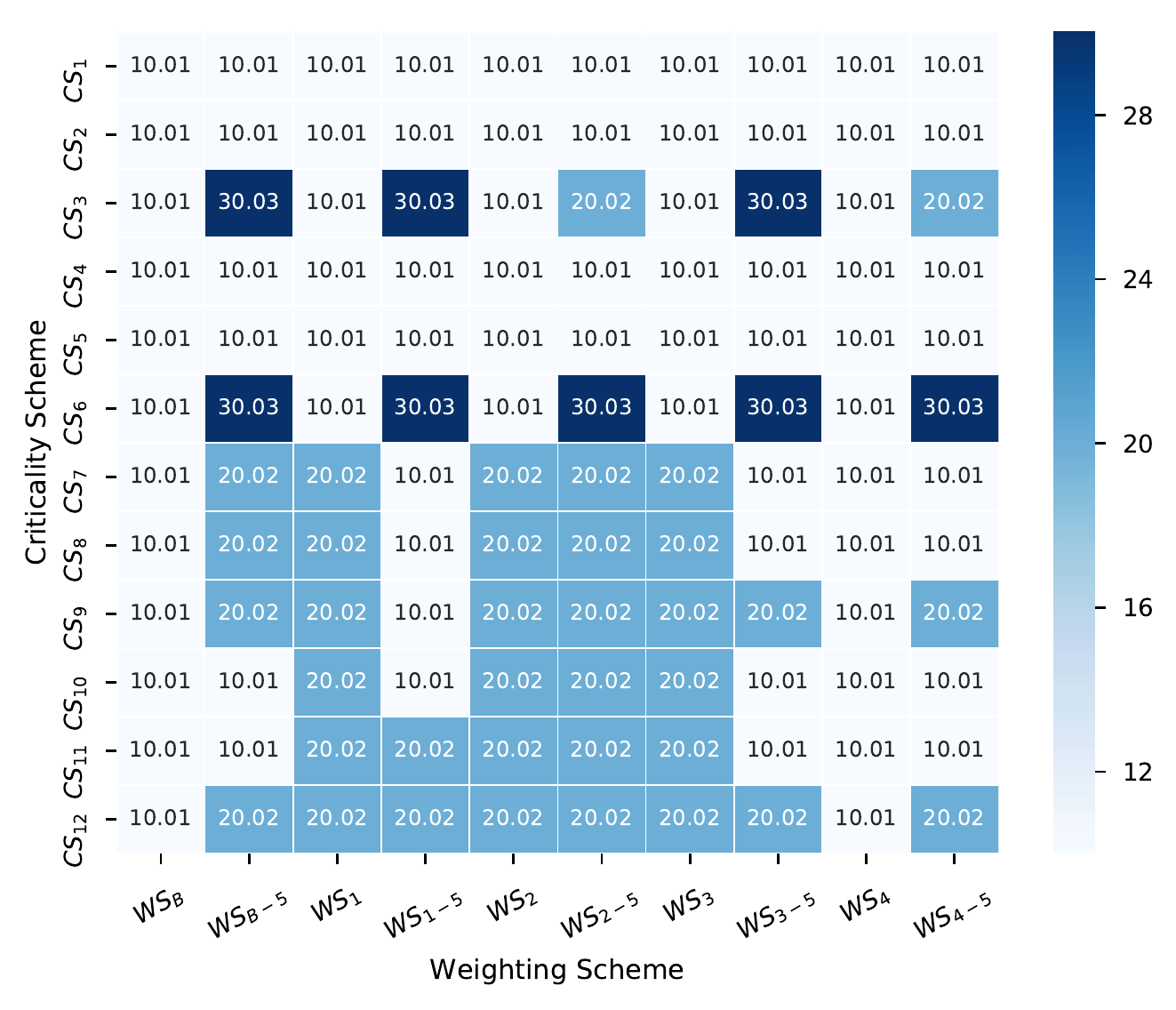}
	\caption{awROCE 0.5\% values}
	\label{fig:fxa_awROCE_0.5}
	\end{subfigure}
\caption{ROCE and awROCE 0.5\% values for the FXa class}
\end{figure}

\begin{table}[H]
\centering
\resizebox{\textwidth}{!}{
\begin{tabular}{ l l l l c c}
\hline
\thead{Fingerprint} & \thead{Bits} & \thead{Radius} & \thead{Use Features} & \thead{ROCE 0.5\%} & \thead{awROCE 0.5\%}\\
\addlinespace[0.1cm]
\hline
MACCS & N/A & N/A & N/A & 9.06 & 30.03 \\
Morgan & 1024.0 & 1.0 & False & 6.04 & 20.02 \\
Morgan & 1024.0 & 1.0 & True & 3.02 & 10.01 \\
Morgan & 1024.0 & 2.0 & False & 9.06 & 20.02 \\
Morgan & 1024.0 & 2.0 & True & 3.02 & 10.01 \\
Morgan & 1024.0 & 3.0 & False & 18.11 & 21.02 \\
Morgan & 1024.0 & 3.0 & True & 3.02 & 10.01 \\
Morgan & 1024.0 & 4.0 & False & \textbf{30.19} & \textbf{32.03} \\
Morgan & 1024.0 & 4.0 & True & 3.02 & 10.01 \\
Morgan & 2048.0 & 1.0 & False & 6.04 & 20.02 \\
Morgan & 2048.0 & 1.0 & True & 3.02 & 10.01 \\
Morgan & 2048.0 & 2.0 & False & 12.08 & 20.35 \\
Morgan & 2048.0 & 2.0 & True & 3.02 & 10.01 \\
Morgan & 2048.0 & 3.0 & False & 9.06 & 20.02 \\
Morgan & 2048.0 & 3.0 & True & 3.02 & 10.01 \\
Morgan & 2048.0 & 4.0 & False & 18.11 & 30.70 \\
Morgan & 2048.0 & 4.0 & True & 3.02 & 10.01 \\
\hline
\end{tabular}
}
\caption{FXa class -- Fingerprint results}
\label{tab:fxa_fingerprint}
\end{table}

%==================================
%	HIVRT
%==================================
\subsubsection{HIVRT}

\begin{figure}[H]
%\centering
	% figure A
	\begin{subfigure}[b]{0.48\textwidth}
	\centering
	\includegraphics[width=\textwidth]{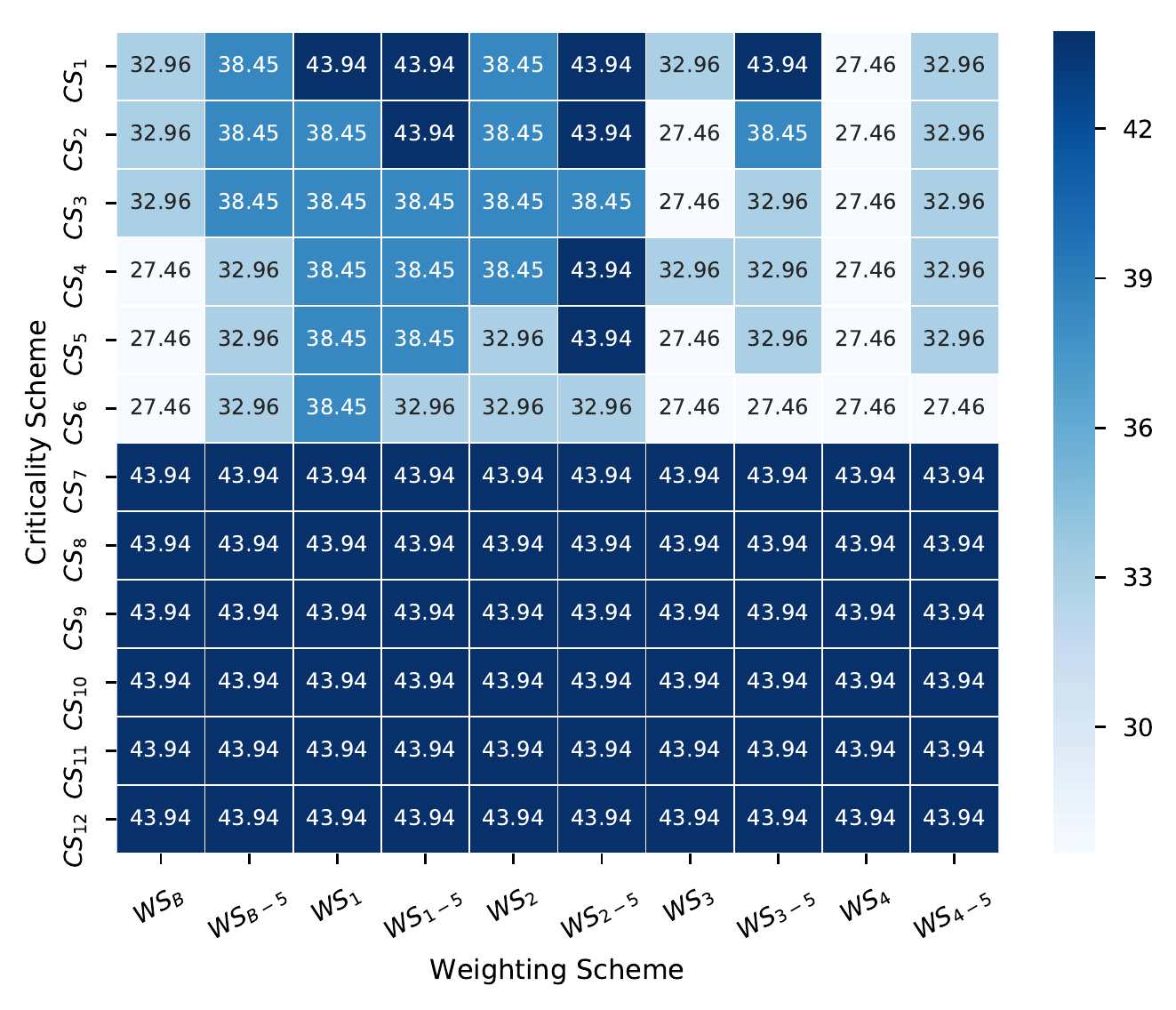}
	\caption{ROCE 0.5\% values}
	\label{fig:hivrt_ROCE_0.5}
	\end{subfigure}
        %\quad
        	% figure B
	\begin{subfigure}[b]{0.48\textwidth}
	\centering
	\includegraphics[width=\textwidth]{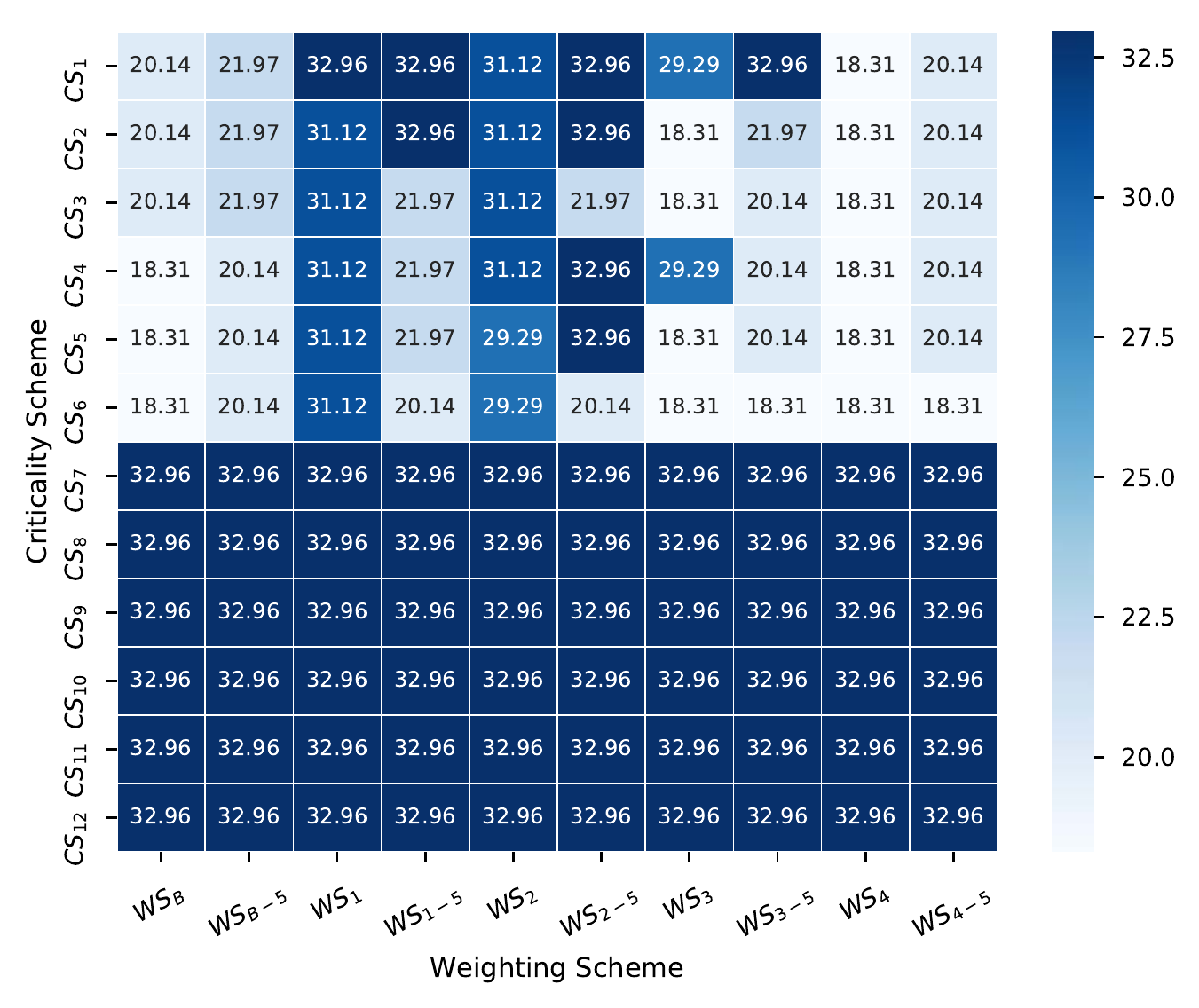}
	\caption{awROCE 0.5\% values}
	\label{fig:hivrt_awROCE_0.5}
	\end{subfigure}
\caption{ROCE and awROCE 0.5\% values for the HIVRT class}
\end{figure}

\begin{table}[H]
\centering
\resizebox{\textwidth}{!}{
\begin{tabular}{ l l l l c c}
\hline
\thead{Fingerprint} & \thead{Bits} & \thead{Radius} & \thead{Use Features} & \thead{ROCE 0.5\%} & \thead{awROCE 0.5\%}\\
\addlinespace[0.1cm]
\hline
MACCS & N/A & N/A & N/A & 27.46 & 18.31 \\
Morgan & 1024.0 & 1.0 & False & 32.96 & 20.14 \\
Morgan & 1024.0 & 1.0 & True & 32.96 & 29.29 \\
Morgan & 1024.0 & 2.0 & False & 38.45 & 31.12 \\
Morgan & 1024.0 & 2.0 & True & 32.96 & 29.29 \\
Morgan & 1024.0 & 3.0 & False & 32.96 & 29.29 \\
Morgan & 1024.0 & 3.0 & True & 32.96 & 29.29 \\
Morgan & 1024.0 & 4.0 & False & 32.96 & 29.29 \\
Morgan & 1024.0 & 4.0 & True & 32.96 & 29.29 \\
Morgan & 2048.0 & 1.0 & False & 32.96 & 20.14 \\
Morgan & 2048.0 & 1.0 & True & 32.96 & 29.29 \\
Morgan & 2048.0 & 2.0 & False & \textbf{43.94} & \textbf{32.96} \\
Morgan & 2048.0 & 2.0 & True & 32.96 & 29.29 \\
Morgan & 2048.0 & 3.0 & False & 38.45 & 31.12 \\
Morgan & 2048.0 & 3.0 & True & 32.96 & 29.29 \\
Morgan & 2048.0 & 4.0 & False & 38.45 & 31.12 \\
Morgan & 2048.0 & 4.0 & True & 32.96 & 29.29 \\
\hline
\end{tabular}
}
\caption{HIVRT class -- Fingerprint results}
\label{tab:hivrt_fingerprint}
\end{table}

%==================================
%	InhA
%==================================
\subsubsection{InhA}

\begin{figure}[H]
%\centering
	% figure A
	\begin{subfigure}[b]{0.48\textwidth}
	\centering
	\includegraphics[width=\textwidth]{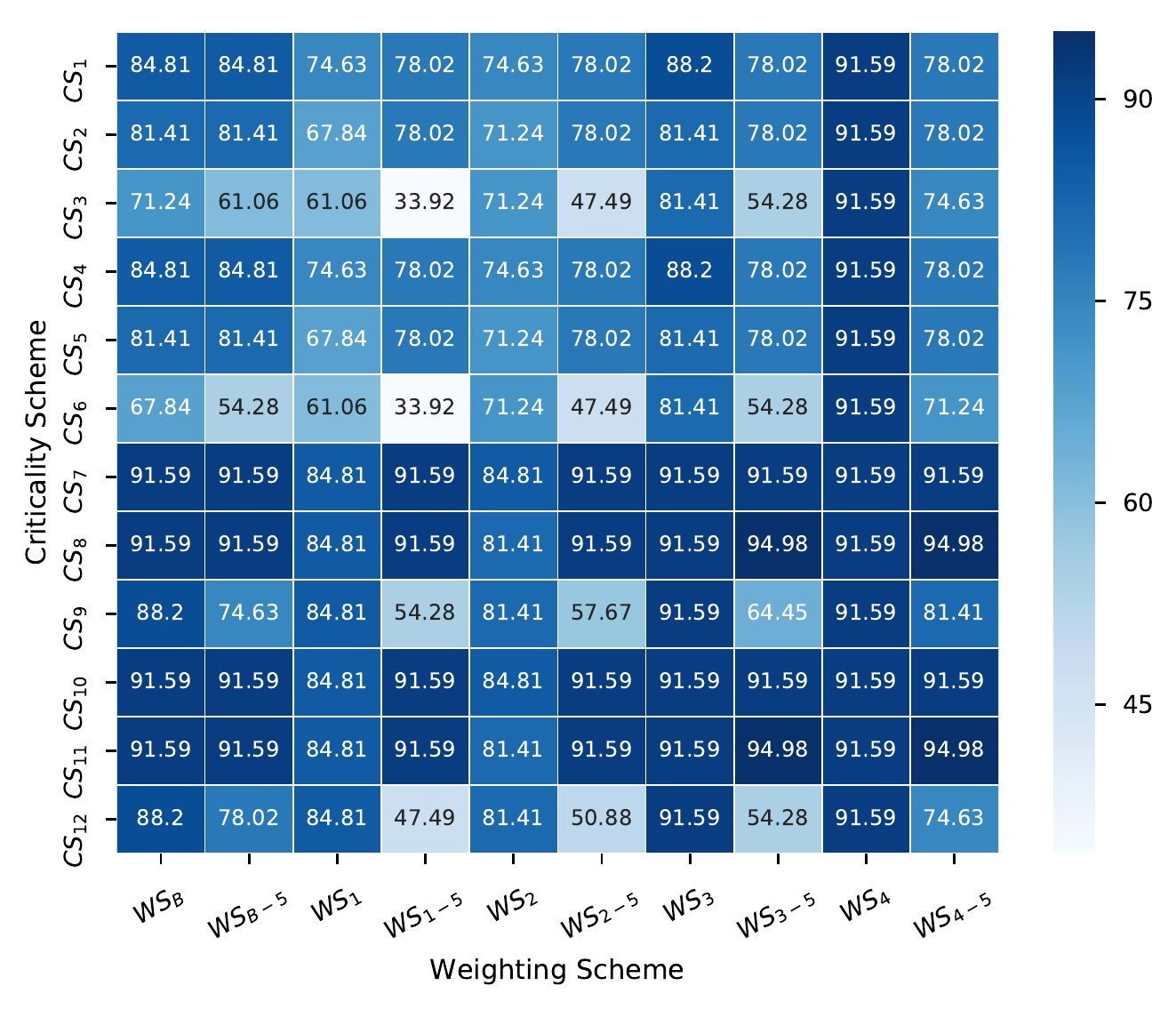}
	\caption{ROCE 0.5\% values}
	\label{fig:inha_ROCE_0.5}
	\end{subfigure}
        %\quad
        	% figure B
	\begin{subfigure}[b]{0.48\textwidth}
	\centering
	\includegraphics[width=\textwidth]{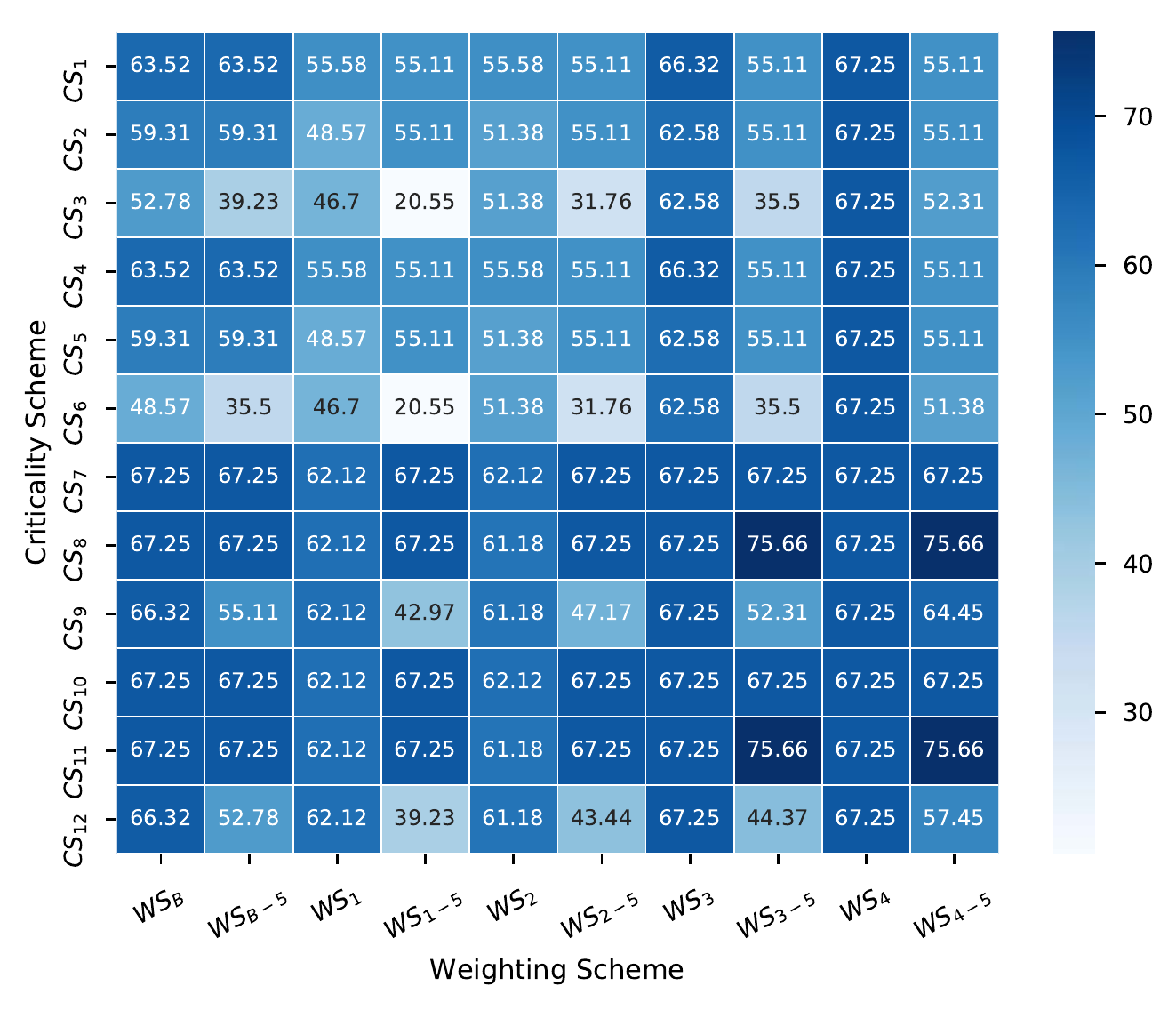}
	\caption{awROCE 0.5\% values}
	\label{fig:inha_awROCE_0.5}
	\end{subfigure}
\caption{ROCE and awROCE 0.5\% values for the InhA class}
\end{figure}

\begin{table}[H]
\centering
\resizebox{\textwidth}{!}{
\begin{tabular}{ l l l l c c}
\hline
\thead{Fingerprint} & \thead{Bits} & \thead{Radius} & \thead{Use Features} & \thead{ROCE 0.5\%} & \thead{awROCE 0.5\%}\\
\addlinespace[0.1cm]
\hline
MACCS & N/A & N/A & N/A & 64.45 & 41.50 \\
Morgan & 1024.0 & 1.0 & False & 84.81 & 57.91 \\
Morgan & 1024.0 & 1.0 & True & 91.59 & 80.33 \\
Morgan & 1024.0 & 2.0 & False & 88.20 & 66.32 \\
Morgan & 1024.0 & 2.0 & True & 91.59 & 74.73 \\
Morgan & 1024.0 & 3.0 & False & 88.20 & 66.32 \\
Morgan & 1024.0 & 3.0 & True & 91.59 & 74.73 \\
Morgan & 1024.0 & 4.0 & False & 88.20 & 66.32 \\
Morgan & 1024.0 & 4.0 & True & 88.20 & 66.32 \\
Morgan & 2048.0 & 1.0 & False & 84.81 & 57.91 \\
Morgan & 2048.0 & 1.0 & True & 94.98 & 83.13 \\
Morgan & 2048.0 & 2.0 & False & 88.20 & 66.32 \\
Morgan & 2048.0 & 2.0 & True & \textbf{98.37} & \textbf{91.54} \\
Morgan & 2048.0 & 3.0 & False & 88.20 & 66.32 \\
Morgan & 2048.0 & 3.0 & True & 91.59 & 74.73 \\
Morgan & 2048.0 & 4.0 & False & 88.20 & 66.32 \\
Morgan & 2048.0 & 4.0 & True & 94.98 & 83.13 \\
\hline
\end{tabular}
}
\caption{InhA class -- Fingerprint results}
\label{tab:inha_fingerprint}
\end{table}

%==================================
%	P38
%==================================
\subsubsection{P38 MAP}

\begin{figure}[H]
%\centering
	% figure A
	\begin{subfigure}[b]{0.48\textwidth}
	\centering
	\includegraphics[width=\textwidth]{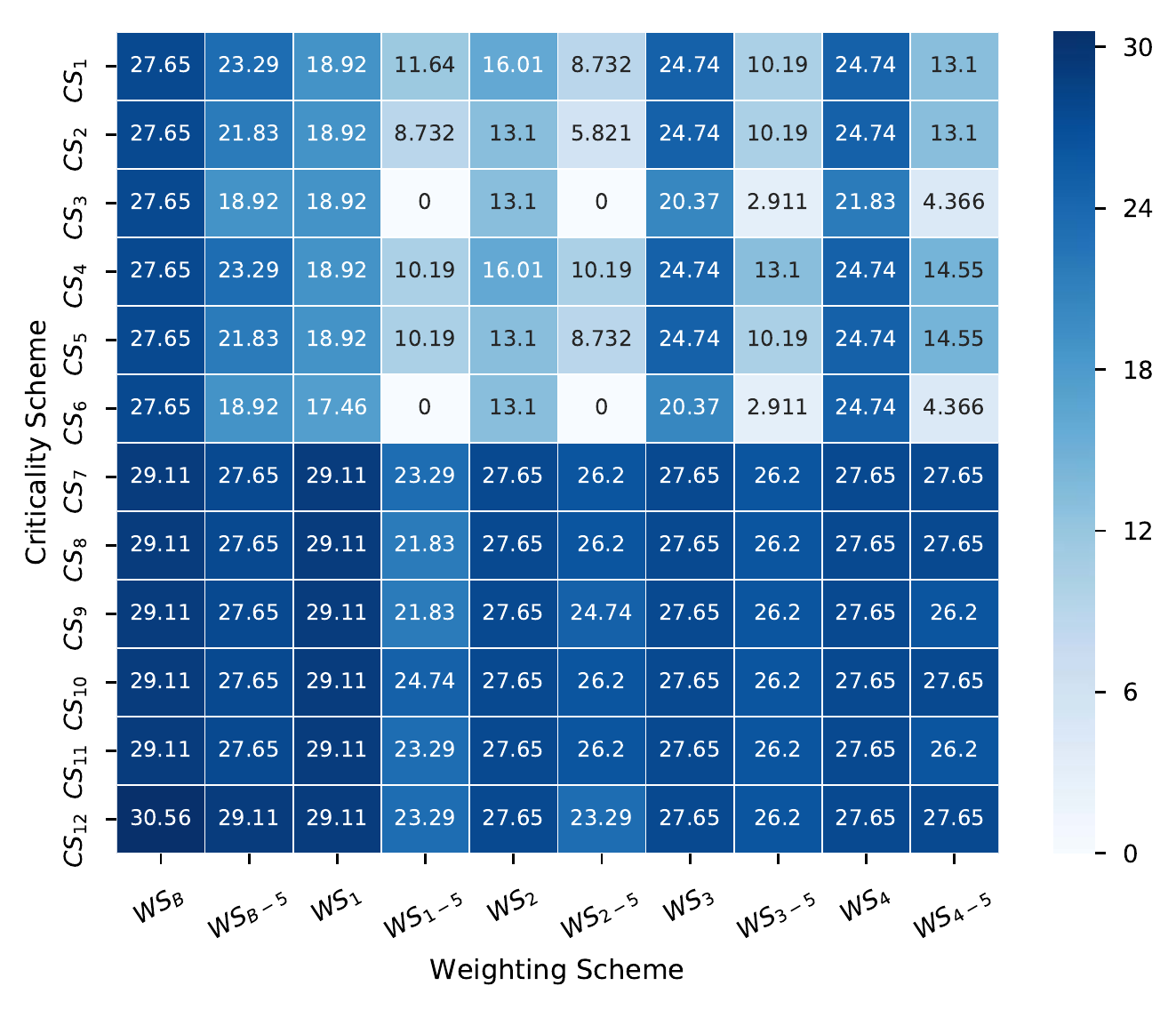}
	\caption{ROCE 0.5\% values}
	\label{fig:p38_ROCE_0.5}
	\end{subfigure}
        %\quad
        	% figure B
	\begin{subfigure}[b]{0.48\textwidth}
	\centering
	\includegraphics[width=\textwidth]{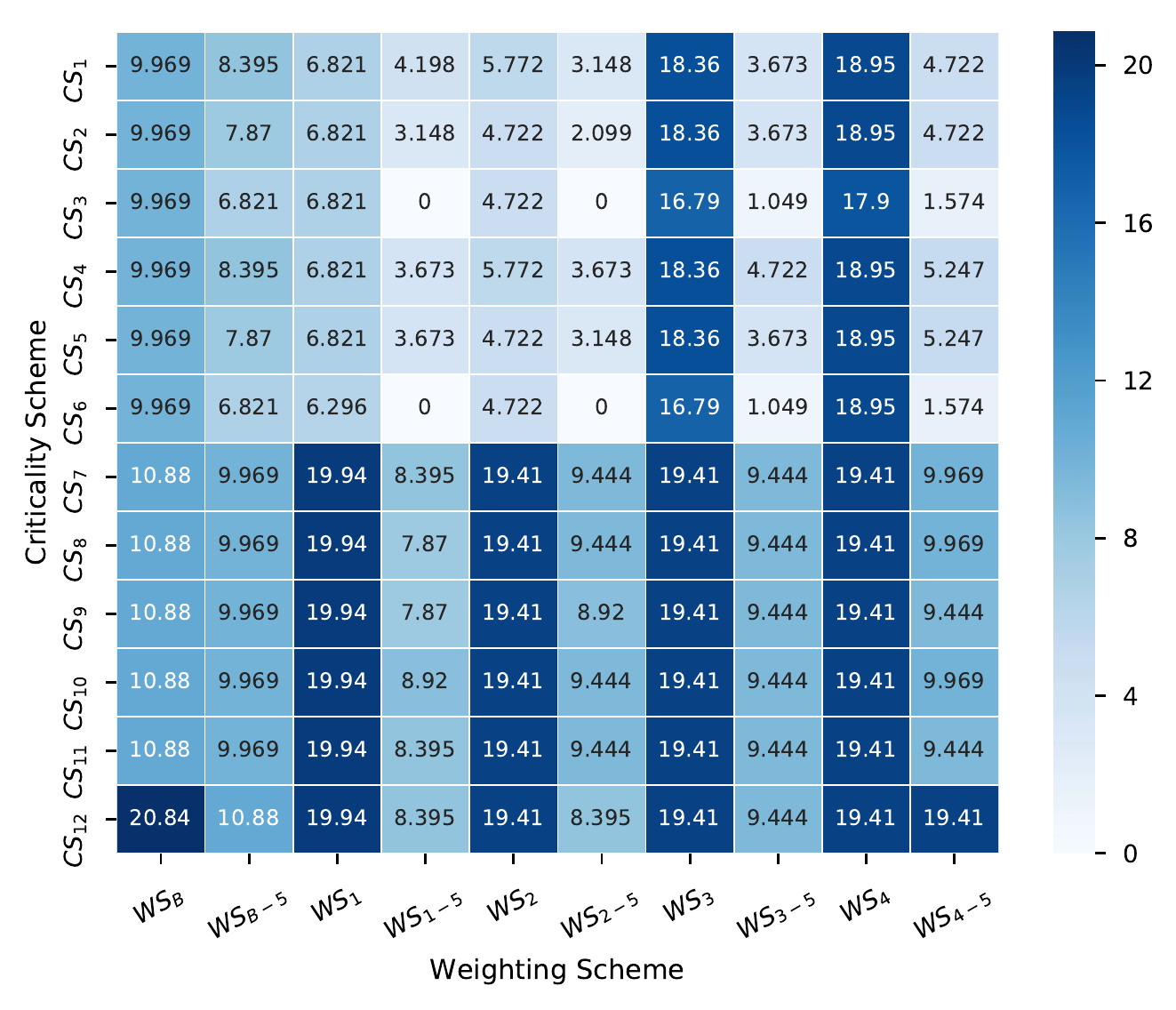}
	\caption{awROCE 0.5\% values}
	\label{fig:p38_awROCE_0.5}
	\end{subfigure}
\caption{ROCE and awROCE 0.5\% values for the P38 MAP class}
\end{figure}

\begin{table}[H]
\centering
\resizebox{\textwidth}{!}{
\begin{tabular}{ l l l l c c}
\hline
\thead{Fingerprint} & \thead{Bits} & \thead{Radius} & \thead{Use Features} & \thead{ROCE 0.5\%} & \thead{awROCE 0.5\%}\\
\addlinespace[0.1cm]
\hline
MACCS & N/A & N/A & N/A & 2.91 & 1.05 \\
Morgan & 1024.0 & 1.0 & False & 14.55 & 5.25 \\
Morgan & 1024.0 & 1.0 & True & 1.46 & 0.52 \\
Morgan & 1024.0 & 2.0 & False & 14.55 & 5.25 \\
Morgan & 1024.0 & 2.0 & True & 8.73 & 3.15 \\
Morgan & 1024.0 & 3.0 & False & 10.19 & 3.67 \\
Morgan & 1024.0 & 3.0 & True & 10.19 & 3.67 \\
Morgan & 1024.0 & 4.0 & False & 10.19 & 3.67 \\
Morgan & 1024.0 & 4.0 & True & 7.28 & 2.62 \\
Morgan & 2048.0 & 1.0 & False & \textbf{16.01} & \textbf{5.77} \\
Morgan & 2048.0 & 1.0 & True & 4.37 & 1.57 \\
Morgan & 2048.0 & 2.0 & False & \textbf{16.01} & \textbf{5.77} \\
Morgan & 2048.0 & 2.0 & True & 8.73 & 3.15 \\
Morgan & 2048.0 & 3.0 & False & 13.10 & 4.72 \\
Morgan & 2048.0 & 3.0 & True & 11.64 & 4.20 \\
Morgan & 2048.0 & 4.0 & False & 13.10 & 4.72 \\
Morgan & 2048.0 & 4.0 & True & 7.28 & 2.62 \\
\hline
\end{tabular}
}
\caption{P38 MAP class -- Fingerprint results}
\label{tab:p38_fingerprint}
\end{table}

%==================================
%	PDE5
%==================================
\subsubsection{PDE5}

\begin{figure}[H]
%\centering
	% figure A
	\begin{subfigure}[b]{0.48\textwidth}
	\centering
	\includegraphics[width=\textwidth]{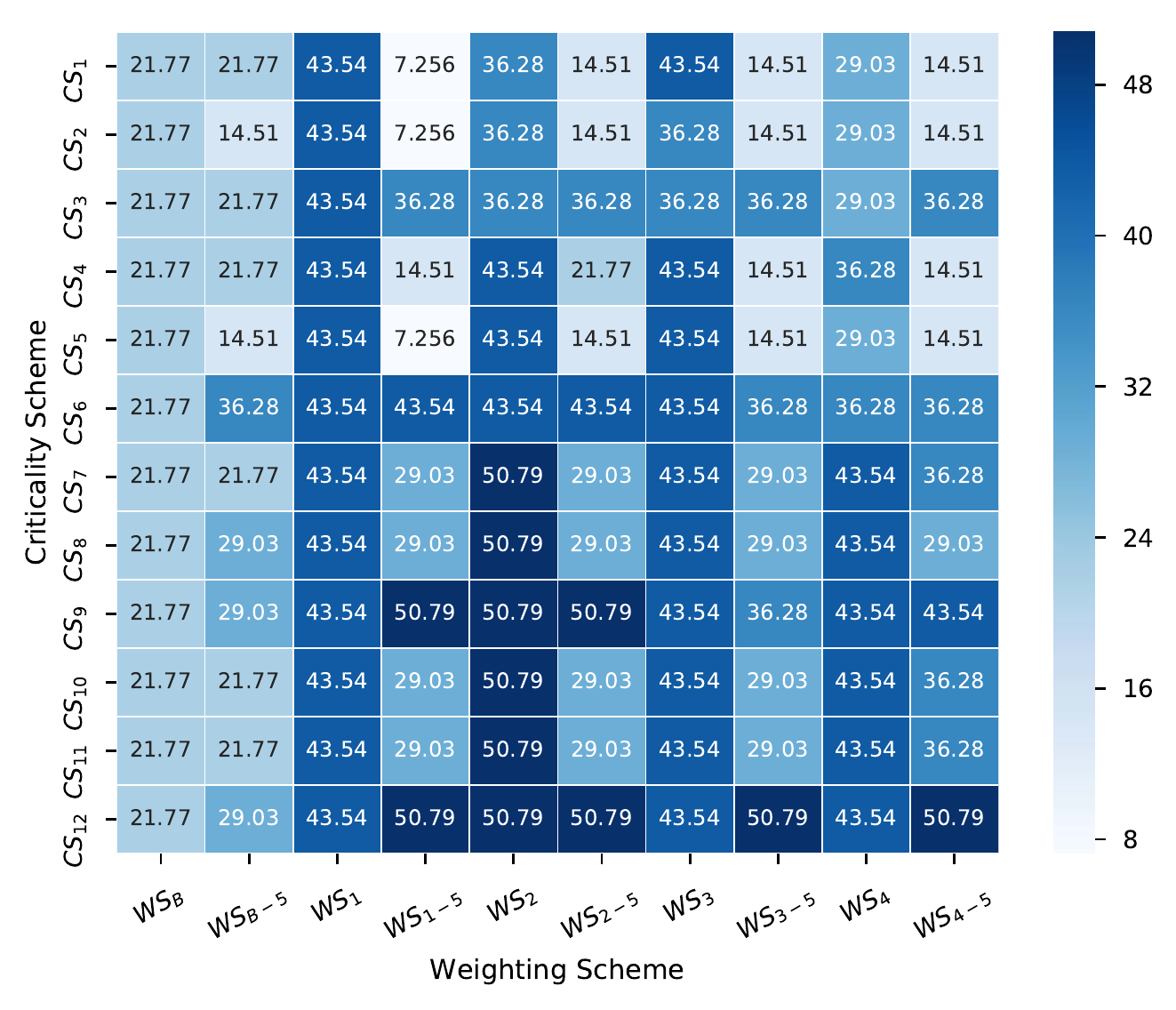}
	\caption{ROCE 0.5\% values}
	\label{fig:pde5_ROCE_0.5}
	\end{subfigure}
        %\quad
        	% figure B
	\begin{subfigure}[b]{0.48\textwidth}
	\centering
	\includegraphics[width=\textwidth]{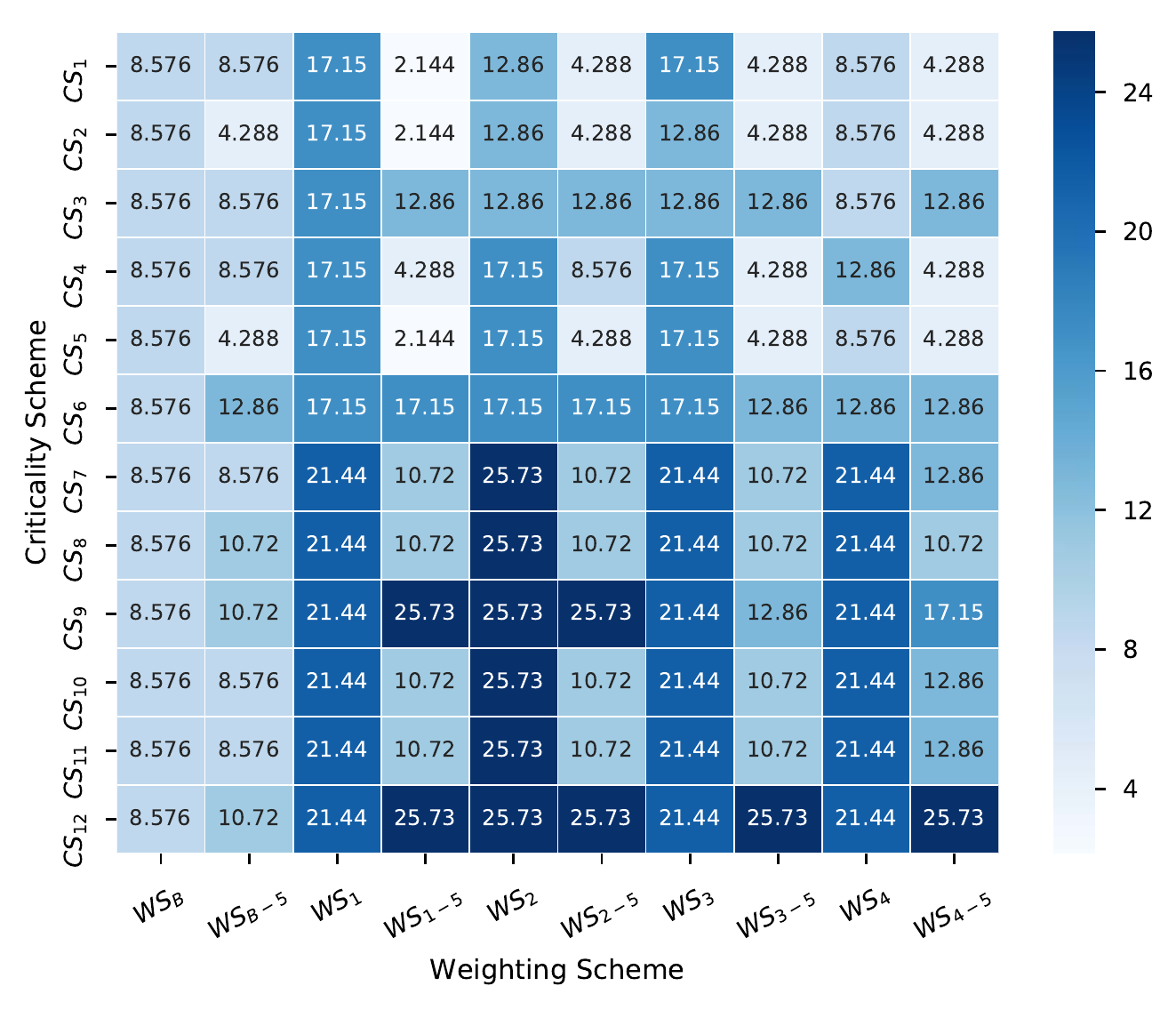}
	\caption{awROCE 0.5\% values}
	\label{fig:pde5_awROCE_0.5}
	\end{subfigure}
\caption{ROCE and awROCE 0.5\% values for the PDE5 class}
\end{figure}

\begin{table}[H]
\centering
\resizebox{\textwidth}{!}{
\begin{tabular}{ l l l l c c}
\hline
\thead{Fingerprint} & \thead{Bits} & \thead{Radius} & \thead{Use Features} & \thead{ROCE 0.5\%} & \thead{awROCE 0.5\%}\\
\addlinespace[0.1cm]
\hline
MACCS & N/A & N/A & N/A & 7.26 & 4.29 \\
Morgan & 1024.0 & 1.0 & False & 29.03 & \textbf{17.15} \\
Morgan & 1024.0 & 1.0 & True & 7.26 & 2.14 \\
Morgan & 1024.0 & 2.0 & False & 21.77 & 8.58 \\
Morgan & 1024.0 & 2.0 & True & 21.77 & 8.58 \\
Morgan & 1024.0 & 3.0 & False & 21.77 & 8.58 \\
Morgan & 1024.0 & 3.0 & True & 21.77 & 8.58 \\
Morgan & 1024.0 & 4.0 & False & 21.77 & 8.58 \\
Morgan & 1024.0 & 4.0 & True & 36.28 & 15.01 \\
Morgan & 2048.0 & 1.0 & False & 29.03 & \textbf{17.15} \\
Morgan & 2048.0 & 1.0 & True & 7.26 & 2.14 \\
Morgan & 2048.0 & 2.0 & False & 21.77 & 8.58 \\
Morgan & 2048.0 & 2.0 & True & 21.77 & 8.58 \\
Morgan & 2048.0 & 3.0 & False & 21.77 & 8.58 \\
Morgan & 2048.0 & 3.0 & True & 36.28 & 12.86 \\
Morgan & 2048.0 & 4.0 & False & 21.77 & 8.58 \\
Morgan & 2048.0 & 4.0 & True & \textbf{43.54} & \textbf{17.15} \\
\hline
\end{tabular}
}
\caption{PDE5 class -- Fingerprint results}
\label{tab:pde5_fingerprint}
\end{table}

%==================================
%	PDGFrb
%==================================
\subsubsection{PDGFrb}

\begin{figure}[H]
%\centering
	% figure A
	\begin{subfigure}[b]{0.48\textwidth}
	\centering
	\includegraphics[width=\textwidth]{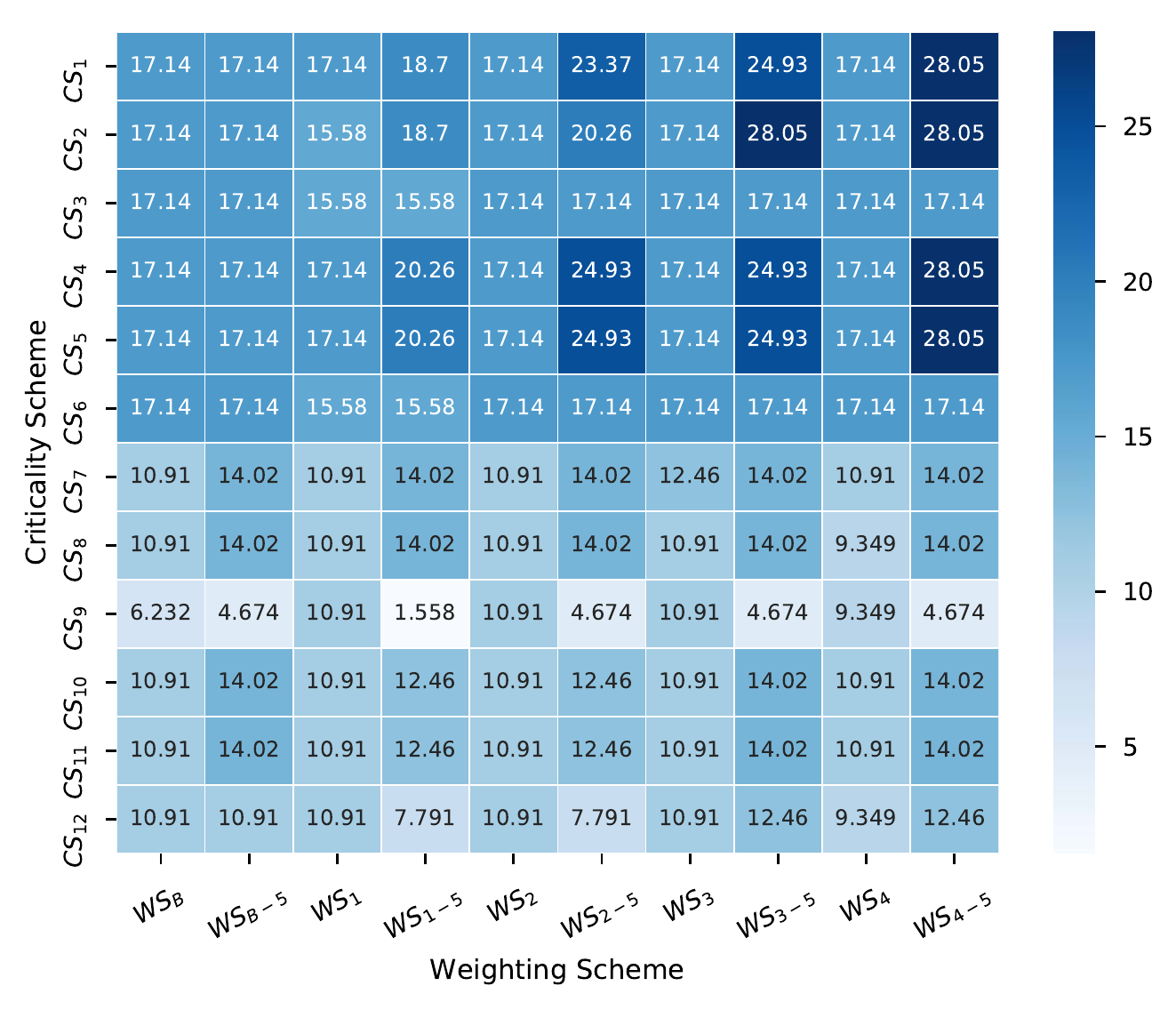}
	\caption{ROCE 0.5\% values}
	\label{fig:pdgfrb_ROCE_0.5}
	\end{subfigure}
        %\quad
        	% figure B
	\begin{subfigure}[b]{0.48\textwidth}
	\centering
	\includegraphics[width=\textwidth]{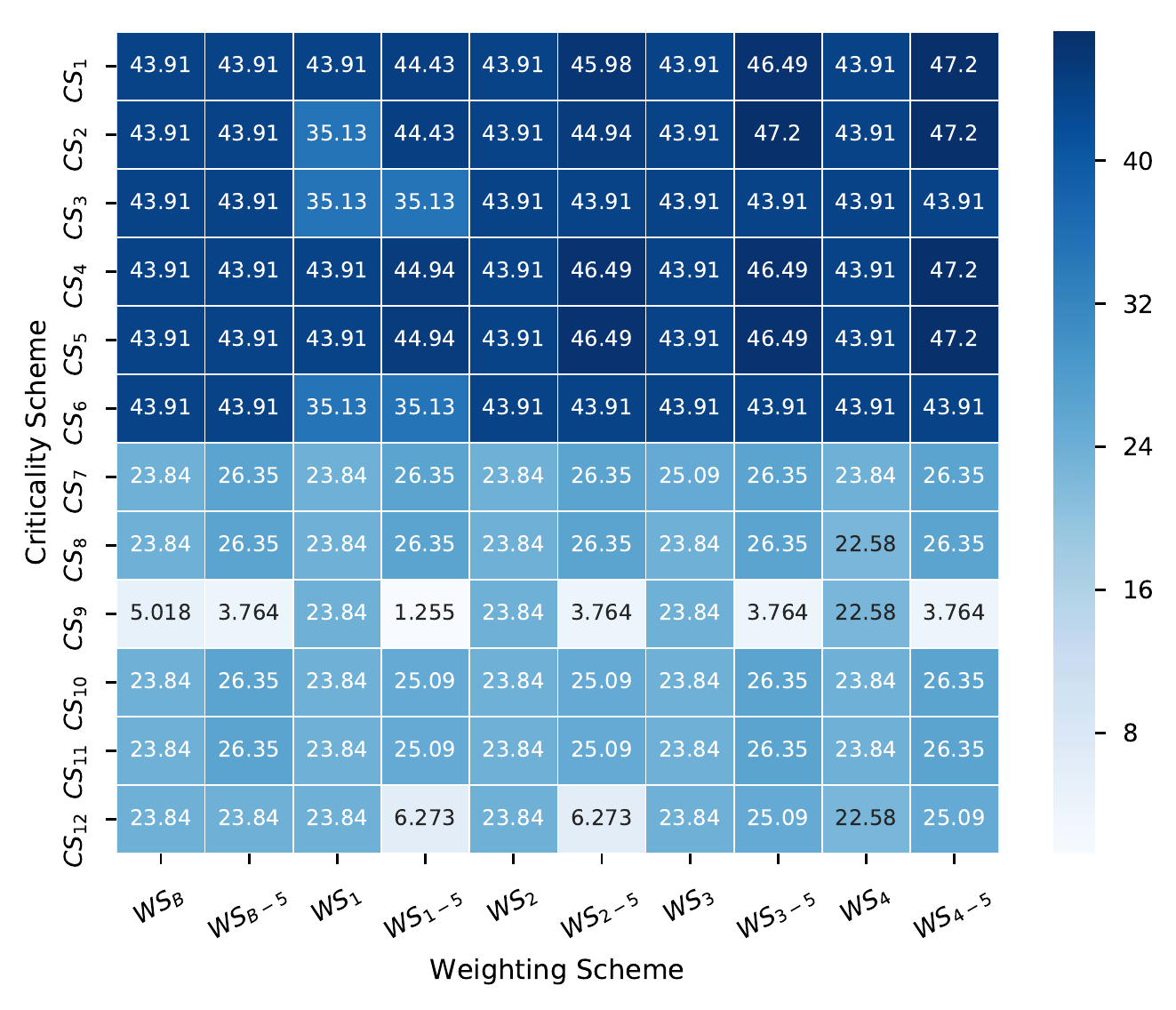}
	\caption{awROCE 0.5\% values}
	\label{fig:pdgfrb_awROCE_0.5}
	\end{subfigure}
\caption{ROCE and awROCE 0.5\% values for the PDGFrb class}
\end{figure}

\begin{table}[H]
\centering
\resizebox{\textwidth}{!}{
\begin{tabular}{ l l l l c c}
\hline
\thead{Fingerprint} & \thead{Bits} & \thead{Radius} & \thead{Use Features} & \thead{ROCE 0.5\%} & \thead{awROCE 0.5\%}\\
\addlinespace[0.1cm]
\hline
MACCS & N/A & N/A & N/A & 14.02 & 23.39 \\
Morgan & 1024.0 & 1.0 & False & 20.26 & 44.94 \\
Morgan & 1024.0 & 1.0 & True & 17.14 & 43.91 \\
Morgan & 1024.0 & 2.0 & False & 21.81 & 46.70 \\
Morgan & 1024.0 & 2.0 & True & 17.14 & 43.91 \\
Morgan & 1024.0 & 3.0 & False & \textbf{24.93} & \textbf{47.73} \\
Morgan & 1024.0 & 3.0 & True & 17.14 & 43.91 \\
Morgan & 1024.0 & 4.0 & False & 21.81 & 45.46 \\
Morgan & 1024.0 & 4.0 & True & 17.14 & 43.91 \\
Morgan & 2048.0 & 1.0 & False & 21.81 & 45.46 \\
Morgan & 2048.0 & 1.0 & True & 17.14 & 43.91 \\
Morgan & 2048.0 & 2.0 & False & 20.26 & 44.94 \\
Morgan & 2048.0 & 2.0 & True & 17.14 & 43.91 \\
Morgan & 2048.0 & 3.0 & False & 18.70 & 44.43 \\
Morgan & 2048.0 & 3.0 & True & 17.14 & 43.91 \\
Morgan & 2048.0 & 4.0 & False & 17.14 & 43.91 \\
Morgan & 2048.0 & 4.0 & True & 17.14 & 43.91 \\
\hline
\end{tabular}
}
\caption{PDGFrb class -- Fingerprint results}
\label{tab:pdgfrb_fingerprint}
\end{table}

%==================================
%	SRC
%==================================
\subsubsection{Src}

\begin{figure}[H]
%\centering
	% figure A
	\begin{subfigure}[b]{0.48\textwidth}
	\centering
	\includegraphics[width=\textwidth]{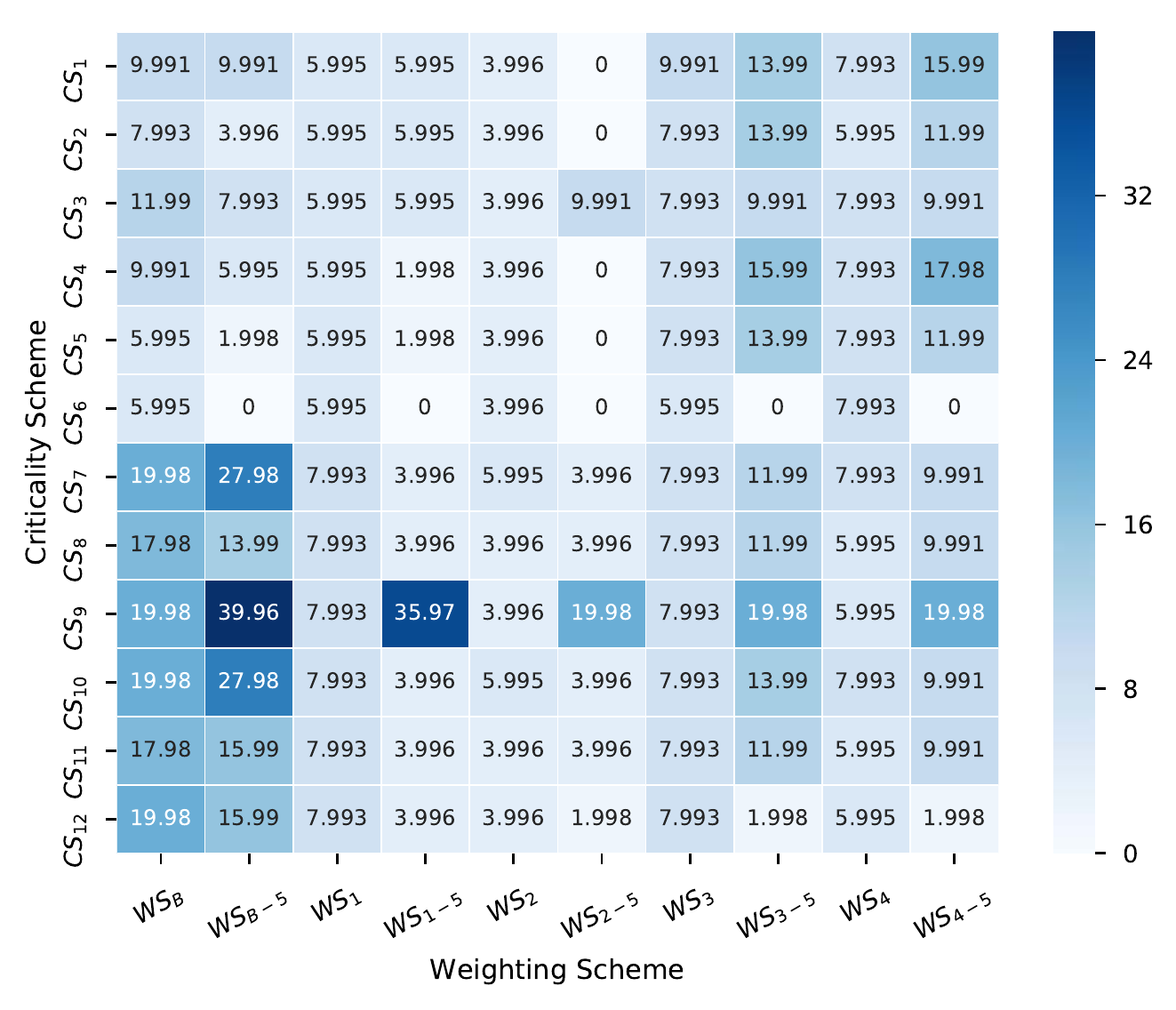}
	\caption{ROCE 0.5\% values}
	\label{fig:src_ROCE_0.5}
	\end{subfigure}
        %\quad
        	% figure B
	\begin{subfigure}[b]{0.48\textwidth}
	\centering
	\includegraphics[width=\textwidth]{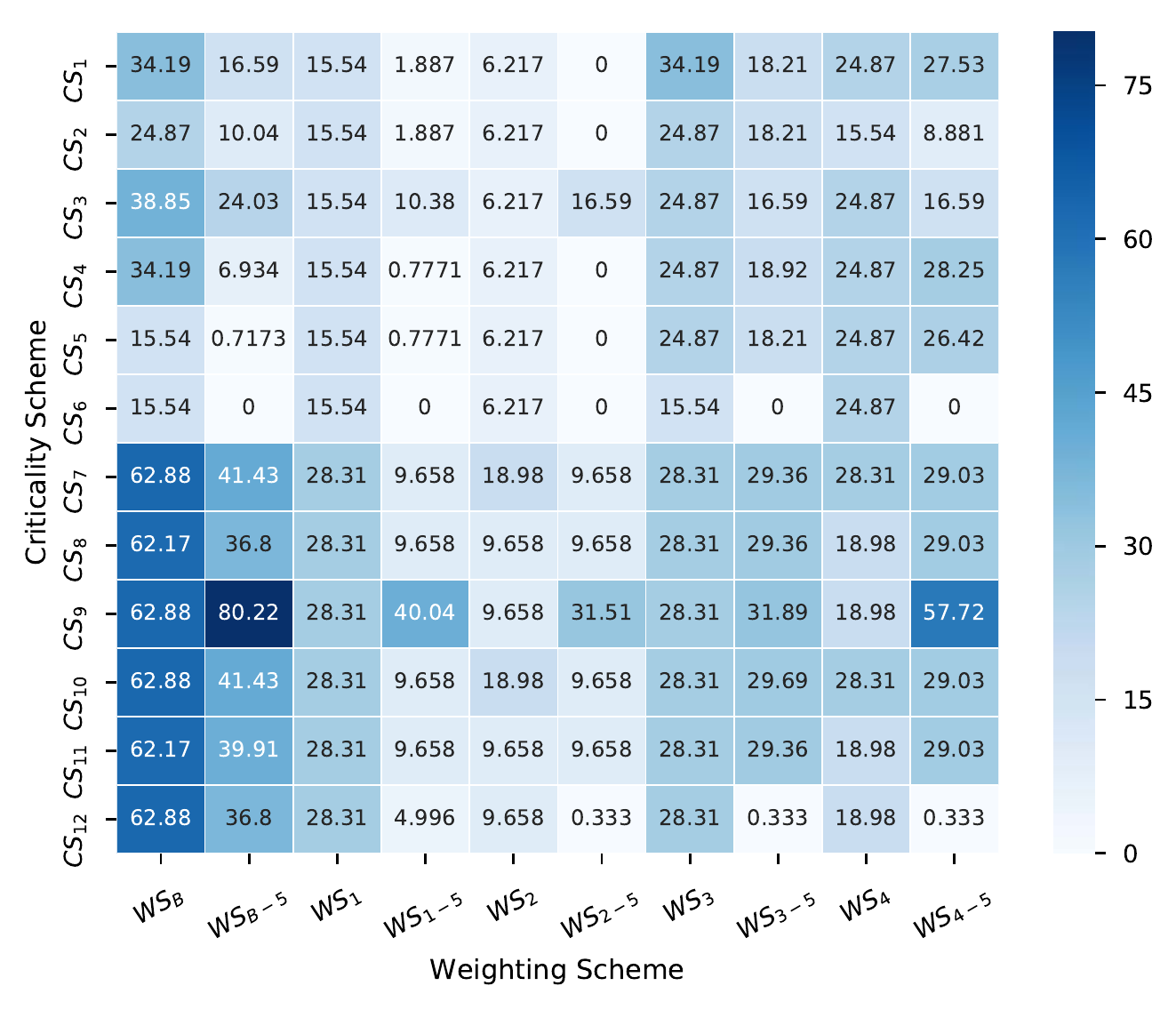}
	\caption{awROCE 0.5\% values}
	\label{fig:src_awROCE_0.5}
	\end{subfigure}
\caption{ROCE and awROCE 0.5\% values for the Src class}
\end{figure}

\begin{table}[H]
\centering
\resizebox{\textwidth}{!}{
\begin{tabular}{ l l l l c c}
\hline
\thead{Fingerprint} & \thead{Bits} & \thead{Radius} & \thead{Use Features} & \thead{ROCE 0.5\%} & \thead{awROCE 0.5\%}\\
\addlinespace[0.1cm]
\hline
MACCS & N/A & N/A & N/A & 0.00 & 0.00 \\
Morgan & 1024.0 & 1.0 & False & 15.99 & 5.35 \\
Morgan & 1024.0 & 1.0 & True & 2.00 & 0.72 \\
Morgan & 1024.0 & 2.0 & False & 23.98 & 8.61 \\
Morgan & 1024.0 & 2.0 & True & 19.98 & 7.41 \\
Morgan & 1024.0 & 3.0 & False & 17.98 & 6.46 \\
Morgan & 1024.0 & 3.0 & True & 19.98 & 7.29 \\
Morgan & 1024.0 & 4.0 & False & 13.99 & 5.02 \\
Morgan & 1024.0 & 4.0 & True & 23.98 & 8.79 \\
Morgan & 2048.0 & 1.0 & False & 25.98 & 8.94 \\
Morgan & 2048.0 & 1.0 & True & 2.00 & 0.72 \\
Morgan & 2048.0 & 2.0 & False & 27.98 & 9.66 \\
Morgan & 2048.0 & 2.0 & True & 21.98 & 8.13 \\
Morgan & 2048.0 & 3.0 & False & \textbf{29.97} & \textbf{9.99} \\
Morgan & 2048.0 & 3.0 & True & 23.98 & 8.85 \\
Morgan & 2048.0 & 4.0 & False & 25.98 & 9.33 \\
Morgan & 2048.0 & 4.0 & True & 19.98 & 7.23 \\
\hline
\end{tabular}
}
\caption{Src class -- Fingerprint results}
\label{tab:src_fingerprint}
\end{table}

%==================================
%	VEGFr2
%==================================
\subsubsection{VEGFr-2}

\begin{figure}[H]
%\centering
	% figure A
	\begin{subfigure}[b]{0.48\textwidth}
	\centering
	\includegraphics[width=\textwidth]{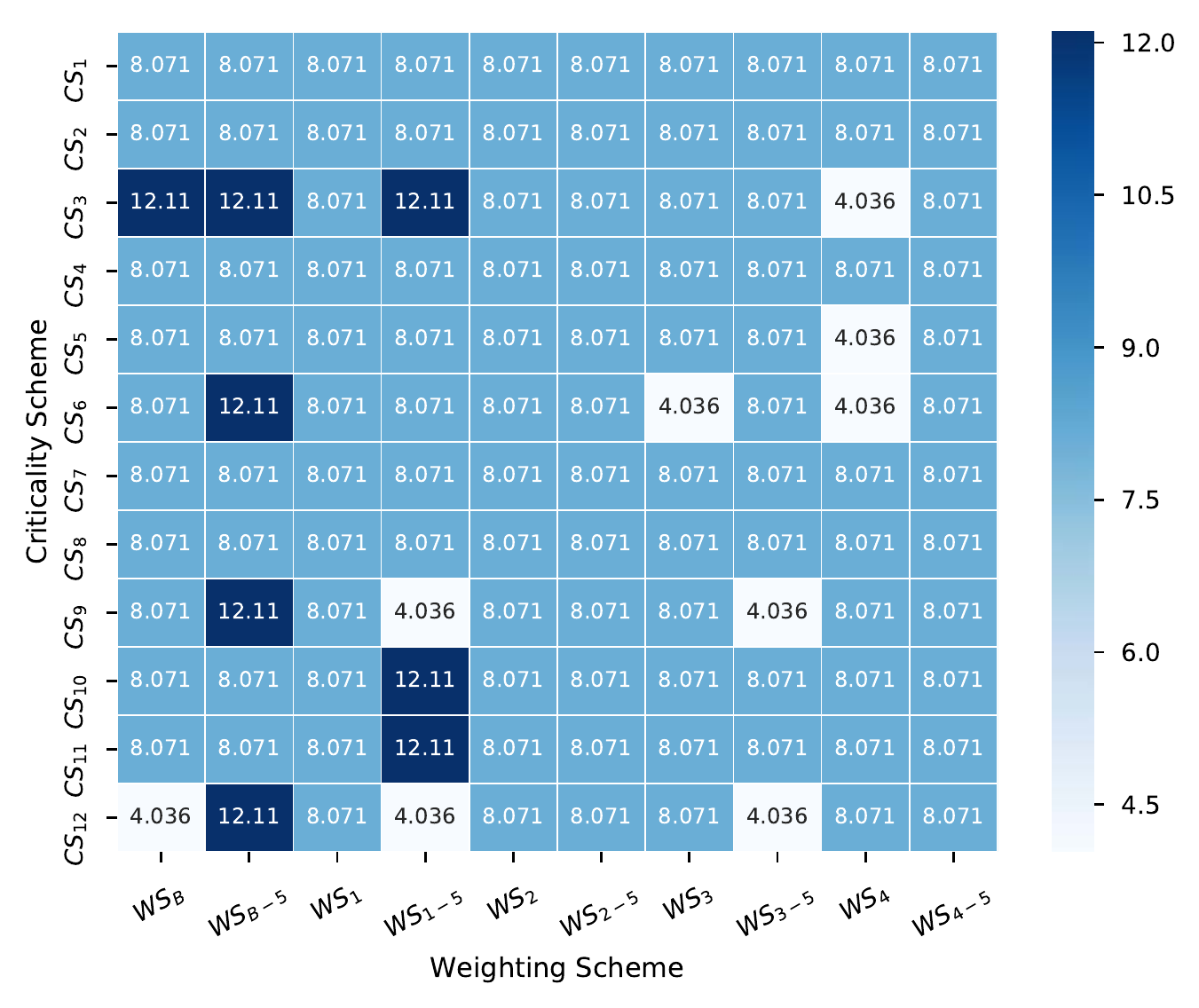}
	\caption{ROCE 0.5\% values}
	\label{fig:vegfr2_ROCE_0.5}
	\end{subfigure}
        %\quad
        	% figure B
	\begin{subfigure}[b]{0.48\textwidth}
	\centering
	\includegraphics[width=\textwidth]{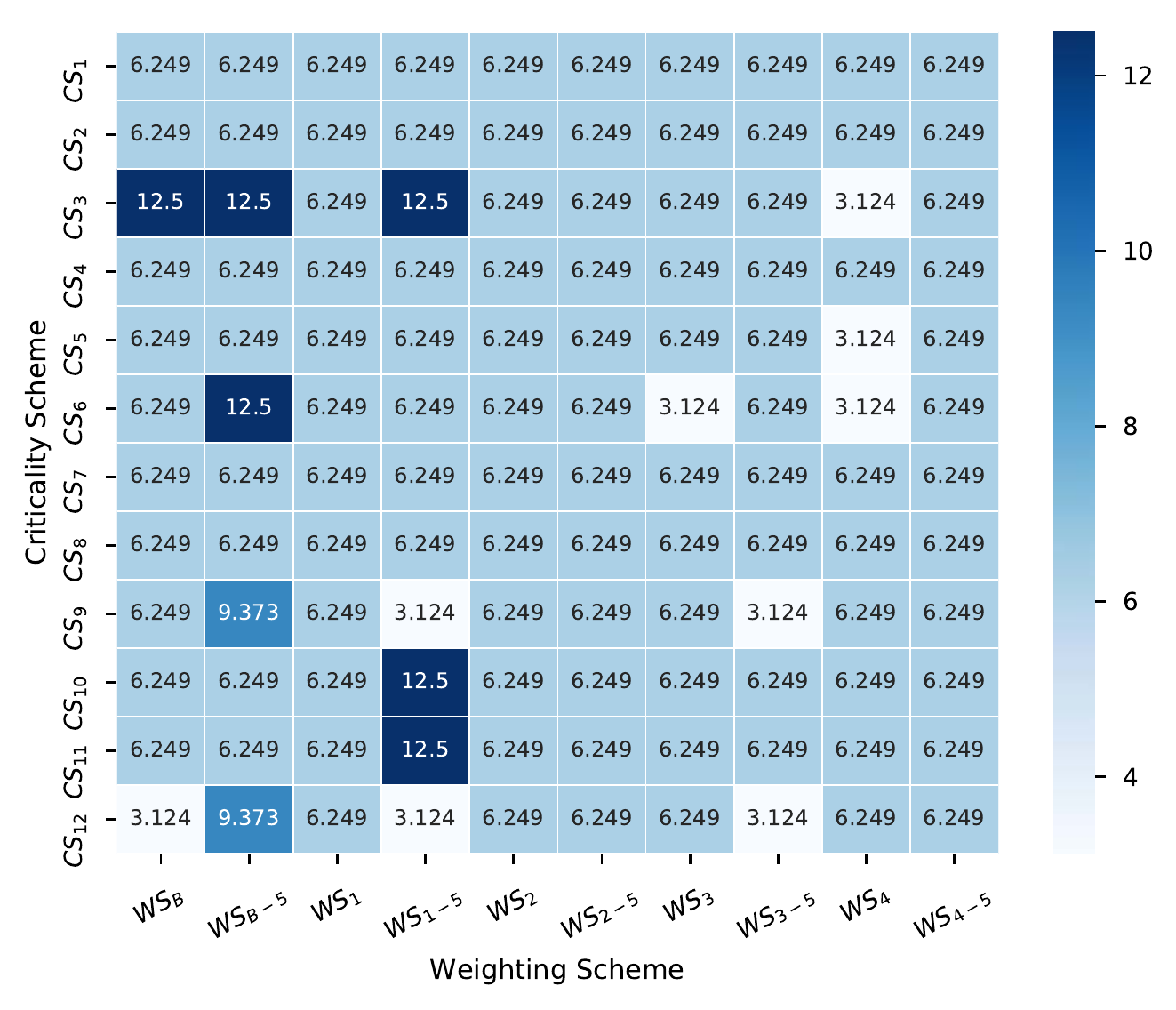}
	\caption{awROCE 0.5\% values}
	\label{fig:vegfr2_awROCE_0.5}
	\end{subfigure}
\caption{ROCE and awROCE 0.5\% values for the VEGFr-2 class}
\end{figure}

\begin{table}[H]
\centering
\resizebox{\textwidth}{!}{
\begin{tabular}{ l l l l c c}
\hline
\thead{Fingerprint} & \thead{Bits} & \thead{Radius} & \thead{Use Features} & \thead{ROCE 0.5\%} & \thead{awROCE 0.5\%}\\
\addlinespace[0.1cm]
\hline
MACCS & N/A & N/A & N/A & \textbf{8.07} & \textbf{6.25} \\
Morgan & 1024.0 & 1.0 & False & \textbf{8.07} & \textbf{6.25} \\
Morgan & 1024.0 & 1.0 & True & \textbf{8.07} & \textbf{6.25} \\
Morgan & 1024.0 & 2.0 & False & \textbf{8.07} & \textbf{6.25} \\
Morgan & 1024.0 & 2.0 & True & \textbf{8.07} & \textbf{6.25} \\
Morgan & 1024.0 & 3.0 & False & \textbf{8.07} & \textbf{6.25} \\
Morgan & 1024.0 & 3.0 & True & \textbf{8.07} & \textbf{6.25} \\
Morgan & 1024.0 & 4.0 & False & \textbf{8.07} & \textbf{6.25} \\
Morgan & 1024.0 & 4.0 & True & \textbf{8.07} & \textbf{6.25} \\
Morgan & 2048.0 & 1.0 & False & \textbf{8.07} & \textbf{6.25} \\
Morgan & 2048.0 & 1.0 & True & \textbf{8.07} & \textbf{6.25} \\
Morgan & 2048.0 & 2.0 & False & \textbf{8.07} & \textbf{6.25} \\
Morgan & 2048.0 & 2.0 & True & \textbf{8.07} & \textbf{6.25} \\
Morgan & 2048.0 & 3.0 & False & \textbf{8.07} & \textbf{6.25} \\
Morgan & 2048.0 & 3.0 & True & \textbf{8.07} & \textbf{6.25} \\
Morgan & 2048.0 & 4.0 & False & \textbf{8.07} & \textbf{6.25} \\
Morgan & 2048.0 & 4.0 & True & \textbf{8.07} & \textbf{6.25} \\
\hline
\end{tabular}
}
\caption{VEGFr-2 class -- Fingerprint results}
\label{tab:vegfr2_fingerprint}
\end{table}

\end{document}